\documentclass[preprint,11pt]{JHEP3} % 10pt is ignored!

\JHEPspecialurl{http://jhep.sissa.it/JOURNAL/JHEP3.tar.gz}

\usepackage{epsfig,multicol,amsmath}

\usepackage{bm}  % for the command \bm (boldface greek letters)
\usepackage{amsmath}     % defines the command \dfrac (fractions in displaystyle)
\usepackage{epsfig,epsf}
\usepackage{cite}

\def\beq{\begin{equation}}
\def\eeq{\end{equation}}
\def\bea{\begin{eqnarray}}
\def\eea{\end{eqnarray}}

\def\eq#1{{Eq.~(\ref{#1})}}
\def\fig#1{{Fig.~\ref{#1}}}
\newcommand{\bas}{\bar{\alpha}_S}

\newcommand{\Lb}{\left(}
\newcommand{\Rb}{\right)}
\newcommand{\h}{\frac{1}{2}}
\parskip=\itemsep               %?

\newcommand{\nn}{\nonumber}

\newcommand{\ga}{\gamma}

\def\pom{{I\!\!P}}

\begin{boldmath}
\title{Azimuthal angle  correlations in hadron-nucleus scattering:  enhanced diagrams}
\end{boldmath}
\author{\Large 
Eugene Levin ${}^{a,b}$  and 
Sebastian Tapia${}^a$\\
 ${}^a$\, Departamento de F\'\i sica,
Universidad T$\acute{e}$cnica Federico Santa Mar\'\i a   and
Centro Cient\'\i fico-Tecnol$\acute{o}$gico de Valpara\'\i so,
Casilla 110-V,  Valparaiso, Chile\\
${}^b$ \, Department of Particle Physics, School of Physics and Astronomy,
Tel Aviv University, Tel Aviv, 69978, Israel\\
}

\abstract
{In this paper we calculate the contribution  to rapidity and angular correlations of the first Pomeron loop diagram  in the dense partonic environment. This diagram is expected to give the largest contribution to the density variation mechanism of  the angular correlations. We show that this diagrams leads to sizable contributions  to the rapidity correlation functions of the order of $\sigma_{in} /\Lb \pi R^2\Rb$ where $\sigma_{in}$ is the inelastic cross section and $R$ is the size of the typical dipole inside the proton  saturation scale. Therefore, the correlations do not depend on the saturation scale. We demonstrated that density variation mechanism does not lead to suppression of the angular dependance of the double inclusive cross section  generating the coefficient in front of $\cos^2 \varphi$ in $A^{1/3}$ larger in the case of hadron-nucleus collision than in hadron-hadron interaction. The angular correlations are suppressed in comparison with the rapidity ones but only due to large multiplicity of the produced gluons. We consider this paper as the first attempt of quantitative description of the density variation mechanism in CGC/saturation approach. }

\dedicated{PACS: 12.38-t, 12.38.Cy,1 2.38.Lg, 13.60.Hd, 24.85.+p, 25.30.Hm}
\preprint{TAUP  \\
{\tt }\\
\today}

\begin{document}
%%%%%%%%%%%%%%%%%%%%%%%%%%%%%%%%%%%%%%%%%%%%%%%%%%%%
\section{ Introduction}
%%%%%%%%%%%%%%%%%%%%%%%%%%%%%%%%%%%%%%%%%%%%%%%%%%%%
One of the most intriguing experimental observation made at the LHC and RHIC, is the same pattern of the azimuthal angle correlations in three type of the interactions: hadron-hadron, hadron-nucleus and nucleus-nucleus collisions. In all three reactions the correlations are observed between two charged hadrons which are separated by the large values of rapidity in the events with large density of the produced particles\cite{CMSPP,STARAA,PHOBOSAA,STARAA1,CMSPA,CMSAA,ALICE}.
We believe that these experiments provide a strong evidence that the underlying physics is the same for all three reactions and it is related the partonic state with high density that has been produced at high energies in all three reactions. Due to causality arguments\cite{CAUSALITY} two hadrons with large difference in rapidity between them could correlate only at  the early stage  of the collision and, therefore,  we expect that the correlations between two particles with large rapidity difference (at least the correlations in rapidity) stem from the partonic state with large parton density.  The parton (gluon) density is governed in QCD by the linear BFKL equation\cite{BFKL,LIREV} which is independent of the type of reaction and leads to the increase of the gluon density. The BFKL evolution describes the emission of gluons at high energy but do not take into account the possible annihilation processes at high energy that stop the density growth and manifest themselves in the gluon saturation\cite{GLR} and in the appearance of the new scale: saturation momentum $Q_s(x)$( where $x$ is the fraction of energy carried by the gluon)\cite{GLR,MUQI,MV,MUCD}. Such dense system of gluons is frequently referred to as the Color Glass Condensate (CGC). For the collision of dilute gluon system with the dense system (say for hadron-nucleus collisions) we can analyze CGC  using the BK-JIMWLK approach \cite{BK,JIMWLK,REV}  which provide us the equations for the non-linear evolution. However, for dilute-dilute scattering (say  hadron-hadron collisions) and for dense -dense scattering (nucleus-nucleus scattering) we can base  our approach to correlations on the  analysis of  large Pomeron loops contribution (see Ref.\cite{AKLL}).

Unlike the rapidity correlations at large values of the rapidity difference which stem from the initial state interactions, the azimuthal angle correlations can be originated by collective flow in the final sate \cite{FINSTATE}.  Nevertheless, in this paper we would like to analyze the same mechanism for both correlations: the initial state interaction in the CGC phase of QCD. However, even in the framework of saturation/CGC approach we are not able  propose the unique mechanism for the azimuthal angle correlations.   At the moment  we have three sources of the azimuthal angle correlation on the market:\footnote{We use classification and terminology suggested in Ref. \cite{KOVT}.
}
Bose enhancement in the wave function\cite{DDGJLR},  local anisotropy\cite{KOLUCOR,KOLUREV} and density variation \cite{LERECOR,RAY}. We cite only the restricted number of papers for each approach. A reader could find more references and more ideas on the origin of the correlation in the review paper of Refs.\cite{KOLUREV,REV1,REV2,REV3,REV4,REV5,REV6}.

\begin{figure}[h]
\begin{center}
 \includegraphics[width=0.4\textwidth]{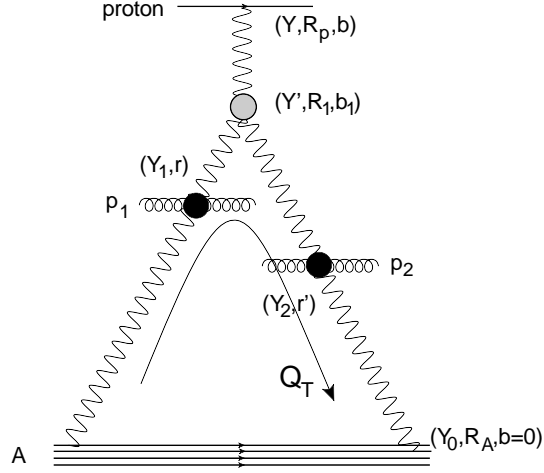}
\end{center}
\caption{ Mueller diagrams\cite{MUDI}: the first fan diagram for two particle correlation.
  Wavy  lines denote the BFKL
 Pomerons.  Helix lines show the gluons. Black blob stands for the Mueller vertex for inclusive production of gluon jet with the transverse momentum $p_{\perp,1}$ ($p_{\perp, 2}$), respectively. }
 \label{1difa}
\end{figure}

The goal of this paper to study in more details the density variation mechanism proposed in Ref.\cite{LERECOR}. In this approach  both rapidity and  azimuthal angle correlations stem from two gluons production  from two parton showers.  This production can be written using Mueller diagrams \cite{MUDI}  (see \fig{1difa} ).  The difference between rapidity and azimuthal angle correlations is only in the form of the Mueller vertices in \fig{1difa}.
 
 %%%%%%%%%%%%%%%%%%%%%%%%%%%%%%%%%%%%%%%%%%%%%%%%%%%%

%%%%%%%%%%%%%%%%%%%%%%%%%%%%%%%%%%%%%%%%%%%%%%%%%%%%%%%%%%%%%%%%%%%%%%%%%%%%%%%%%%%%%%%%%%%%%%%%%%%%%%%%
For rapidity correlations such vertex can be considered  being independent on $Q_T$ while for the
azimuthal angle correlation this vertex is proportional to $\Lb\vec{Q}_T \cdot p_{1,\bot}\Rb^2$ or $
\Lb\vec{Q}_T \cdot p_{2,\bot}\Rb^2$. The integration over the direction of $\vec{Q}_T$  leads to the term $\Lb \vec{p}_{1,\bot}\cdot \vec{p}_{2 \bot}\Rb^2$ which is proportional to $\cos 2 \varphi$ resulting in the azimuthal angle correlations. The strength of the term  $\cos 2 \varphi$ is proportional to $\langle Q_T^2\rangle^2$ where averaging is taken over the wave function of one parton shower which is described by the BFKL Pomeron. In other words since $\vec{Q}_T = i \nabla_b$ where $b$ is the impact factor for the scattering process the magnitude of the azimuthal angle correlation depends on the gradient of the parton density. As it is shown in Ref.\cite{LERECOR}  for hadron-nucleus scattering the diagram of \fig{1difa} generates the following azimuthal angle correlations:
\bea \label{COR}
 R\Lb b,\varphi, Y_1,Y_2\Rb \,\,&=& \,\, \sigma_{in} \frac{d \sigma}{ d Y_1\,d Y_2\,d^2 p_{1 \bot}\,d p_{2\bot}}
 \Bigg{/}\frac{d \sigma}{ d Y_1\,d^2 p_{1 \bot}}\frac{d \sigma}{ d Y_2\,d p_{2\bot}}\,-\,1\nn\\
 &=&p^2_{1\bot}\,p^2_{2\bot} 2\,\Big(\langle \frac{1}{q^4}\rangle|_{\mbox{proton}}\Big)^2\nabla_b^2\nabla^2_b\,S^2_{A}(b) \Lb 2\,+\,\cos 2 \varphi\Rb
\eea
where 
  \beq \label{SA}
    S_A\Lb b \Rb\,\,=\,\,\int d l \, \rho\Lb l, b \Rb  ~~~~~~\mbox{with normalization}~~~\int d^2 b \,S_A\Lb b \Rb\,=\,A
    \eeq
    where   $\rho\Lb l, b\Rb$ is the nucleon density and $l$ is the longitudinal coordinate.
   
   One can see that \eq{COR} gives  small correlations since $\nabla^2 S_A\Lb b\Rb \propto 1/R^2_A$.  In other words the correlation turns out to be small since $Q_T \approx 1/R_A$ in the diagram of \fig{1difa}. In this diagram the variation of the gluon density is reduced to the variation of the density of the nucleons in the nucleus which characterize by the large correlation length of the order of the nucleus radius. On the other hand, in the CGC phase of QCD the natural correlation length is of the order of $1/Q_s \propto 1/A^{1/6}$ which  translates to $Q_T \,\approx \,Q_s$.  The  diagrams, in which we can expect that $Q_T$ will be about $Q_s$ , are the enhanced diagrams\footnote{The importance of enhanced diagram for the azimuthal angle correlations was noted  first in Refs.\cite{KOLUCOR,LERECOR}.}. The simplest 
one is shown in \fig{1di}.  One can see without any detailed calculations that the typical value of $Q_T$ will be equal to $1/r$, $1/R_1$ or $1/R_2$ , all of which are not related to the radius of the nucleus.   
   
   In this paper we will calculate the simplest enhanced diagrams for the hadron-nucleus and hadron-hadron interactions
   and will try to understand the main features of this diagram which would affect the azimuthal angle correlations.  In section 2 we introduce the main ingredients and calculate the first  Pomeron loop diagram for the hadron-hadron interaction. We show that this calculation are quite different from the calculation of the same diagram but for its contribution to the total cross section. We discuss in details the integration over the loop momentum which leads to the same values of the dipole sizes in the triple Pomeron vertices. Section 3 is devoted to the calculation of Green's function of the BFKL Pomeron in the saturation environment. The equations are written and solved. In section 4 we summarize our calculations for hadron-nucleus scattering. In conclusions we present our main result and discussed their naturalness.
    
\section{Correlations: first enhanced diagrams contribution}
\begin{figure}[h]
\begin{minipage}{11cm}
\begin{center}
 \includegraphics[width=0.40\textwidth]{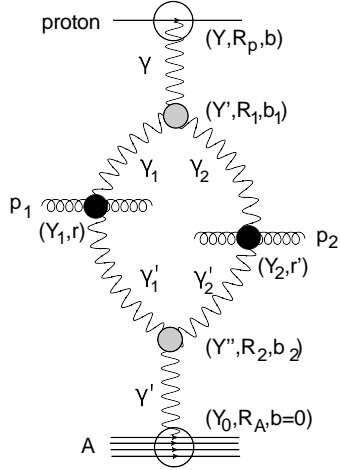}
\end{center}
\end{minipage}
 \begin{minipage}{6.5cm}
\caption{ The first enhanced (loop) diagram for two particle correlation.
  Wavy  lines denote the BFKL
 Pomerons.  Helix lines show the gluons. Black blob stands for the Mueller vertex ($a_\pom$)  for inclusive production of gluon jet with the transverse momentum $p_{\perp,1}$ ($p_{\perp, 2}$), respectively.  The gray blobs show the triple Pomeron vertices ($G_{3\pom}$ ). The vertices of interaction with the proton and the nucleus  ($  g_p$ and $g_A$) are shown by circles.}
 \label{1di}
\end{minipage}
\end{figure}

We start the analysis of the azimuthal angle correlation considering the first enhanced diagram, shown in \fig{1di}.
As we have discussed in the introduction we are looking for the mechanism of correlation based on large densities gradients\cite{LERECOR,KOVT} and the enhanced diagrams lead to the gradients of the order of $Q_s$. Calculating the first loop diagram we wish to demonstrate this fact and  to discuss the scale of the correlations.

The contribution of \fig{1di} to the double inclusive cross section for scattering of two dipole with the sizes $R_p$ and $R_A$\footnote{We need to stipulate that $R_A$ is not the radius of a nucleus but the size of the typical dipole inside the nucleus which is of the same order as $R_p$.}, takes the form\cite{BRN} (see \fig{1di} for notations):
\bea 
&&\frac{ p^2_{\bot,1}  p^2_{\bot,2} d \sigma}{d Y_1 d Y_2 d^2 p_{\bot,1} d^2p_{\bot,2}}\,\,= \,\,\frac{4 \pi^2 \bas^4}{N^2_c}\int d^2  b \int d^2 b_1 \int d^2 b_2 \int ^Y_{Y_1} d Y' \int^{Y_2}_{Y_0} d Y'' \label{1DI}\\
&&\int \frac{d^2 R_1}{R^4_1} \int \frac{ d^2 R_2}{R^4_2}\,N_\pom\Lb Y - Y'; R_p,R_1,\vec{b} - \vec{b}_1\Rb  \int d^2 R'_1\int d^2 R'_2 \,\, K\Lb R_1,R'_1\Rb\nn\\
&& N^{\mbox{\tiny incl}}_\pom\Lb 
Y' - Y", R'_1, R'_2, \vec{b}_1 - \vec{b}_2| p_{\bot,1}, Y_1\Rb \, N^{\mbox{\tiny incl}}_\pom\Lb 
Y' - Y", \vec{R}_1 - \vec{R}'_1, \vec{R}_2 - \vec{R}'_2, \vec{b}_1 - \vec{b}_2| p_{\bot,2}, Y_2\Rb   \nn\\
&& K\Lb R_2,R'_2\Rb\,N_\pom\Lb Y'' - Y_0; R_2,R_A,\vec{b}_2\Rb \nn
 \eea
where $N_\pom\Lb Y ; r_1, r_2,\vec{b} \Rb$ is the amplitude of the dipole-dipole scattering with the dipole sizes $r_1$ and $r_2$, at  \,\,rapidity $Y$ and at impact parameter $b$ due to the exchange of the BFKL Pomeron \cite{BFKL,LIREV,REV}. $ N^{\mbox{\tiny incl}}_\pom\Lb 
Y, r_1, r_2, \vec{b}| p_{\bot}, Y_1\Rb$ is the BFKL Pomeron contribution to  the inclusive cross section of the gluon jet  production with rapidity $Y_1$ and transverse momentum $p_\bot$ in the dipole-dipole scattering of two dipoles with the sizes $r_1$ and $r_2$ at rapidity Y and at impact parameters $b$. The triple Pomeron vertex $K\Lb r,r'\Rb$ takes a familiar form:
\beq \label{K}
K\Lb r,r'\Rb\,\,\,=\,\,\frac{r^2}{r'^2\,\Lb \vec{r}\,-\,\vec{r}^{\,'}\Rb^2}
\eeq

%%%%%%%%%%%%%%%%%%%%%%%%%%%%%%%%%%%%%%%%%%%%%%%%%%%%

%%%%%%%%%%%%%%%%%%%%%%%%%%%%%%%%%%%%%%%%%%%%%%%%%%%%
All ingredients of \eq{1DI} we consider separately below, starting with $N_\pom$.

%%%%%%%%%%%%%%%%%%%%%%%%%%%%%%%%%%%%%%%%%%%%%%%%%

\subsection{The BFKL Pomeron: generalities}
%%%%%%%%%%%%%%%%%%%%%%%%%%%%%%%%%%%%%%%%%%%%%%%%%%%%
The general solution to the BFKL equation for the scattering amplitude of two dipoles with the sizes $r_1$ and $r_2$ has been derived in Ref.\cite{LIREV} and it takes the form
\bea \label{GENSOL}
&&N_\pom\Lb r_1,r_2; Y, b \Rb\,\,= \\
&&\,\,\sum_{n=0}^{\infty}
\int \frac{d \gamma}{2\,\pi\,i}\,\phi^{(n)}_{in}(\gamma; r_2)
\,\,d^2\, R_1 \,\,d^2\,R_2\,\delta(\vec{R}_1 - \vec{R}_2 - 
\vec{b})\,
e^{\omega(\gamma, n )\,Y}
\,E^{\gamma,n}\Lb r_1, R_1\Rb\,E^{1 - \gamma,n}\Lb r_2, R_2 \Rb\nn
\eea
with
\beq \label{OMEGA}
\omega(\gamma, n)\,\,=\,\,\bas \chi(\gamma, n)\,\,  =\,\,\bas \Lb 2 \psi\Lb 1\Rb \,-\,\psi\Lb \gamma + |n|/2\Rb\,\,-\,\,\psi\Lb 1  - \gamma + |n|/2\Rb\Rb;
\eeq
where $~\psi\Lb \gamma\Rb \,\,=\,\,d \ln \Gamma\Lb \gamma\Rb/d \gamma $ and $\Gamma\Lb \gamma\Rb$ is  Euler gamma function. Functions $ E^{n, \gamma} \Lb \rho_{1a},\rho_{2a}\Rb$ are given by the following equations.
\begin{align}\label{EFUN}
  E^{n, \gamma} \Lb \rho_{1a},\rho_{2a}\Rb \,=\, \Lb
  \frac{\rho_{12}}{\rho_{1a} \, \rho_{2a}}\Rb^{1 - \gamma + n/2}
  \, \Lb \frac{\rho^*_{12}}{\rho^*_{1a} \, \rho^*_{2a}}
  \Rb^{1 - \gamma - n/2},
\end{align}
In \eq{EFUN} we use the complex numbers to characterize the point on the plane
\begin{align}\label{COMNUM}
  \rho_i = x_{i,1} + i \, x_{i,2};\,\,\,\,\,\,\, \rho^*_i = x_{i,1} -  i \, x_{i,2}
\end{align}
where the indices $1$ and $2$ denote  two transverse axes. Notice that
\beq \label{NOT}
\rho_{12}\,\rho^*_{12}\,\,=\,\,r^2_i ;~~~~~~\rho_{1 a}\,\rho^*_{1 a}\,=\,\Lb\vec{R}_i\,-\,\frac{1}{2}\vec{r}_{i}\Rb^2~~~~~~\rho_{2 a}\,\rho^*_{2a}\,=\,\Lb\vec{R}_i\,+\,\frac{1}{2}\vec{r}_{i}\Rb^2
\eeq
At large values of $Y$ the main contribution stems from the first term with $n =0$.  For this term \eq{EFUN} can be re-written in the form
\beq \label{E}
E^{\gamma,0}\Lb r_i,R_i\Rb \,\,=\,\,\left( \,\frac{r^2_{i}}{(\vec{R}_i
\,+\,\frac{1}{2}\vec{r}_{i})^2\,\,
(\vec{R}_i\,-\,\frac{1}{2}\vec{r}_{i})^2}\,\right)^{1 - \gamma}\,\,.
\eeq

The integrals over $R_1$ and $R_2$ were taken in Refs.\cite{LIREV,NAPE} and at $n=0$ we have
\bea \label{H}
&&H^\ga\Lb w, w^*\Rb\,\,\equiv\,\,\int d^2\,R_1\,E^{\ga,0}\Lb r_{1},R_1\Rb\, E^{1 - \ga, 0}\Lb r_{2}, \vec{R}_1
\,-\,\vec{b}\Rb\,= \\
&&
\,\frac{ (\gamma - \h)^2}{( 
\gamma (1 - \gamma)
)^2} \Big\{b_\ga\,w^\gamma\,{w^*}^\gamma\,F\Lb\gamma, \gamma, 2\gamma, w\Rb\,
F\Lb\gamma, \gamma, 2\gamma, w^*\Rb
\,+ \nn\\
&&  b_{1 - \ga} w^{1 -
\gamma}{w^*}^{1-\gamma}
F\Lb 1 - \gamma, 1 -\gamma, 2 - 2\gamma, w\Rb\,F\Lb 1 - \gamma,1 -\gamma,2 
-2\gamma, w^*\Rb \Big\}\nn
\eea
where $F$ is hypergeometric function \cite{RY}. In  \eq{H}
$w\,w^*$ is equal to
\beq \label{W}
w\,w^*\,\,=\,\,\frac{r^2_{1}\,r^2_{2}}{\Lb\vec{b} - \h\Lb\,\vec{r}_{1}\,
- \,\vec{r}_{2}\Rb\Rb^2
\,\Lb\vec{b} \,+\, \h \Lb\,\vec{r}_{1} \,- \,\vec{r}_{2}\Rb\Rb^2}
\eeq
and  $b_\ga$ is equal to
\beq \label{BGA}
b_{\ga} \, = \, \pi^3 \, 2^{4(1/2 - \ga)} \, \frac{\Gamma \Lb\ga \Rb}{\Gamma \Lb 1/2 - \ga \Rb}
  \, \frac{\Gamma \Lb 1 - \ga  \Rb}{\Gamma \Lb 1/2 + \ga \Rb}.
\eeq

Finally, the solution at large $Y$ takes the form
\beq \label{FINSOL}
N_\pom\Lb r_1,r_2; Y, b \Rb\,\,= \,\,
\int \frac{d \gamma}{2\,\pi\,i}
\,
e^{\omega(\gamma, 0 )\,Y}\,H^\ga\Lb w, w^*\Rb
\eeq

In the vicinity of the saturation scale $N_\pom$ takes the form (see Refs. \cite{MUTR,IIM})
\bea 
&&N_\pom\Lb r_1,r_2; Y, b \Rb\,\,= \,\,\frac{\Lb \ga_{cr} - \h\Rb^2}{\ga_{cr} ( 1 - \ga_{cr})}\,b_{\ga_{cr}}\Big(w w^* e^{\kappa Y})^{1 - \gamma_{cr}}\nn\\
&&\,\,=\,\,\frac{\Lb \ga_{cr} - \h\Rb^2}{\ga_{cr} ( 1 - \ga_{cr})}\,b_{\ga_{cr}}\Lb \Lb \frac{r^2_{1}\,r^2_{2}}{\Lb\vec{b} - \h\Lb\,\vec{r}_{1}\,
- \,\vec{r}_{2}\Rb\Rb^2
\,\Lb\vec{b} \,+\, \h \Lb\,\vec{r}_{1} \,- \,\vec{r}_{2}\Rb\Rb^2}\Rb e^{\bas \frac{\chi\Lb \ga_{cr}\Rb}{1 - \ga_{cr}}\,Y}\Rb^{1 - \gamma_{cr}} \label{GSNB}\\
&&\xrightarrow{r_2 \gg r_1}~~~~ \phi_0\,\Big( r^2_1 Q^2_s\Lb r_2,b; Y\Rb\Big)^{1 - \gamma_{cr} }~~~~\mbox{with}~~~~~Q^2_s\Lb r_2,b; Y\Rb\,\,= \frac{\,r^2_{2}\,e^{\bas \frac{\chi\Lb \ga_{cr}\Rb}{1 - \ga_{cr}}\,Y}}{\Lb\vec{b} - \h\vec{r}_{2}\Rb^2
\,\Lb\vec{b} \,+\, \h \vec{r}_{2}\Rb^2}\label{GSNB1}
\eea
where (see Refs.\cite{GLR,MUTR,MUPE})
\beq \label{GACR}
\frac{\chi\Lb \ga_{cr}\Rb}{1 - \ga_{cr}}\,\,=\,\,- \frac{d \chi\Lb \ga_{cr}\Rb}{d \ga_{cr}}
\,\,\,\,\,\mbox{where}\,\,\,\,\,\,\chi\Lb \ga\Rb\,=\,2 \psi\Lb 1 \Rb \,-\,\psi\Lb \ga\Rb \,-\,\psi\Lb 1 - \ga\Rb \,\leftarrow \mbox{kernel of the BFKL equation}
\eeq
We denote  below  by $\bar{\gamma} = 1 - \gamma_{cr}$ and  will use \eq{GSNB} and \eq{GSNB1} in the momentum transferred   representation,viz.
\beq \label{QRE}
N_\pom\Lb r_1,r_2; Y, Q_T \Rb\,\,\,=\,\,\int d^2 b \,\,e^{i \vec{Q}_T \cdot \vec{b}}\,\,\,N_\pom\Lb r_1,r_2; Y, b \Rb\,
\eeq

Integral of \eq{QRE} with $ N_\pom\Lb r_1,r_2; Y, b \Rb$ from \eq{GSNB}  can be taken using the complex number description for the point on the plane (see \eq{COMNUM} and \eq{NOT}).   The integral takes the form\cite{LIREV,NAPE}
\beq \label{NPQ10}
N_\pom\Lb r_1,r_2; Y, Q_T \Rb =\,\Lb r^2_1\,r^2_2 \Rb^{\bar{\gamma}} \,e^{\bas \, \chi\Lb \ga_{cr}\Rb\,Y}\int d \rho_b  \,\,e^{i \rho^*_Q \rho_b}\Lb  \frac{1}{\rho^2_b - \rho^2_{12}}\Rb^{\gamma}\int d \rho^*_b  \,\,e^{i \rho_Q \rho^*_b} \Lb \frac{1}{\rho^{*2}_b - \rho^{*2}_{12}}\Rb^{\gamma}
\eeq

%%%%%%%%%%%%%%%%%%%%%%%%%%%%%%%%%%%%%%%%%%%%%%%%%%%%
\FIGURE[h]{
\begin{minipage}{7cm}{
\centerline{\epsfig{file=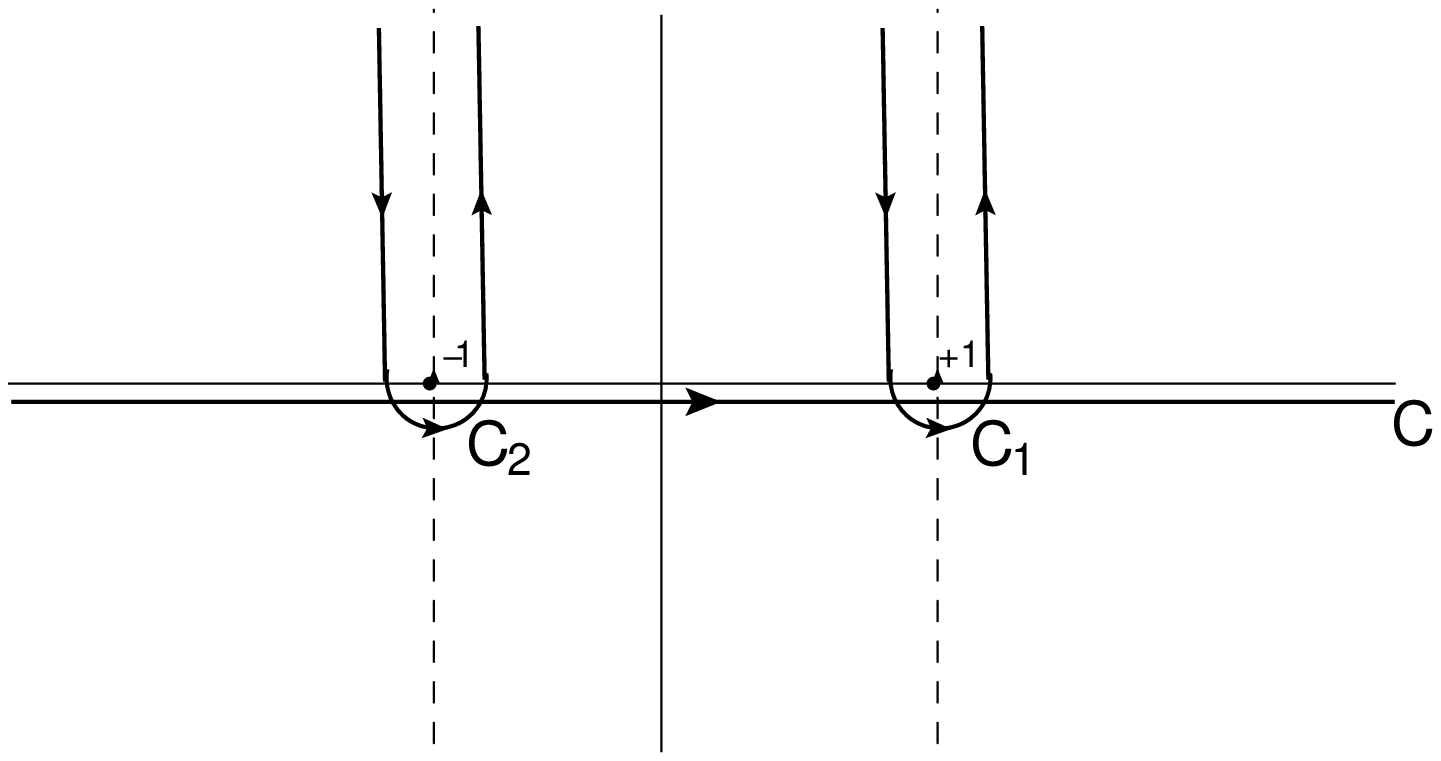,width=60mm}}
\caption{ Contours of  integration in \protect\eq{HAN}.}
 \label{cont}  }
 \end{minipage}
}

%%%%%%%%%%%%%%%%%%%%%%%%%%%%%%%%%%%%%%%%%%%%%%%%%%%%

Using new variables $t= \rho_b/\rho_{12}$ and $t^{*} =  \rho^{*}_b/\rho_{12}$ and the integral representation of Hankel
functions (see formulae {\bf 8.422(1,2)} in Ref. \cite{RY})
\beq \label{HAN}
H^{(1,2)}_\nu\Lb z \Rb\,\,=\,\,\frac{\Gamma\Lb \h - \nu\Rb}{\pi i \Gamma\Lb \h\Rb} \Lb \h z \Rb^\nu \oint_{C_{1,2}}d t e^{i z t}\Lb t^2 - 1\Rb^{\nu - \h}
\eeq
where contours $C_1$ and $C_2$ are shown in \fig{cont}, we obtain that
\beq\label{NPQ1}
N_\pom\Lb r_1,r_2; Y, Q_T \Rb \,\,=\,\,C^2(\gamma) \,r^2_{12}  \,e^{\bas \, \chi\Lb \ga\Rb\,Y}
\,\Lb\frac{ r^2_1 r^2_2}{r^4_{12}}\Rb^{\gamma}\,\Lb Q^2 r^2_{12}\Rb^ { - \h + \gamma} ~
J_{\h - \gamma}\Lb \rho^*_Q \rho_{12}\Rb J_{\h - \gamma}\Lb \rho_Q \rho^{*}_{12}\Rb
\eeq
where  $ 2 J_\nu(z) = H^{(1)}\Lb z\Rb + H^{(2)}_\nu\Lb z \Rb$; $\vec{r}_{12} = \h\Lb \vec{r}_1 - \vec{r}_2\Rb$ and 
\beq \label{CGA}
C(\gamma)\,\,=\,\,2^{-\frac{5}{2} + \gamma} \pi  \frac{\Gamma\Lb \h \Rb}{\Gamma\Lb  \gamma\Rb  }
\eeq
Two limits will be useful for further presentation:
\bea
N_\pom\Lb r_1,r_2; Y, Q_T \Rb \,\,&\xrightarrow{Q_T \to 0}&\,\,\,C^2(\gamma)\,r^2_{12}  \,e^{\bas \, \chi\Lb \ga\Rb\,Y}
\,\Lb\frac{ r^2_1 r^2_2}{r^4_{12}}\Rb^{\gamma}\label{NPQ2}\\
&\xrightarrow{Q^2_T\,r^2_{12}\,\,\gg\,\,1 }&\,\,\frac{2}{\pi}\,C^2(\gamma) \,r^2_{12}  \,e^{\bas \, \chi\Lb \ga\Rb\,Y}
\,\Lb\frac{ r^2_1 r^2_2}{r^4_{12}}\Rb^{\gamma}\,\Lb Q^2 r^2_{12}\Rb^ { - 1 + \gamma} \cos^2\Lb \pi \gamma/2\Rb e^{i \vec{Q} \cdot \vec{r}_{12}}\label{NPQ3}
\eea

\eq{NPQ2} can be re-written at $r_1 \ll r_2$ in the form :
\bea \label{NPQ4}
N_\pom\Lb r_1,r_2; Y, Q_T \Rb \,\,&\xrightarrow{Q_T \to 0, r_1 \,\ll\,r_2}&\,\,\,C^2(\gamma)\,r^2_{12} \Lb r^2_1 Q^2_s\Lb Y, r_2\Rb \Rb^{ \gamma} ~~~~\mbox{with}~~Q^2_s\,\,=\,\,\frac{1}{r^2_2}\,e^{\bas \,\frac{ \chi\Lb \ga\Rb}{\gamma}\,Y};\nn\\
& \xrightarrow{Q_T  r_2\,\gg\,1, r_1 \,\ll\,r_2}&\,\,\frac{2}{\pi}\,
C^2(\gamma) \,\cos^2\Lb \pi \gamma/2\Rb e^{i \vec{Q} \cdot \vec{r}_{12}}  \,e^{\bas \, \chi\Lb \ga\Rb\,Y}\frac{1}{Q^2_T}\Lb Q^2_T \,r^2_1\Rb ^{\gamma}
\eea

One can see   that we can write a simple interpolation formula which we will bear in our mind in our estimates.
\beq \label{NPQ5}
\hspace{-0.3cm}N_\pom\Lb r_1,r_2; Y, Q_T \Rb = C^2(\gamma)\,r^2_{12} \Lb r^2_1 Q^2_s\Lb Y, r_2\Rb \Rb^{ \gamma} \,\Lb 1 \,+\,a(\gamma)\,Q^2_T \,r^2_{12}\Rb^{-1 + \gamma}~\mbox{with}~a^{-1 + \gamma}\Lb \gamma\Rb= \frac{2}{\pi} \cos^2\Lb \pi \gamma/2\Rb 
\eeq
For $N_\pom$ in the vicinity of the saturation scale $\gamma\,=\,\bar{\gamma} = 1 - \gamma_{cr}$.

 %%%%%%%%%%%%%%%%%%%%%%%%%%%%%%%%%%%%%%%%%%%%%%%%%%%%   
    \subsection{Inclusive production in dipole-dipole scattering: the BFKL Pomeron contribution}
        %%%%%%%%%%%%%%%%%%%%%%%%%%%%%%%%%%%%%%%%%%%%%%%%%%%%%%%%%
    In this subsection we calculate the cross section of the inclusive production of gluon jet with the transverse momentum $p_\bot$ at rapidity $Y_1$ in the collision of two dipoles with the sizes $r_1$ and $r_2$ at rapidity $Y$ and at impact parameter $b$.  The general formula which shows the $k_T$-factorization \cite{KTF}, has been derived in Ref.\cite{KTINC} and it takes the form
    \bea \label{INC}
\frac{d \sigma}{d^2 b\, d Y_1 \,d^2 p_{\bot}}&= & \\
& & \frac{2C_F}{\alpha_s (2\pi)^4}\,\frac{1}{p^2_\bot}\int \!\!d^2 \vec B \,d^2 \vec r_\bot\,e^{i \vec{p}_\bot\cdot \vec{r}_\bot}\,\,\nabla^2_\bot\,N^G\Lb Y_1; r_\bot,r_{1}; b \Rb\,\,\nabla^2_\bot\,N^G\Lb  Y - Y_1; r_\bot, r_2; |\vec b-\vec B| \Rb\nn
\eea
  where 
  \beq \label{INC1}
N^G\Lb Y; r_\bot,r_i; b \Rb\,\,=\,\,2 \,N\Lb Y; r_\bot , r_i ; b \Rb\,\,-\,\,N^2\Lb Y; r_\bot, r_i; b \Rb,
\eeq  
   Evaluating \eq{1DI} we need to know such cross section only for the BFKL Pomeron exchange        for which \eq{INC1} reduces to the following equation
   \beq \label{INC2}
   N^G_\pom \Lb Y; r_\bot,r_i; b \Rb\,\,=\,\,2 \,N_\pom\Lb Y; r_\bot , r_i ; b \Rb   
   \eeq
  Plugging \eq{INC2} in \eq{INC} we have
  \bea \label{INC3}
 &&  N^{\mbox{\tiny incl}}_\pom\Lb Y, r_1, r_2, b, p_{\bot}, Y_1\Rb = \\
   && \frac{8\,C_F}{\alpha_s (2\pi)^4}\,\frac{1}{p^2_\bot}\int \,d^2 \vec B \,d^2 \vec r_\bot\,e^{i \vec{p}_\bot\cdot \vec{r}_\bot}\,\,\nabla^2_\bot\,N_\pom\Lb Y_1; r_\bot,r_{1}; B \Rb\,\,\nabla^2_\bot\,N_\pom\Lb  Y - Y_1; r_\bot, r_2; |\vec b-\vec B| \Rb\nn
 \eea 
 It is worthwhile mentioning that $b$ is the difference of the impact parameters between scattering dipoles while $B$ is the impact parameter of the produced gluon with respect to  the dipole with size $r_1$.
 
 In the vicinity of the saturation scale $N_\pom$ takes the form of \eq{GSNB} and $ \nabla^2_\bot\,N_\pom\Lb Y_1; r_\bot,r_{1}; b \Rb$ is equal to
 \bea \label{INC4}
\nabla^2_\bot\,N_\pom\Lb Y_1; r_\bot,r_{1}; b \Rb \,&=&\\
&=&\,\phi_0\,\bar{\gamma}^2\,\Lb  \frac{r^2_{1}\,r^2_{2}}{\Lb\vec{b} - \vec{r}_{12}\Rb^2
\,\Lb\vec{b} \,+\, \vec{r}_{12}\Rb^2} e^{\bas \frac{\chi\Lb \ga_{cr}\Rb}{1 - \ga_{cr}}\,Y}\Rb^{\bar{\gamma}}\,\Lb \frac{2 \vec{r}}{r^2} \,-\, \frac{\vec{b} - \vec{r}_{12}}{\Lb\vec{b}\, -\, \vec{r}_{12}\Rb^2}\,+\, \frac{\vec{b} \,+\, \vec{r}_{12}}{\Lb\vec{b} \,+\, \vec{r}_{12}\Rb^2}\Rb^2\nn
\eea

However, it turns out that it is more convenient to use $ \nabla^2_\bot\,N_\pom\Lb Y_1; r_\bot,r_{1}; b \Rb$ in momentum representation, namely,
\beq \label{INC5}
\int d^2 b e^{i \vec{Q}_T \cdot \vec{b}}~\nabla^2_\bot\,N_\pom\Lb Y_1; r_\bot,r_{1}; b \Rb \,=\,\nabla^2_\bot\,N_\pom\Lb Y_1; r_\bot,r_{1}; Q_T \Rb  
\eeq

Using \eq{NPQ1} for $N_\pom\Lb Y_1; r_\bot,r_{1}; Q_T \Rb  $ and $\nabla^2_r\,\,=\,\,4 \partial_\rho \partial_{\rho^*}$ we obtain( denoting $r_\bot \equiv r_0$)
\bea \label{INC6}
\nabla^2_\bot\,N_\pom\Lb Y_1; r_\bot,r_{1}; Q_T \Rb\,&=&\,4\,\phi_0\,C^2(\gamma) \,r^2_{01}  \,e^{\bas \, \chi\Lb \ga\Rb\,Y}
\,\Lb\frac{ r^2_1 r^2_0}{r^4_{01}}\Rb^{\gamma}\,\Lb Q^2 r^2_{01}\Rb^ { - \h + \gamma} \\
&\times& \Bigg\{ \Big(\frac{\gamma}{\rho_r} - \frac{\h + \gamma}{\rho_{01}} \Big)\,J_{\h - \gamma}\Lb\rho^*_Q \rho_{01}\Rb + \h\rho^*_Q\Lb J_{-\h - \gamma}\Lb \rho^*_Q \rho_{01}\Rb\,-\,J_{-\frac{3}{2} - \gamma}\Lb \rho^*_Q \rho_{01}\Rb\Rb\Bigg\}\nn\\
&\times& \Bigg\{\Big( \frac{\gamma}{\rho^*_r} - \frac{\h + \gamma}{\rho^*_{01}} \Big)\, J_{\h - \gamma}\Lb\rho_Q \rho^*_{01}\Rb +  \h\rho_Q\Lb J_{-\h - \gamma}\Lb \rho_Q \rho^*_{01}\Rb\,-\,J_{-\frac{3}{2} - \gamma}\Lb \rho_Q \rho^*_{01}\Rb\Rb\Bigg\}\nn
\eea   

We need  to estimate
\beq \label{INC7}
N^{\mbox{\tiny incl}}_\pom\Lb Y, r_1, r_2, Q_T, p_{\bot}, Y_1\Rb \,=\,\int d^2 b ~e^{i \vec{Q}_T \cdot \vec{b}}\,N^{\mbox{\tiny incl}}_\pom\Lb Y, r_1, r_2, b, p_{\bot}, Y_1\Rb
\eeq
 for calculation of the diagrams of \fig{1di}.  From \eq{INC3} and \eq{INC6} we obtain
 \bea\label{INC8}
&& N^{\mbox{\tiny incl}}_\pom\Lb Y, r_1, r_2, Q_T, p_{\bot}, Y_1\Rb \,=\\
  &&~~~~~~~~~~~~~~~~~ \, \frac{8\,C_F}{\alpha_s (2\pi)^4}\,\frac{1}{p^2_\bot}\int d^2 \vec r_\bot\,e^{i \vec{p}_\bot\cdot \vec{r}_\bot}\,\,\nabla^2_\bot\,N_\pom\Lb Y_1; r_\bot,r_{1}; Q_T \Rb\,\,\nabla^2_\bot\,N_\pom\Lb  Y - Y_1; r_\bot, r_2; Q_T \Rb\nn~
 \eea
 For $Q_T \to 0$ \eq{INC7} reduces to the following equation (see \eq{INC6})
 \bea \label{INC9}
 && N^{\mbox{\tiny incl}}_\pom\Lb Y, r_1, r_2, Q_T=0, p_{\bot}, Y_1\Rb \,=\,\\
 &&~~~~~~~~~~~~~~~~~~~~~~~~=
 \, \frac{8\,C_F}{\alpha_s (2\pi)^4}\,C^4\Lb \bar \ga\Rb \bar{\ga}^2\,\frac{\phi^2_0}{p^2_\bot}\int \frac{d^2  r_\bot}{r^4_\bot}\,e^{i \vec{p}_\bot\cdot \vec{r}_\bot}\,\,r^2_1\,r^2_2\Big(r^2 \,Q^2_s\Lb r_1, Y - Y_1\Rb\Big)^{\bar \ga} \Big(r^2 \,Q^2_s\Lb r_2, Y_1\Rb\Big)^{\bar \ga}\nn\\
  &&~~~~~~~~~~~~~~~~~~~~~~~~=
 \, \frac{8\,C_F}{\alpha_s (2\pi)^4}\,\frac{\bar{\phi^2_0}}{p^2_\bot}\int \frac{d^2  r_\bot}{r^4_\bot}\,e^{i \vec{p}_\bot\cdot \vec{r}_\bot}\,\,r^2_1\,r^2_2\Big(r^2 \,Q^2_s\Lb r_1, Y - Y_1\Rb\Big)^{\bar \ga} \Big(r^2 \,Q^2_s\Lb r_2, Y_1\Rb\Big)^{\bar \ga}\nn\\ &&~~~~~~~~~~~~~~~~~~~~~~~~~~=
 \,\bar{\phi^2_0}\frac{ 8\,C_F}{\alpha_s (2\pi)^4}\,2^{-3 +4 \bar{\ga}}\,r^2_1\,r^2_2\,\frac{\Gamma\Lb 1 - 2 \bar{\ga}\Rb}{\Gamma\Lb 2 - 2 \bar{\ga}\Rb }\Big(\frac{Q^2_s\Lb r_1, Y - Y_1\Rb}{p^2_\bot}\Big)^{\bar \ga} \Big(\frac{Q^2_s\Lb r_2, Y_1\Rb}{ p^2_\bot}\Big)^{\bar \ga}\nn\\ 
  &&\mbox{where} ~Q^2_s\Lb r_1, Y - Y_1\Rb \,\,=\,\,\frac{1}{r^2_1}\exp\Big( \bas \frac{\chi\Lb \gamma_{cr}\Rb}{1 - \gamma_{cr}}\, \Lb Y  - Y_1\Rb \Big) ~ \mbox{and}~ Q^2_s\Lb r_1,  Y_1\Rb \,\,=\,\,\frac{1}{r^2_2}\exp\Big( \bas \frac{\chi\Lb \gamma_{cr}\Rb}{1 - \gamma_{cr}}\, Y_1 \Big) \nn
 \eea
  
  %%%%%%%%%%%%%%%%%%%%%%%%%%%%%%%%%%%%%%%%%%%%%%%%%%% 
 \subsection{Inclusive production in dipole-dipole scattering:  non-linear equation}
 %%%%%%%%%%%%%%%%%%%%%%%%%%%%%%%%%%%%%%%%%%%%%%%%%%% 
 The main features of the inclusive production which we will use in our calculation of the first loop diagram, is  the suppression of this production inside the saturation region.  It follows directly from the general expression of \eq{INC} since
 $N^G\Lb Y_1; r_\bot,r_{1}; b \Rb\,\xrightarrow{ r^2_\bot Q^2_s \gg 1} \,1$ while  $\nabla^2_\bot N^G\Lb Y_1; r_\bot,r_{1}; b \Rb\,\xrightarrow{ r^2_\bot Q^2_s \gg 1} \,0$. Honestly, these features
  do not appear in the simple diagram of \fig{1di}, they manifest themselves only  in the diagrams of \fig{1dienv}
  which describes the Pomeron loop in the dense environment.  However,   we believe that it is needed to discuss
  these features now since they are essential for our calculations.
  
 In  the saturation  region the amplitude shows the geometric scaling behaviour \cite{GS} and the solution of the BFKL equation deeply inside of this region takes the following form\cite{LT}
 
 \beq \label{INC10}
 N\Lb z = \ln\Lb r^2 Q^2_s\Rb\Rb\,\,=\,\,1 \,\,-\,\,\exp\Big( - \frac{z^2}{2 \,\kappa}\,\Big);~~ N^G\Lb z = \ln\Lb r^2 Q^2_s\Rb\Rb\,\,=\,\,1 \,\,-\,\,\exp\Big( - \frac{z^2}{ \kappa}\Big)  
 \eeq 
 with $\kappa =  \frac{\chi\Lb \gamma_{cr}\Rb}{1 - \gamma_{cr}} $  and 
 \beq \label{INC11}
-\,\nabla^2_\bot  N^G\Lb z = \ln\Lb r^2 Q^2_s\Rb\Rb \,\,=\,\,\frac{8 (-\kappa \,+\,2 z^2)}{\kappa^2}\exp\Big( - \frac{z^2}{\kappa}\Big)
\eeq

One can see that $\nabla^2_\bot  N^G\Lb z = \ln\Lb r^2 Q^2_s\Rb\Rb$ falls down and only $z \,\approx \,1$ is essential.

In the vicinity of the saturation scale $N \,=\,1 - \exp\Big( - \phi_0 \Lb  r^2 Q^2_s\Rb^{1 - \gamma_{cr}}\Big)$ (see Ref. \cite{LT})  which leads to
\beq \label{INC12}
\,\nabla^2_\bot  N^G\Lb  T =  2  \phi_0  \Lb r^2 Q^2_s\ \Rb^{1 - \gamma_{cr}}\Rb\,\,=\,\,\Lb 1 - \gamma_{cr}\Rb^2\frac{4 T}{r^2}\Lb T - \h \frac{1 - 2 \ga_{cr}}{1 - \ga_{cr}}\Rb e^{-T}
\eeq
%%%%%%%%%%%%%%%%%%%%%%%%%%%%%%%%%%%%%%%%%%%%%%%%
\begin{figure}[h]
\begin{center}
 \includegraphics[width=0.5 \textwidth]{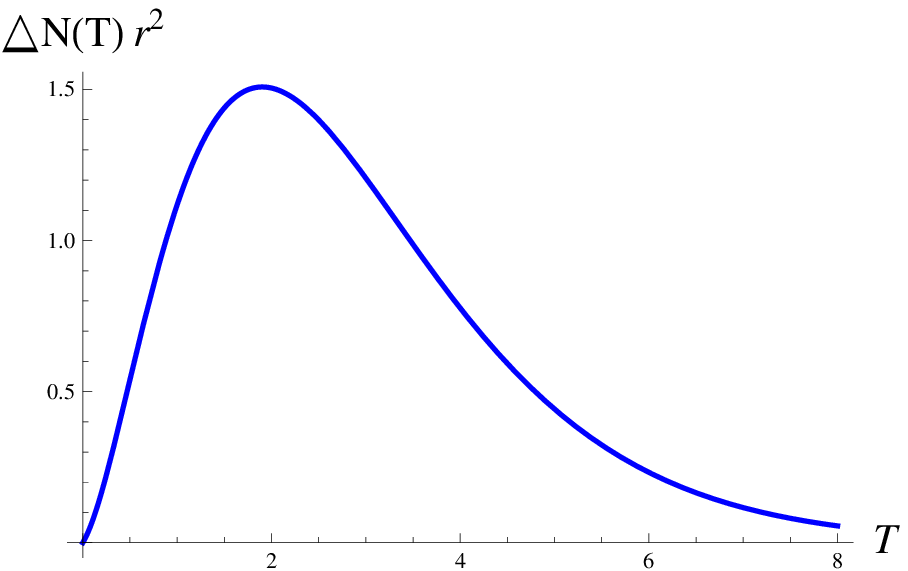}
\end{center}
\caption{ 
$ r^2\nabla^2_\bot  N^G\Lb  T =  2  \phi_0 r^2 Q^2_s\ \Rb$ of \protect\eq{INC12}  versus $T$ }
 \label{dn}
\end{figure}

One can see that $\nabla^2 N^G$ has a maximum at $T \sim 2$ and the integral over $T$ is equal to 8.39$\Lb 1 - \gamma_{cr}\Rb^2/r^2$.
%%%%%%%%%%%%%%%%%%%%%%%%%%%%%%%%%%%%%%%%%%%%%%%%%%%%
  %%%%%%%%%%%%%%%%%%%%%%%%%%%%%%%%%%%%%%%%%%%%%%%%%%% 
 \subsection{Rapidity correlations}
 %%%%%%%%%%%%%%%%%%%%%%%%%%%%%%%%%%%%%%%%%%%%%%%%%%%%
In this section we calculate the rapidity correlations that stem from the diagram of \fig{1di}.
 \eq{GSNB} in the momentum representation takes the form
 \bea \label{RC0}
&&\frac{d \sigma}{d Y_1 d Y_2 d^2 p_{\bot,1} d^2p_{\bot,2}}\,\,= \,\,\frac{4 \pi^2 \bas^4}{N^2_c}\int ^Y_{Y_1} d Y' \int^{Y_2}_{Y_0} d Y'' \label{RC1}\\
&&\int \frac{d^2 R_1}{R^4_1} \int \frac{ d^2 R_2}{R^4_2}\,N_\pom\Lb Y - Y'; R_p,R_1,Q_T=0\Rb  \int d^2 R'_1\int d^2 R'_2 \,\, K\Lb R_1,R'_1\Rb\nn\\
&& \int d^2 Q_T\, N^{\mbox{\tiny incl}}_\pom\Lb 
Y' - Y", R'_1, R'_2, Q_T| p_{\bot,1}, Y_1\Rb \, N^{\mbox{\tiny incl}}_\pom\Lb 
Y' - Y", \vec{R}_1 - \vec{R}'_1, \vec{R}_2 - \vec{R}'_2, Q_T| p_{\bot,2}, Y_2\Rb   \nn\\
&& K\Lb R_2,R'_2\Rb\,N_\pom\Lb Y'' - Y_0; R_2,R_A,Q_T=0\Rb \nn
 \eea 
  Our main goal to find the largest contribution. As has been discussed in the previous section all Pomerons in the loop have the largest contributions in the vicinity of the saturation scale while the upper and low Pomerons can contribute inside of the saturation domain. Recall that for the Pomerons in the loop $\gamma \to \bar{\gamma} =1 - \gamma_{cr}$ in \eq{NPQ1} - \eq{NPQ5}.

\begin{figure}[h]
\begin{center}
 \includegraphics[width=0.5 \textwidth]
{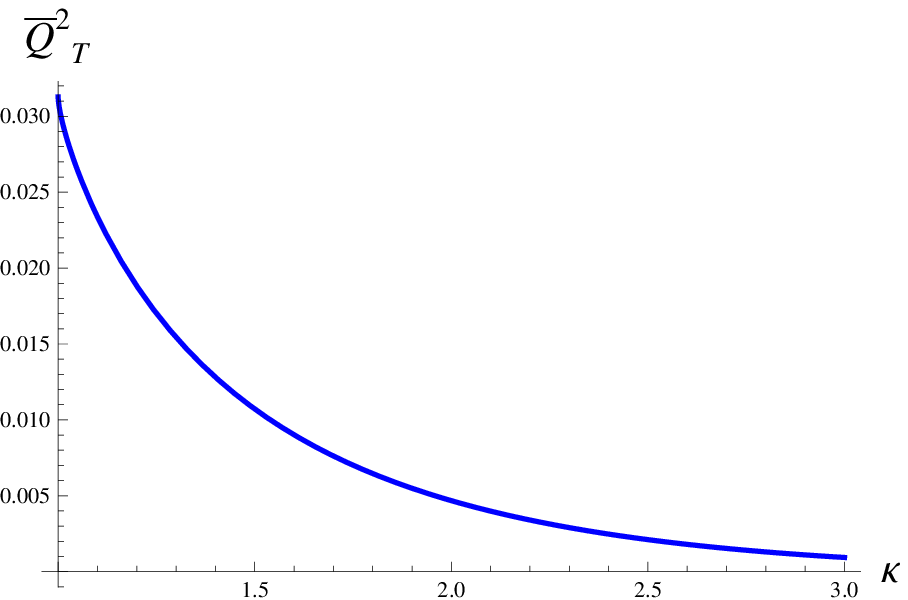}
\end{center}
\caption{ $ \bar{Q}^2_T \Lb \kappa\Rb$ defined in \protect\eq{RCN30} versus $\kappa = R_1/R_2$.}
 \label{avq}
\end{figure}

 Let us concentrate our effort on $Q_T$ integration considering three kinematic regions:
 \begin{enumerate}
 \item \quad $Q_T R_1 \ll 1$ and $Q_T R_2 \ll 1$. We will see below that both $R'_1$ and $| \vec{R}_1 - \vec{R}^{\,'}_1|$ are of the order of $R_1$ as well as $R'_2$ and $| \vec{R}_2 - \vec{R}^{\,'}_2|$ are of the order of $R_2$. Using \eq{NPQ2} one can see that $\nabla^2_\bot\,N_\pom\Lb Y  -Y'; r_\bot, R_{1}; Q_T \Rb  $ as well as others $\nabla^2_\bot\,N_\pom$ entering \eq{INC7} and \eq{RC1} do not depend on $Q_T$. Hence the integral over $d^2 Q_T$ diverges in this region and the upper limit of integration gives the main contribution.
  \item \quad $Q_T R_1 \sim 1$ and $Q_T R_2 \ll  1$. For  hadron-nucleus collisions we expect that the main contributions would come from  the saturation region in which  $R_2$ is proportional to $1/Q_s(A; Y'')$ while $R_1 \propto 1/Q_s(\mbox{proton}; Y - Y')$ and we expect that $R_2 \ll R_1$ since $Q_s(A; Y'') \gg  Q_s(\mbox{proton}; Y - Y')$.
 In this region $\nabla^2_\bot\,N_\pom\Lb Y_1 - Y''; r_\bot, R_{2}; Q_T \Rb $ and  $\nabla^2_\bot\,N_\pom\Lb Y_2 -Y''; r'_\bot, R_{2}; Q_T \Rb $ do not depend on $Q_T$ while  $\nabla^2_\bot\,N_\pom\Lb Y' - Y_1; r_\bot, R_{1}; Q_T \Rb $ and  $\nabla^2_\bot\,N_\pom\Lb Y' - Y_2; r'_\bot, R_{2}; Q_T \Rb $  are in the region of \eq{NPQ3} and give contribution proportional to $
  Q^{-2 (1 - \bar{\gamma})}$   each. Therefore, the integral has a form $d^2 Q_T\,   Q^{-4 (1 - \bar{\gamma})}_T  $. Recalling that $\bar{\gamma} =  1 - \gamma_{cr} = 0.63$ one can see  that the integral diverges in this region.
    \item \quad $Q_T R_1 \gg 1$ and $Q_T R_2 \gg  1$. The integrant has the $Q_T$ dependance which is  $\Lb Q_T^2
\Rb^{-4 ( 1 - \bar{\gamma})}$ and the integral converges. Therefore the main contribution stems from the region when  
$Q_T \propto 1/R_2 $ and it is proportional to $\Lb R^2_2\Rb^{3 - 4 \bar{\gamma}}$.

   \item \quad $Q_T R_1 \gg 1$ and $Q_T R_2 \gg  1$.  The integral over $Q_T$ takes the form (assuming that $r \ll R_1$ and/or $R_2$)
   \bea \label{INTQ}
 \hspace{-2cm}  \int_{Q_T > 1 /R_2} \!\!\!\frac{ d^2 Q_T}{4 \pi^2} \Lb Q_T^2\Rb^{-4 ( 1 - \bar{\gamma})}e^{ 2 i \vec{Q}_T \cdot( \vec{ R}_1 + \vec{R}_2)} &\rightarrow&\,
 \frac{1}{4 \pi^2}  \int \frac{d^2 Q_T\,J_0\Lb 2  Q_T | \vec{R}_1 + \vec{R}_2|\Rb}{ \Lb Q^2_T +  1/R^2_2\Rb^{4 ( 1 - \bar{\gamma})} }\,=\nn\\
   &  \,=\,&\frac{1}{2 \pi \Gamma\Lb 4 - 4 \bar{\ga}\Rb}
      \Lb | \vec{R}_1 + \vec{R}_2| R_2\Rb^{3 - 4\bar{\ga}}K_{3 - 4 \bar{\ga}}\Lb 2 \frac{| \vec{R}_1 + \vec{R}_2| }{R_2}\Rb
      \eea  
    
     \end{enumerate} 
Concluding this discussion we see that the typical $Q_T \approx 1/R^2_2$  where $R_2$ is the size of the smallest dipole in triple Pomeron vertices.
One can also see  that \eq{INTQ} leads to $| \vec{R}_1 + \vec{R}_2| \approx R_2$ or $R_1 \to R_2$.
Based on these two features we can re-write \eq{RC0} in the following way

 \bea \label{RCN1}
&&\frac{d \sigma}{d Y_1 d Y_2 d^2 p_{\bot,1} d^2p_{\bot,2}}\,\,= \,\,\frac{4 \pi^2 \bas^4}{N^2_c}\,\langle Q^2_T\rangle\int ^Y_{Y_1} d Y' \int^{Y_2}_{Y_0} d Y'' \int \frac{d^2 R_1}{R^4_1} \int \frac{ d^2 R_2}{R^4_2}\,\\
&&N_\pom\Lb Y - Y'; R_p,R_1, Q_T=0\Rb  \int d^2 R'_1\int d^2 R'_2 \,\, K\Lb R_1,R'_1\Rb N^{\mbox{\tiny incl}}_\pom\Lb 
Y' - Y", R'_1, R'_2,Q_T=0 | p_{\bot,1}, Y_1\Rb \, \nn\\
&&N^{\mbox{\tiny incl}}_\pom\Lb 
Y' - Y", \vec{R}_1 - \vec{R}'_1, \vec{R}_2 - \vec{R}'_2, Q_T=0 | p_{\bot,2}, Y_2\Rb
 K\Lb R_2,R'_2\Rb\,N_\pom\Lb Y'' - Y_0; R_2,R_A,Q_T=0\Rb \nn
 \eea 
where we estimate the value of $\langle Q^2_T\rangle $ using \eq{NPQ5} 
   \bea\label{RCN2}
  \langle Q^2_T\rangle\,\, &=&\,\, \frac{1}{2 \pi}\int Q_T d Q_T \frac{J_0\Lb 2 Q_T |\vec{R}_1 + \vec{R}_2|\Rb}{\Lb 1 +  a\Lb \bar{\gamma}\Rb Q^2_T R^2_2\Rb^{4(1 - \bar{\gamma})}} \\
  &=& \frac{1}{2 \pi \Gamma\Lb 4 ( 1 - \bar{\gamma})\Rb}\,\frac{1}{ a\Lb \bar{\gamma}\Rb R^2_2} \Lb\frac{\Lb \vec{R}_1 + \vec{R}_2\Rb^2}{a R^2_2}\Rb^{3/2 - 2 \bar{\gamma}}\!\!\!\!\!\!\!\!\!\! \!\!\!K_{-3 + 4 \bar{\gamma}}\Lb 2 \frac{| \vec{R}_1 + \vec{R}_2|}{\sqrt{ a\Lb \bar{\gamma}\Rb}R_2}\Rb\nn
  \eea 
 Integrating  over angle $\varphi$ we obtain 
 \beq \label{RCN30}
 \bar{Q}^2_T \Lb \kappa\Rb\,\,=\,\,\int d \varphi \,  \langle Q^2_T\rangle\Lb\varphi,\kappa\Rb 
\eeq
where $\kappa=R_1/R_2$.  This function is shown in \fig{avq}. One can see the steep decrease for $\kappa > 1$ and, therefore, we can replace $  \langle Q^2_T\rangle$ by
\beq \label{RCN31}
\langle\langle Q^2_T\rangle\rangle \,\,=\,\,\int d \varphi \,d \kappa\, \,  \langle Q^2_T\rangle\Lb\varphi,\kappa\Rb \,\,=\,\,0.0154\, \delta\Lb  \kappa - 1\Rb
\eeq
%%%%%%%%%%%%%%%%%%%%%%%%%%%%%%%%%%%%%%%%%%%%%%%%
 %%%%%%%%%%%%%%%%%%%%%%%%%%%%%%%%%%%%%%%%%%%%%%%% 

Notice that each $N_\pom\Lb Y_1; r_\bot,r_{i}; R_2 \Rb$ in \eq{RCN2} takes the form (see \eq{H}, \eq{W} and \eq{BGA})
\beq \label{RCN3}
N_\pom\Lb Y; r_\bot,r_{i}; R_2 \Rb\,\,=\,\,\int^{\epsilon + i \infty}_{\epsilon - i \infty} \frac{d \gamma}{2 \pi i} \frac{\Lb \gamma - \h\Rb^2}{\Lb\gamma (1 - \gamma)\Rb}b_\gamma\Lb \frac{r^2}{r^2_i}\Rb^\gamma e^{\omega\Lb \gamma\Rb Y}
\eeq
One can see that in \eq{RCN1}  we have
 \beq \label{RCN4}
 \int d^2 R'_1\, K\Lb R_1,R'_1\Rb  \Lb \frac{r^2}{R'^2_1}\Rb^{\gamma_1}    \Lb \frac{r'^2}{|\vec{R}_1 - \vec{R}^{\,'}_1|^2}\Rb^{\gamma_2}  \,\,=\,\,C_1\Lb \gamma_1, \gamma_2\Rb  \Lb \frac{r^2}{R^2_1}\Rb^{\gamma_1}    \Lb \frac{r'^2}{R^2_1}\Rb^{\gamma_2}
 \eeq
 where 
 \beq \label{RCN5}
 C_1\Lb \gamma_1, \gamma_2\Rb\,\,=\,\,\pi\frac{B\Lb - \gamma_1,-\gamma_2\Rb}{\Lb 1 + \gamma_1 + \gamma_2\Rb \,B\Lb 1 + \gamma_1, 1 + \gamma_2\Rb}\,\,=\,\,\pi\,\frac{\Gamma\Lb - \gamma_1\Rb\,\Gamma\Lb - \gamma_2\Rb\,\Gamma\Lb 1 + \gamma_1 + \gamma_2\Rb}{ \Gamma\Lb 1 + \gamma_1\Rb\,\Gamma\Lb  1 +  \gamma_2\Rb\,\Gamma\Lb - \gamma_1 - \gamma_2\Rb} 
  \eeq
 where $B\Lb x, y \Rb$ is the Euler beta-function (see formula {\bf 8.38} of Ref.\cite{RY}). We introduce the Feynman parameters, using formula {\bf 3.198} of Ref.\cite{RY}, to take the integral of \eq{RCN4}.

\begin{figure}[h]
\begin{center}
 \includegraphics[width=0.7 \textwidth]{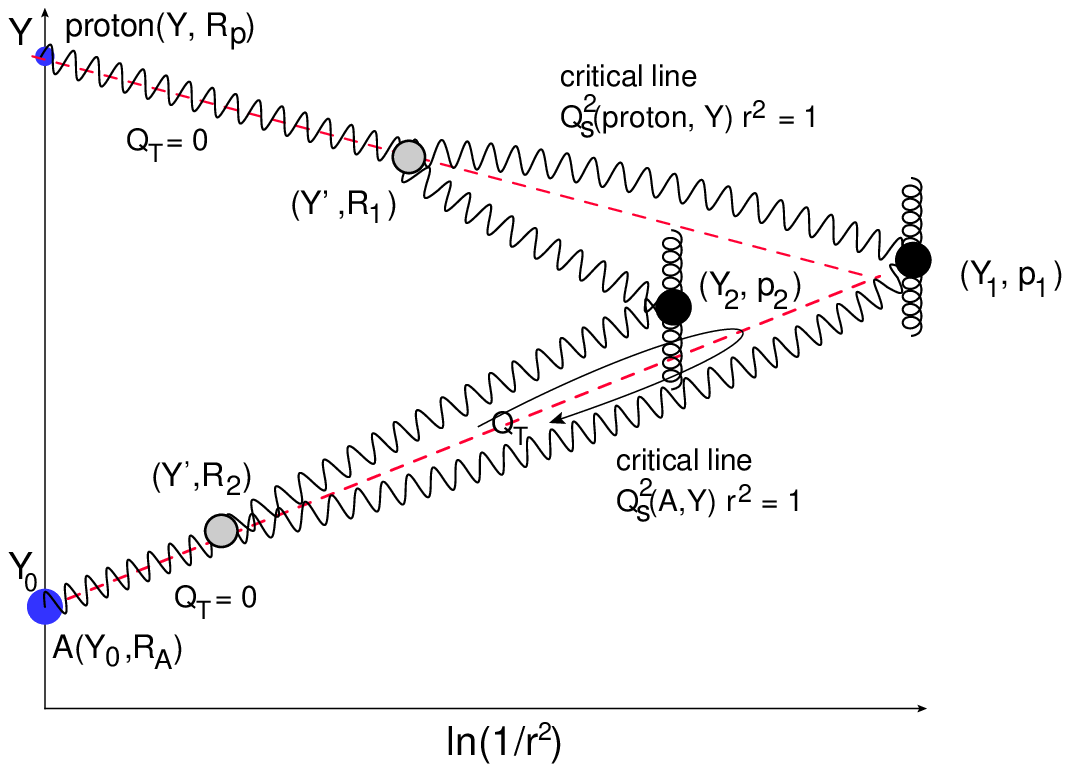}
\end{center}
\caption{ The first enhanced (loop) diagram for two particle correlation.
  Wavy  lines denote the BFKL
 Pomerons.  Helix lines show the gluons. Black blob stands for the Mueller vertex for inclusive production of gluon jet with the transverse momentum $p_{\perp,1}$ ($p_{\perp, 2}$), respectively. Blue blobs describe the interaction of the BFKL Pomeron with the Êproton and the  nucleus. Dashed red line corresponds to the  critical line for proton-nucleus scattering. }
 \label{1disat}
\end{figure}

 The upper and lower Pomerons enter at $Q_T = 0$ and both of them have the form the same as in \eq{NPQ5}.
 Plugging \eq{NPQ5}, \eq{INC9}, \eq{RCN31} and \eq{RCN5} in  \eq{RCN1}  we have the following expression
\bea \label{RCN6}
\hspace{-0.5cm}&&\frac{d \sigma}{d Y_1 d Y_2 d^2 p_{\bot,1} d^2p_{\bot,2}}\,\,=\,\,0.0154\frac{4 \pi^2 \bas^4}{ N^2_c}\frac{8\,C_F}{\bas (2\pi)^4}\,\frac{1}{p^2_{\bot,1}}\frac{8\,C_F}{\bas (2\pi)^4}\,\frac{1}{p^2_{\bot,2}}\, \int \frac{\pi d r^2}{r^4} J_0\Lb r p_{\bot,1}\Rb\,\int \frac{\pi d r'^2}{r'^4} J_0\Lb r' p_{\bot,2}\Rb\nn\\
&&~~~~~~~~~~~~R^2_p \,R^2_p\,{\cal U}\Lb\bar{ \gamma} \Rb\int_{r^2}^{R^2_p}  d R^2_1 \,  \delta\Lb \frac{R_1}{R_2 }- 1\Rb\,\frac{d R^2_2}{R^2_2} \int ^Y_{Y_1} d Y' \int^{Y_2}_{Y_0} d Y''\,\Phi\Lb r,r';R_1,R_2,Y,Y',Y";Y_1,Y_2\Rb\\
&&\Phi\Lb r,r';R_1,R_2,Y,Y',Y";Y_1,Y_2\Rb\,=\,\Bigg( T\Lb R_1,R_p,Y - Y'\Rb\,T\Lb R_2,R_p,Y''\Rb\,T\Lb r,R_1,Y' - Y_1\Rb\, \nn\\
&&~~~~~~~~~~~~~~~~~~~~~~~~~~~~~~~~~~~~~~~~~~
\times T\Lb r,R_2,Y_1 - Y''\Rb\,T\Lb r',R_1,Y' - Y_2\Rb\,T\Lb r',R_2,Y_2 - Y''\Rb\Bigg)^{\bar{\gamma}}\label{PHI}
\eea
where
 \beq \label{T}
 T\Lb r_1,r_2; Y\Rb\,\,=\,\,\frac{r^2_1}{r^2_2} \exp\Lb \frac{\omega\Lb \gamma_{cr}\Rb}{1 - \gamma_{cr}} \, Y\Rb 
 ~~~~\mbox{for}~~~~r_1 < r_2
 \eeq
and     
   \beq \label{U}
   {\cal U}\Lb\bar{\gamma}\Rb\,\,=\,\,\bar{\phi^4_0} \,\Bigg(\frac{\Lb \bar{\gamma} - \h\Rb^2}{\bar{\gamma}\Lb 1 - \bar{\gamma}\Rb} b_{\bar{\gamma}}\Bigg)^2\, C^2\Lb 1+\bar{ \gamma},1+\bar{ \gamma}\Rb 
   \eeq
We can trust \eq{RCN6} only if all  Pomerons approaching the saturation scale, i.e. 
\bea
 T\Lb R_1,R_p,Y - Y'\Rb \,\,\to \,1;&~~~~~&~~~~~T\Lb R_2,R_p,Y'' \Rb\,\,\to \,1;\label{COND1}\\ 
T\Lb r, R_1,Y' - Y_1\Rb \,\,\to \,1;&~~~~~&~T\Lb r, R_2,Y_1 - Y'' \Rb\,\,\to \,1;\label{COND2}\\
T\Lb r', R_1,Y' - Y_2\Rb \,\,\to \,1;&~~~~~&T\Lb r', R_2,Y_2 - Y''\Rb \,\,\to \,1;\label{COND3}
 \eea
 The maximal contribution to the diagram  with the BFKL Pomerons which contributions can be trusted in perturbative QCD stem from the region where we have the sign of equality  for all Pomerons , i.e.  in the equations: \eq{COND1}, \eq{COND2} and \eq{COND3}. For such kinematic region 
   we  calculate the diagram of \fig{1di} in the region where $R^2_p Q^2_s\Lb p; Y\Rb \approx 1$ as well as $ R^2_2 Q^2_s\Lb A; Y''\Rb \approx 1 $ and $ R^2_1 Q^2_s\Lb \mbox{proton}; Y - Y'\Rb \approx 1 $.  We illustrate with \fig{1disat} this region of integration drawing the first enhanced diagrams in the two dimensional plane $( \ln(1/r^2), Y)$. 
 
 However, as we will see below, we cannot keep all Pomerons in the vicinity of the saturation scale.   On the other hand we have to keep all Pomerons in the Pomeron loop in the kinematic region close to the saturation scale since, as we have discussed, only in this kinematic region the inclusive production gives the largest contributions. The best choice will be if
  the upper and low Pomerons will be inside the saturation domain where their amplitudes reach the unitarity limit $N_\pom \to 1$.  Finally, we are looking for the contribution in the following kinematic region:
   \bea
 T\Lb R_1,R_p,Y - Y'\Rb \,\,\leq \,1;&~~~~~&~~~~~T\Lb R_2,R_p,Y'' \Rb\,\,\leq \,1;\label{COND11}\\ 
T\Lb r, R_1,Y' - Y_1\Rb \,\,\approx \,1;&~~~~~&~T\Lb r, R_2,Y_1 - Y'' \Rb\,\,\approx \,1;\label{COND12}\\
T\Lb r', R_1,Y' - Y_2\Rb \,\,\approx \,1;&~~~~~&T\Lb r', R_2,Y_2 - Y''\Rb \,\,\approx \,1;\label{COND13}
 \eea   
  \eq{COND12} we can re-write in the following way:
  \bea \label{CONDRED}   
   T\Lb R_1,R_p,Y - Y'\Rb\,\, &=&\,\,\frac{1}{ T\Lb r, R_1,Y' - Y_1\Rb}\frac{r^2}{R^2_p}\,e^{\frac{\omega\Lb \gamma_{cr}\Rb}{1 - \gamma_{cr}} \,\Lb Y - Y_1\Rb}\,\,=\,\,\frac{1}{ T\Lb r, R_1,Y' - Y_1\Rb}\,r^2 \,Q^2_s\Lb p, Y - Y_1\Rb;\nn\\
T\Lb R_2,R_p,Y''\Rb\,\,&=&\,\,\frac{1}{ T\Lb r, R_2,Y_1 - Y''\Rb}\frac{r^2}{R^2_p}\,e^{\frac{\omega\Lb \gamma_{cr}\Rb}{1 - \gamma_{cr}} \,\Lb  Y_1\Rb}\,\,=\,\,\frac{1}{ T\Lb r, R_1,Y' - Y_1\Rb}\,r^2 \,Q^2_s\Lb p, Y_1\Rb;   
  \eea 
with
\beq \label{QS}
Q^2_s\Lb p; Y \Rb\,\,=\,\,\frac{1}{R^2_p}\exp\Big( \bas \frac{\chi\Lb \gamma_{cr}\Rb}{1 - \gamma_{cr}}\, Y \Big) ~~\mbox{and} ~~  Q^2_s\Lb A; Y \Rb\,\,=\,\,\frac{A}{R^2_A}\exp\Big( \bas \frac{\chi\Lb \gamma_{cr}\Rb}{1 - \gamma_{cr}}\,Y\,\Big)
\eeq 
    %%%%%%%%%%%%%%%%%%%%%%%%%%%%%%%%%%%%%%%%%%%%%%%%%%%%

%%%%%%%%%%%%%%%%%%%%%%%%%%%%%%%%%%%%%%%%%%%%%%%%%%%%
In \eq{CONDRED} we introduce the saturation momentum for nucleus ($Q_s\Lb A; Y\Rb$). Strictly speaking in the simple diagram, that we consider,  $N_\pom^{\mbox{\tiny incl}}\Lb Y_1, r; R_A\Rb $ is proportional to $A 
 N_\pom^{\mbox{\tiny incl}}\Lb Y_1, r; R_p\Rb$. However, we will show in the next section that more complicated diagrams lead to $N_\pom^{\mbox{\tiny incl}}\Lb Y_1, r; R_A\Rb \propto \Lb r^2  Q^2_s\Lb A;  Y_1 - 0\Rb\Rb^{\bar \ga}$.
 
\begin{figure}[h]
\begin{tabular}{c c c}
 \includegraphics[width=0.45\textwidth]{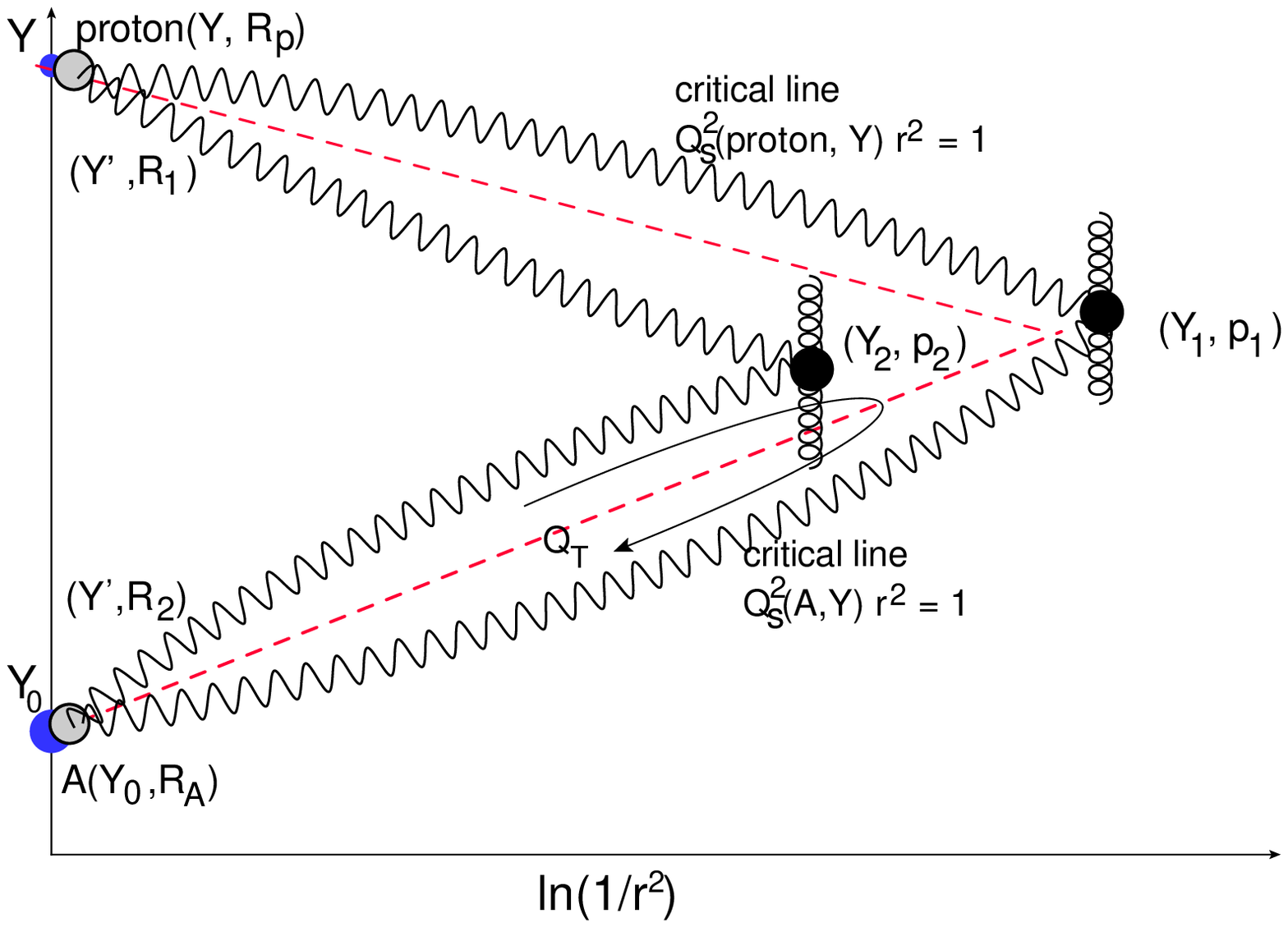} &~~~~~~~& \includegraphics[width=0.4 \textwidth]{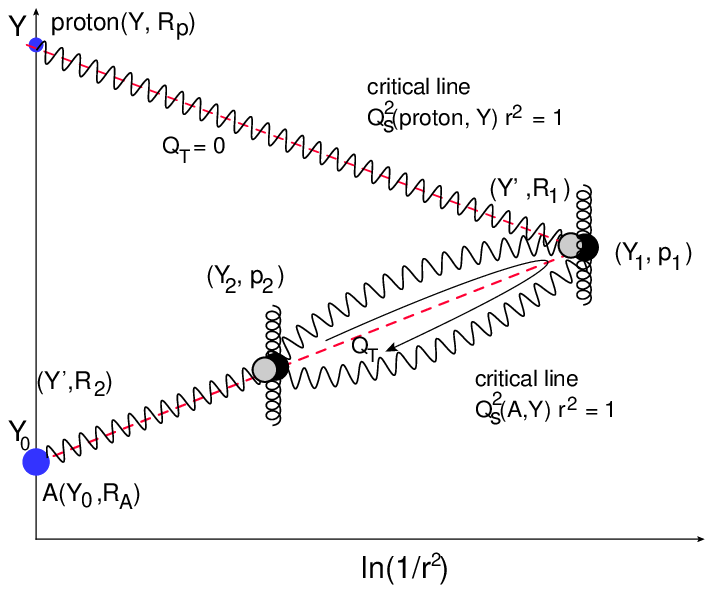}\\
 \fig{1dicont}-a & & \fig{1dicont}-b \\
 \end{tabular}
\caption{ The first enhanced (loop) diagram for two particle correlation after integration over rapidities.
  Wavy  lines denote the BFKL
 Pomerons.  Helix lines show the gluons. Black blob stands for the Mueller vertex for inclusive production of gluon jet with the transverse momentum $p_{\perp,1}$ ($p_{\perp, 2}$), respectively. Blue blobs describe the interaction of the BFKL Pomeron with the Êproton and the  nucleus. Dashed red line corresponds to the  critical line for proton-nucleus scattering. }
 \label{1dicont}
\end{figure}

Taking into account \eq{CONDRED} \eq{PHI} can be re-written in the form
\bea\label{PHI1}
&&\Phi\Lb r,r';R_1,R_2,Y,Y',Y";Y_1,Y_2\Rb\,= \\
&&~~~~~~~~~~~~~~~~~\,\Bigg(r^2 \,Q^2_s\Lb A, Y_1\Rb\,r^2 \,Q^2_s\Lb p, Y-Y_1\Rb \,T\Lb r',R_1,Y' - Y_2\Rb\,T\Lb r',R_2,Y_2 - Y''\Rb\Bigg)^{\bar{\gamma}}\nn
\eea
One can see that we can satisfy conditions of \eq{COND13} only if
\beq \label{RA}
\frac{R^2_1}{R^2_2}\,\,=\,\,\exp\Lb \frac{\omega\Lb \gamma_{cr}\Rb}{1 - \gamma_{cr}} \Lb  Y" + Y' - 2 Y_2\Rb\Rb
\eeq
Since this ration is equal to 1, one sees that $Y'' + Y' \,= \,2 Y_2$. Finally 
\bea \label{PHI2}
&&\int_{r^2}^{R^2_p}  d R^2_1 \,  \delta\Lb \frac{R_1}{R_2 }- 1\Rb\,\frac{d R^2_2}{R^2_2} \int ^Y_{Y_1} d Y' \int^{Y_2}_{Y_0} d Y''\,\Phi\Lb r,r';R_1,R_2,Y,Y',Y";Y_1,Y_2\Rb\,\,=\\
&&\int_{r^2}^{R^2_p}d R^2_2\int^{Y}_{Y_1}
d  Y' \, \frac{1}{ \omega\Lb \gamma_{cr}\Rb}\Bigg(r^2 \,Q^2_s\Lb A, Y_1\Rb\,r^2 \,Q^2_s\Lb p, Y-Y_1\Rb \,\frac{r'^4}{R^4_2} \exp\Big( \bas \frac{\chi\Lb \gamma_{cr}\Rb}{1 - \gamma_{cr}}\, \Lb Y' - Y''\Rb\Big)^{\bar{\gamma}}\,\,=\nn\\
&&\int_{r^2}^{R^2_p}d R^2_2 \frac{1}{2 \omega^2\Lb \gamma_{cr}\Rb}\Big\{ e^{2 \omega\Lb \gamma_{cr}\Rb\Lb Y - Y_2\Rb}\,-\,e^{2 \omega\Lb \gamma_{cr}\Rb\Lb Y_1 - Y_2\Rb}\Big\}\Bigg(r^2 \,Q^2_s\Lb A, Y_1\Rb\,r^2 \,Q^2_s\Lb p, Y-Y_1\Rb \,\frac{r'^4}{R^4_2}\Bigg)^{\bar{\gamma}}\,=\nn\\
&&\int_{r^2}^{R^2_p} d R^2_2 \frac{1}{2 \omega^2\Lb \gamma_{cr}\Rb}\Bigg\{\Bigg(r^2 \,Q^2_s\Lb A, Y_1\Rb\,r^2 \,Q^2_s\Lb p, Y-Y_1\Rb \,r'^2 \,Q^2_s\Lb R_2;Y - Y_2\Rb\,r'^2 Q^2_s\Lb R_2; Y_2 - Y''\Rb\Bigg)^{\bar{\gamma}}\,-\nn\\
&&~~~~~~~~~~~~~~~~~~~~~\Bigg(r^2 \,Q^2_s\Lb A, Y_1\Rb\,r^2 \,Q^2_s\Lb p, Y-Y_1\Rb \,r'^2 \,Q^2_s\Lb R_2;Y_1 - Y_2\Rb\,r'^2 Q^2_s\Lb R_2; Y_1 - Y_2\Rb\Bigg)^{\bar{\gamma}}\Bigg\}\label{PHI3}
\eea
where $Q_s\Lb R_2; Y\Rb$ is the saturation scale of \eq{QS} where $R^2_p$ or $R^2_A$ ia replaced by $R^2_2$,

The first term in \eq{PHI3} corresponds to $Y' = Y$  and it is shown in \fig{1dicont}-a for $Y_1\,=\,Y_2 \,=\, \h Y$, while the second term stems from $Y' = Y_1$ (see \fig{1dicont}-b).

One can see from  \eq{COND12} and \eq{COND3}   that we can have four BFKL Pomerons in the vicinity of the saturation region only for  $Y_1$ and $Y_2$ that satisfy the equations: $Q^2_s\Lb p; Y - Y_1\Rb = Q^2_s\Lb A; Y_1 \Rb$
and  $Q^2_s\Lb p; Y - Y_2\Rb = Q^2_s\Lb A; Y_2 \Rb$. This region is shown in \fig{1dicont}-a and we expect the largest contributions at  $p_{\bot,1}^2 \sim 1/\Lb R^2_p \exp\Lb\bas \frac{\chi\Lb \bar{\gamma}\Rb}{\bar{ \gamma}}\,\Lb Y - Y_1\Rb\Rb\Rb$ and $p_{\bot,1}^2 \sim  p^2_{\bot,2} \exp\Lb-\bas \frac{\chi\Lb \bar{\gamma}\Rb}{\bar{ \gamma}}\,\Lb Y_1 - Y_2\Rb\Rb$ .

In the vicinity of the saturation scale $N_\pom$ takes the form\footnote{The expression for $Q_s(Y)$ should be changed from $ \ln\Lb Q^2_s\Lb Y \Rb/Q^2_s\Lb Y = Y_0\Rb\Rb = \frac{ \chi\Lb \gamma_{cr}\Rb}{ 1 -  \gamma_{cr}} \Lb Y - Y_0\Rb$ to more complicated expression (see Refs.\cite{MUTR,MUPE,REV}).} (see Refs. \cite{IIM,IIMU,REV})
\beq \label{RCN9}
N_\pom\Lb r_1,r_2, Y, Q_T=0; Y\Rb\,\,=\,\phi_0\,r^2_2 \Lb r^2_1\,Q^2_s\Lb r_2, Y\Rb\Rb^{\bar{\gamma}}\exp\Big( - \frac{z^2}{2\omega''_{\ga \ga}\Lb \ga=\ga_{cr},0\Rb \,Y}\Big)
\eeq 
where $z\,=\,\ln\Big(r^2_1\,Q^2_s\Lb r_2, Y\Rb\Big)$.

\begin{figure}[h]
\begin{center}
 \includegraphics[width=0.4 \textwidth]{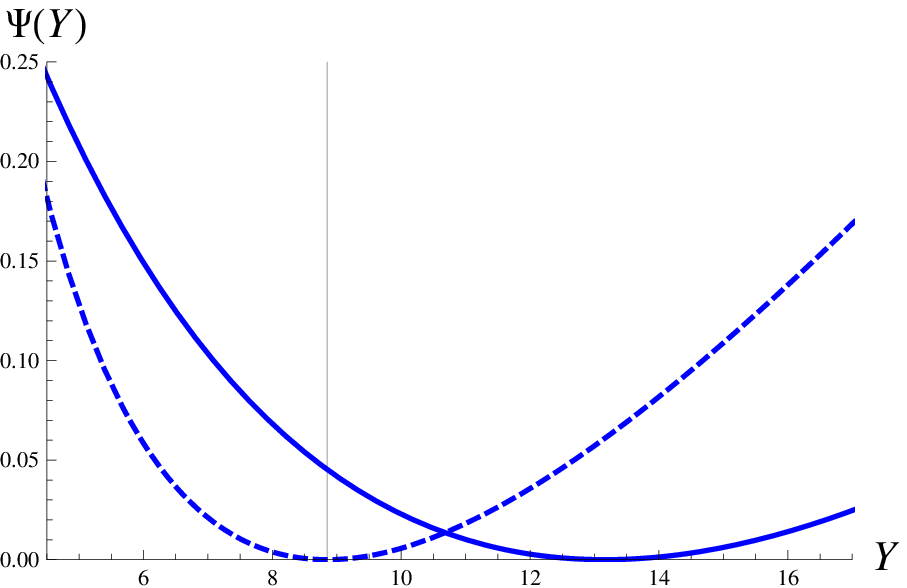}
\end{center}
\caption{ Function $\Psi$ of \protect\eq{RCN102} versus rapidity at $W = 7 TeV$. The vertical line show the $Y_1 = \h Y$. The solid  line shows $\Psi$ for proton-gold scattering while the dashed line corresponds to the proton-proton scattering.}
 \label{psi}
\end{figure}

We  calculate the diagram of \fig{1dicont}-a  assuming \eq{RCN9} for $N_\pom$.  This assumption can be justified only in the limited kinematic region where $\ln \Big(Q^2_s\Lb A; Y_1\Rb\Big{/}Q^2_s\Lb p; Y - Y_1\Rb\Big)\ll\,\sqrt{ 2 \pi \omega''_{\ga \ga}\Lb \ga=\ga_{cr},0\Rb Y_i}$ where $Y_i $ is a minimum from $Y- Y_1$ and $Y_1$.

\eq{RCN9} allows us to estimate the range in rapidities in which we can trust our evaluation of this diagram. As has been mention only at rapidity $Y_1$ from the equation $Q_s\Lb p ; Y - Y_1\Rb = Q_s\Lb A; Y_1\Rb$ all four Pomerons in the loop can be considered in the saturation region, However,  we really used the form of the amplitude from \eq{RCN9} but replacing
$ \exp\Big( - \frac{z^2}{2\omega''_{\ga \ga}\Lb \ga=\ga_{cr},0\Rb \,Y}\Big)$ by unity.  Rewriting this factor in terms of the saturation scale and deviation from it the factor for the  Pomeron exchange with rapidity $ Y - Y_1$ in the loop takes the form

\beq \label{RCN101}
\exp\Big( - \frac{z^2}{2\omega''_{\ga \ga}\Lb \ga=\ga_{cr},0\Rb \,Y}\Big)\,\,=\,\,\exp\Big( - \frac{\chi\Lb \gamma_{cr}\Rb}{ 1 - \gamma_{cr}}\frac{\ln^2\Lb r^2 Q^2_s\Lb p, Y - Y_1\Rb\Rb}{2\chi''_{\ga \ga}\Lb \ga=\ga_{cr},0\Rb \,\ln\Lb Q_s^2\Lb p, Y - Y_1\Rb/Q^2_s\Lb p, 0\Rb\Rb}\Big)
\eeq
As has been discussed $r^2 = 1/Q^2_s\Lb A, Y_1\Rb$, therefore,  we can replace the Pomeron exchange by $\Lb r^2 Q^2_s\Lb p, Y-Y_1\Rb\Rb^{1 - \gamma_{cr}}$ if
\beq \label{RCN102}
e^\Psi\,\,\,=\,\,\,\exp\Big( - \frac{\chi\Lb \gamma_{cr}\Rb}{ 1 - \gamma_{cr}}\frac{\ln^2\Lb  Q^2_s\Lb p, Y - Y_1\Rb/Q^2_s\Lb A; Y_1\Rb\Rb}{2\chi''_{\ga \ga}\Lb \ga=\ga_{cr},0\Rb \, \,\ln\Lb Q_s^2\Lb p, Y - Y_1\Rb/Q^2_s\Lb p, 0\Rb\Rb}\Big)\,\,\to\,\,1\,\,\,\mbox{or} \,\,\,\Psi \,\,\ll\,\,1
\eeq
In \fig{psi} we plotted $\Psi$  for the LHC energy $W\,=\,  7TeV$ using the KLN parameterization for the saturation scale (see Ref.\cite{KLN} ). One can see that we can use our approach in the wide range of rapidities.

 %%%%%%%%%%%%%%%%%%%%%%%%%%%%%%%%%%%%%%%%%%%%%%%%%%%%

%%%%%%%%%%%%%%%%%%%%%%%%%%%%%%%%%%%%%%%%%%%%%%%%%%%%

The last integration over $R_2$ in \eq{PHI3} brings  factor $1/(2\bar{\gamma} - 1)$ and two limits of integrations:  $R_2 = R_p$ and $R_2 = r$ correspond to \fig{1dicont}-a and \fig{1dicont}-b, respectively,

Integrating over $r$ and $r'$ we obtain the following expression for the double inclusive cross section
\bea \label{RCN11}
\frac{d \sigma}{d Y_1 d Y_2 d^2 p_{\bot,1} d^2p_{\bot,2}}\,\,&=&\,\,N^2_0\,\bar{\phi^4_0}\,,\frac{8\,C_F}{\alpha_s (2\pi)^4}\,\frac{8\,C_F}{\alpha_s (2\pi)^4}\,\,R^4_p  R^2_A \\
&\times& \Lb \frac{Q_s^2\Lb p;  Y - Y_1\Rb}{p^2_{\bot,1}}\Rb^{\bar{\ga}}\Lb \frac{Q_s^2\Lb p; Y - Y_2\Rb}{p^2_{\bot,2}}\Rb^{\bar{\ga}}\Lb \frac{Q_s^2\Lb A; Y_1 - Y_0\Rb}{p^2_{\bot,1}}\Rb^{\bar{\ga}}\Lb \frac{Q_s^2\Lb A; Y - Y_2\Rb}{p^2_{\bot,2}}\Rb^{\bar{\ga}}   \nn
     \eea 
 where we assumed that all $z_i \ll\,\sqrt{ 2 \pi \omega''_{\ga \ga}\Lb \ga=\ga_{cr},0\Rb Y_i}$ .  
 It should be stressed that \eq{RCN11} is valid for $p_{i,\bot} \sim Q_s$.
 
 We introduce factor       $\phi_0$ (see \eq{GSNB1}) which is the value of the $N_\pom\Lb r^2 Q^2_s\Rb$ at $r^2 Q^2_s = 1$.  Generally speaking, $N_0$  is the non-perturbative amplitude of Pomeron-nucleon scattering at low energy. We estimate the value of this amplitude by the contribution of the BFKL Pomeron  in the saturation region replacing Green's functions for upper and low Pomerons in \fig{1di}  by unity.
 \beq \label{N0}
 N^2_0\,\,=\,\,n^2_d\,\,0.0154\frac{4 \pi^3 \bas^4}{N^2_c}\,\, C^2\Lb 1+\bar{ \gamma},1+\bar{ \gamma}\Rb\,\frac{1}{4\,\Lb  \bas\,\chi\Lb \ga_{cr}\Rb\Rb^2}
 \eeq
 where $n_d$ is the number of the dipoles in the nucleon.  In doing this estimates we assumed that the colorless dipoles are correct degrees of freedom for the non-perturbative QCD. For $\bas = 0.2$  $N^2_0\,=\,1.53 \,n^2_d$.
 We  also  included in $N^2_0$ all $\pi$'s and numerical factors that stem from integrations over $R_2$,$R_1$ and $Y'$ in \eq{PHI2}.
  
  For hadron-hadron collisions $R_A -R_p$ and
integrating over $p_{\bot, 1}$ and $p_{\bot, 2}$ we see that  $p_{\bot,1}^2 \sim Q_s\Lb p; Y_1 - Y_0\Rb$ and $p_{\bot,1}^2 \sim  p^2_{\bot,2} \exp\Lb-\bas \frac{\chi\Lb \bar{\gamma}\Rb}{\bar{ \gamma}}\,\Lb Y_1 - Y_2\Rb\Rb$  and  using  \eq{RCN11} we obtain
\bea \label{RCN12}
&&\int d^2 p_{\bot,1} d^2p_{\bot,2} \frac{d \sigma}{d Y_1 d Y_2 d^2 p_{\bot,1} d^2p_{\bot,2}}\,\,=\\
 &&~~~~~~~~~~~~~~~\,\,N^2_0\,\phi^4_0\frac{8\,C_F}{\alpha_s (2\pi)^4}\,\frac{8\,C_F}{\alpha_s (2\pi)^4}\, R^4_p R^2_p  Q^2_s\Lb p;  Y_1\Rb 
 Q^2_s\Lb p;  Y_2\Rb\,\Lb  \frac{Q_s^2\Lb p; Y- Y_1\Rb}{Q^2_s\Lb p; Y_1\Rb}\Rb^{2\bar{\ga}}\nn  
  \eea 

\begin{figure}[h]
\begin{center}
 \includegraphics[width=0.5\textwidth]{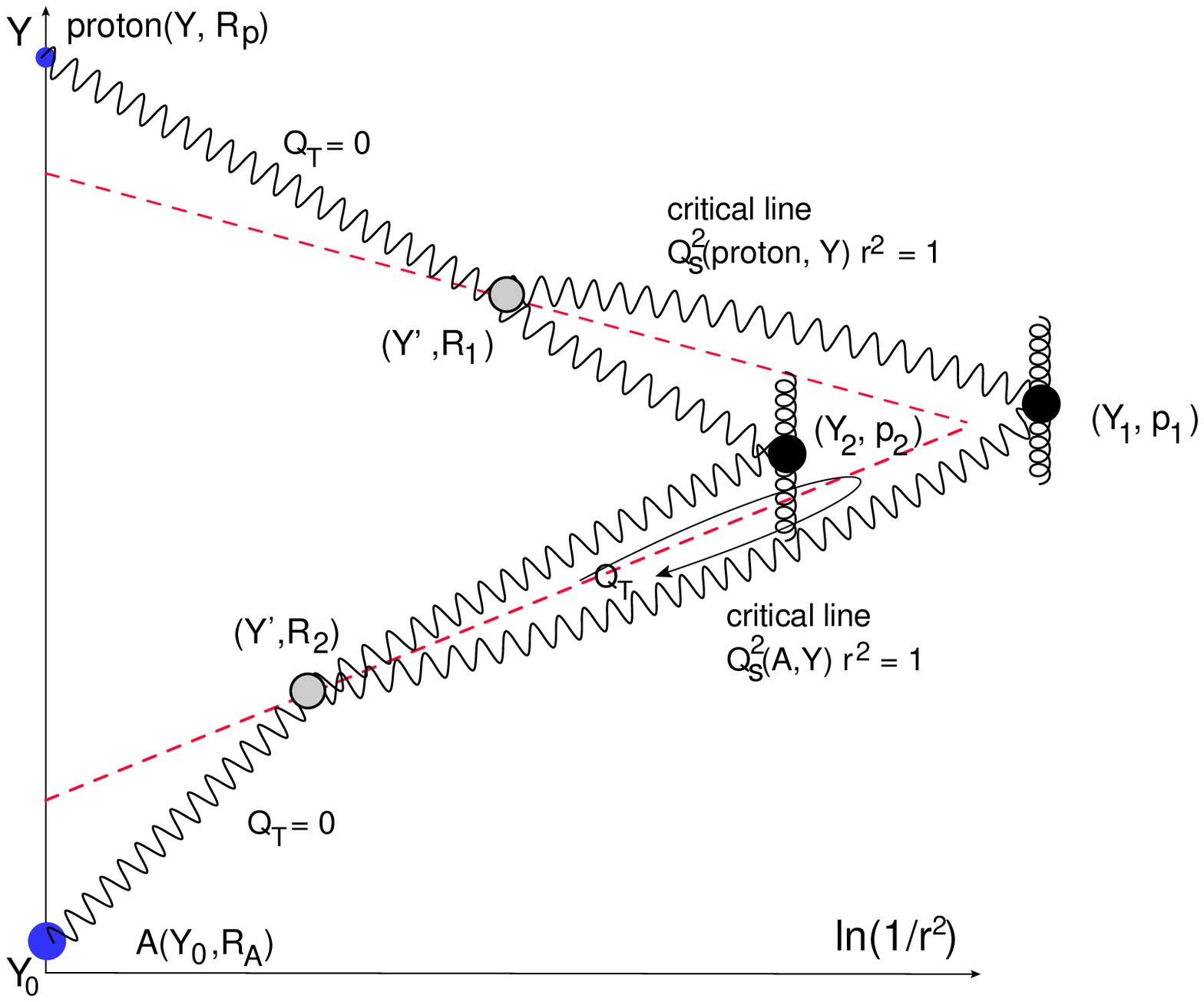} 
   \end{center}
\caption{ The first enhanced (loop) diagram for two particle correlation: the largest contribution in which the upper and lower Pomeran are in the saturation region.
  Wavy  lines denote the BFKL
 Pomerons.  Helix lines show the gluons. Black blob stands for the Mueller vertex for inclusive production of gluon jet with the transverse momentum $p_{\perp,1}$ ($p_{\perp, 2}$), respectively. Blue blobs describe the interaction of the BFKL Pomeron with the Êproton and the  nucleus. Dashed red line corresponds to the  critical line for proton-nucleus scattering. }
 \label{satcont}
\end{figure}

For
the second term of $\Big\{\dots\Big\}$ in \eq{PHI2} the main contribution stems from $R_2 = R_1 \to  r \approx r'$  and the reduced diagram is shown in \fig{1dicont}-b which describes the rapidity correlations inside one parton shower. \fig{1dicont}-a gives the contribution in the  restricted kinematic range of rapidities as has been discussed above.  The largest contribution stems from the diagram of \fig{satcont} in which the upper and lower Pomerons are in the saturation region. However, only two of four Pomerons in the loop can be near to the saturation  momentum and this diagram is needed to be calculate using new vertices   has been estimated in Ref. \cite{KOJA}. It means that this diagram has the same structure as shown in \fig{1dicont}-b and can be calculated in framework of the BFKL Pomeron calculus only for $Y_1 - Y_2 \gg 1$.  Therefore,  we can neglect this contribution for the most interesting case of long range rapidity correlations.

 %%%%%%%%%%%%%%%%%%%%%%%%%%%%%%%%%%%%%%%%%%%%%%%%%%%%
%%%%%%%%%%%%%%%%%%%%%%%%%%%%%%%%%%%%%%%%%%%%%%%%%%%
 The lesson we learned   from our calculations, is that the integration over $Y'$ and $Y''$ reduces to $Y' \to Y$ and $Y'' \to 0$ in the Mueller enhanced  diagram in the same way as for calculation of the total cross section (see Ref. \cite{LMP}.
 
 %%%%%%%%%%%%%%%%%%%%%%%%%%%%%%%%%%%%%%%%%%%%%%%%%%%%
Using \eq{RCN12} and \eq{INC9} we can estimate the rapidity correlation function,i.e.
\bea \label{RAPCOR}
&&R\Lb Y_1,Y_2; Y \Rb\, =\,  \frac{\frac{1}{\sigma_{in}}\,\int \frac{d \sigma}{d Y_1 d Y_2 d^2 p_{\bot,1} d^2p_{\bot,2}}  d p^2_{\bot,1}   d p^2_{\bot,1}}{\frac{1}{\sigma_{in}}\,\int \frac{d \sigma}{d Y_1  d^2 p_{\bot,1}}  d p^2_{\bot,1} \,  \frac{1}{\sigma_{in}}\,\int \frac{d \sigma}{d Y_2  d^2 p_{\bot,1}}  d p^2_{\bot,2}}\,\, - \,\,1\,=\, \tilde{N}^2_0\,I_Y\Lb \bar{\gamma}\Rb\frac{\sigma_{in}\Lb Y\Rb}{\pi R^2_p} \,-\,1 \label{RAPCOR}\\
&&R\Lb Y_1,p_{1.\bot};Y_2, p_{2,\bot}; Y \Rb\, =\, \frac{\frac{1}{\sigma_{in}}\,\frac{d \sigma}{d Y_1 d Y_2 d^2 p_{1,\bot} d^2 p_{2,\bot}}}{\frac{1}{\sigma_{in}}\,\frac{d \sigma}{d Y_1  d^2 p_{\bot,1}}   \,  \frac{1}{\sigma_{in}}\, \frac{d \sigma}{d Y_2  d^2 p_{2,\bot}}}\,\, - \,\,1\,=\,
 \tilde{N}^2_0\,I_Y\Lb \bar{\gamma}\Rb\frac{\sigma_{in}\Lb Y\Rb}{\pi R^2_p} \,-\,1 \label{RAPCORP}
\eea
where $I_Y\Lb \bar{\ga}\Rb$ is the coefficient that will be written in \eq{II} below. $\tilde{N} = N/n_d$ and $R$ is the size of the dipole inside of the hadron. We will discuss below both of this 
ingredients  as well as the inelastic cross section $\sigma_{in}\Lb Y \Rb$ in the next section.

%%%%%%%%%%%%%%%%%%%%%%%%%%%%%%%%%%%%%%%%%%%%%%%%%%%%
 \subsection{Azimuthal angle  correlations}
 %%%%%%%%%%%%%%%%%%%%%%%%%%%%%%%%%%%%%%%%%%%%%%%%%%%%
  As we have discussed in the introduction the azimuthal angle correlation arises from the terms $\Lb \vec{p}_{\bot,1} \cdot \vec{Q}_T\Rb^2$ and $ \Lb \vec{p}_{\bot,2} \cdot \vec{Q}_T\Rb^2$ after integration over $Q_T$ in the Pomeron loop in the diagram of \fig{1di} since \\$\int d^2 Q_T    \Lb \vec{p}_{\bot,1} \cdot \vec{Q}_T\Rb^2  \Lb \vec{p}_{\bot,2} \cdot \vec{Q}_T\Rb^2 \to \Lb  \vec{p}_{\bot,1} \cdot      \vec{p}_{\bot,2} \Rb^2$.    Such terms in the coordinate representation that we are using here, stem from the terms  $\Lb \vec{r}_{12} \cdot \vec{Q}_T\Rb^2$ and $ \Lb \vec{r}^{\,'}_{12} \cdot \vec{Q}_T\Rb^2$ in $N_\pom\Lb r_1, r_2, Y, Q_T\Rb$ and $N_\pom\Lb r'_1, r'_2, Y, Q_T\Rb$   (see \eq{NPQ1}). These terms come from $J_{\h - \bar{\gamma}}\Lb \rho^{*}_Q\,\rho_{12}\Rb \,J_{\h - \bar{\gamma}}\Lb \rho_Q\,\rho^{*}_{12}\Rb$. For small $Q_T$ we can see how there terms appear expending $J_{\h - \bar{\gamma}}$.

   Indeed,
   \bea \label{AC1}
   &&J_{\h - \bar{\gamma}}\Lb \rho^*_Q\,\rho_{12}\Rb \,J_{\h - \bar{\gamma}}\Lb \rho_Q\,\rho^*_{12}\Rb = \\
 &&=\,\,  \Lb \frac{1}{2^{\h - \bar{\gamma}}\,\Gamma\Lb \frac{3}{2} - \bar{\gamma}\Rb}\Rb^2 \Big\{ 1 + \frac{1}{2 ( - 3 + 2 \bar{\gamma})}
   Q^2_T  r^2_{12} e^{ 2 i ( \phi - \psi)}  \Big\} \Big\{1 + \frac{1}{2 ( - 3 + 2 \bar{\gamma})}
   Q^2_T r^2_{12} e^{ -2 i ( \phi - \psi)}  \Big\}\nn\\
   &&\to\,\, \Lb \frac{1}{2^{\h - \bar{\gamma}}\,\Gamma\Lb \frac{3}{2} - \bar{\gamma}\Rb}\Rb^2   \Big\{ 1 +\frac{1}{ ( - 3 + 2 \bar{\gamma})} Q^2_T r^2_{12} \cos\Lb 2  ( \phi - \psi)\Rb + \Lb \frac{1}{ 2 ( - 3 + 2 \bar{\gamma})}\Rb^2 Q^4_T r^4_{12}\Big\}\nn
   \eea
   \bea
     &&\to\,\,  \Lb \frac{1}{2^{\h - \bar{\gamma}}\,\Gamma\Lb \frac{3}{2} - \bar{\gamma}\Rb}\Rb^2   \Big\{ 1 +\frac{1}{ ( - 3 + 2 \bar{\gamma})}\Lb  2 \Lb \vec{Q}_T \cdot \vec{r}_{12}\Rb^2  - Q^2_T r^2_{12}\Rb   + \Lb \frac{1}{ 2 ( - 3 + 2 \bar{\gamma})}\Rb^2 Q^4_T r^4_{12}\Big\} \nn
     \eea  
 In \eq{AC1} we use the representation of  complex numbers in the polar coordinates, for example, $\rho_Q = Q e^{i \phi}$ and $\rho^{*}_Q = Q e^{-i \phi}$. The same type of contributions come from \eq{INC6}.

  For $Q_T r_{12} \gg 1$ we can see the same features since
  
  \bea
&&N_\pom\Lb r_1,r_2; Y, Q_T \Rb \,\,\xrightarrow{Q^2_T\,r^2_{12}\,\,\gg\,\,1 }\label{AC2}
\\
&&~~~~~~~~~~~~~~~~~~~~~~~~~~~~\,\,\phi_0\,C^2(\bar{\gamma}) \,\frac{2}{\pi}\,
r^2_{12}  \,e^{\bas \, \chi\Lb \ga_{cr}\Rb\,Y}
\,\Lb\frac{ r^2_1 r^2_2}{r^4_{12}}\Rb^{\bar{\gamma}}\,\Lb Q^2_T r^2_{12}\Rb^ { - 1 + \bar{\gamma}} \cos^2\Lb \pi \bar{\gamma}/2\Rb e^{i \vec{Q}_T \cdot \vec{r}_{12}}\nn\\
&& \Big\{ (1/8)(\bar{\gamma}\,(\bar{\gamma} - 2) ( 1 - \bar{\gamma}^2))  + Q^2_T r^2_{12}\,e^{i 2 (\phi - \psi)}\Big\} \Big\{(1/8)(\bar{\gamma}\,(\bar{\gamma} - 2) ( 1 - \bar{\gamma}^2)) + Q^2_T r^2_{12}\,e^{- i 2 (\phi - \psi)}\Big\}\Big{/}Q^4_T r^4_{12}\nn\\
&& =\,\,\,\,\phi_0\,C^2(\bar{\gamma}) \, \,\frac{2}{ \pi}\,r^2_{12}  \,e^{\bas \, \chi\Lb \ga_{cr}\Rb\,Y}
\,\Lb\frac{ r^2_1 r^2_2}{r^4_{12}}\Rb^{\bar{\gamma}}\,\Lb Q^2_T r^2_{12}\Rb^ { - 3 + \bar{\gamma}} \cos^2\Lb \pi \bar{\gamma}/2\Rb e^{i \vec{Q}_T \cdot \vec{r}_{12}} \nn\\
&&\Bigg\{ \Big[(1/8)(\bar{\gamma}\,(\bar{\gamma} - 2) ( 1 - \bar{\gamma}^2)) \Big]^2 +  (1/4)(\bar{\gamma}\,(\bar{\gamma} - 2) ( 1 - \bar{\gamma}^2))  \Lb 2 \Lb \vec{Q}_T \cdot \vec{r}_{12}\Rb^2 - Q^2_T r^2_{12}\Rb + Q_T^4 r^4_{12}\Bigg\}\nn\eea

However the largest contribution stems at $r \ll R_1 $ and $r \ll R_2$  from \eq{INC6} which can be re-written as

\bea \label{AC3}
\nabla^2_\bot\,N_\pom\Lb Y_1; r_\bot,r_{1}; Q_T \Rb\,\,&\to& \\
 &\,&\,4\,\phi_0\,C^2(\bar{\gamma}) \,r^2_{01}  \,e^{\bas \, \chi\Lb \ga_{cr}\Rb\,Y}
\,\Lb\frac{ r^2_0 r^2_1}{r^4_{01}}\Rb^{\bar{\gamma}}\,\Lb Q^2 r^2_{01}\Rb^ { - \h + \bar{\gamma}}\frac{\bar{\gamma}^2}{ r^2}\, J_{\h - \bar{\gamma}}\Lb \rho^*_Q \rho_{01}\Rb \, J_{\h - \bar{\gamma}}\Lb \rho_Q \rho^*_{01}\Rb\nn
\eea
where $r_0 \equiv r_\bot$.

Notice, that at $Q_T \to 0$ \eq{AC3} reduces to
\bea\label{AC31}
&&\nabla^2_\bot\,N_\pom\Lb Y_1; r_\bot,r_{1}; Q_T \Rb\,\,\to\\
&& \,\,4\,\phi_0\,C^2(\bar{\gamma}) \,r^2_{01}  \,e^{\bas \, \chi\Lb \ga_{cr}\Rb\,Y}
\,\Lb\frac{ r^2_0 r^2_1}{r^4_{01}}\Rb^{\bar{\gamma}}\,\Lb \frac{2^{-\h \bar{\gamma}}}{\Gamma\Lb 3/2 - \bar{\gamma}\Rb}\Rb^2\frac{\bar{\gamma}^2}{ r^2}\Big\{ 1 - \frac{1}{2 ( 3 - 2 \bar{\gamma})}\Lb \rho^2_Q \rho^{*2}_{01}\,\,+\,\, \rho^{*2}_Q \rho^{2}_{01}\Rb\Big\} \nn\\
&&= \,\,4\,\phi_0\,C^2(\bar{\gamma}) \,r^2_{01}  \,e^{\bas \, \chi\Lb \ga_{cr}\Rb\,Y}
\,\Lb\frac{ r^2_0 r^2_1}{r^4_{01}}\Rb^{\bar{\gamma}}\,\Lb \frac{2^{-\h \bar{\gamma}}}{\Gamma\Lb 3/2 - \bar{\gamma}\Rb}\Rb^2\frac{\bar{\gamma}^2}{ r^2}\Bigg\{ 1 - \frac{1}{ ( 3 - 2 \bar{ \gamma})}\, \Lb \vec{Q}_T \vec{r}_{01} \Rb^2 \Bigg\}\nn\\
&&= \,\,4\,\phi_0\,C^2(\bar{\gamma}) \,r^2_{01}  \,e^{\bas \, \chi\Lb \ga_{cr}\Rb\,Y}
\,\Lb\frac{ r^2_0 r^2_1}{r^4_{01}}\Rb^{\bar{\gamma}}\,\Lb \frac{2^{-\h \bar{\gamma}}}{\Gamma\Lb 3/2 - \bar{\gamma}\Rb}\Rb^2\frac{\bar{\gamma}^2}{ r^2}\Bigg\{ 1 - \frac{1}{ ( 3 - 2 \bar{ \gamma})}\,\Lb \vec{Q}_T \vec{r} \Rb^2\Bigg\}\nn \eea

\begin{figure}[h]
\begin{center}
 \includegraphics[width=0.5\textwidth]{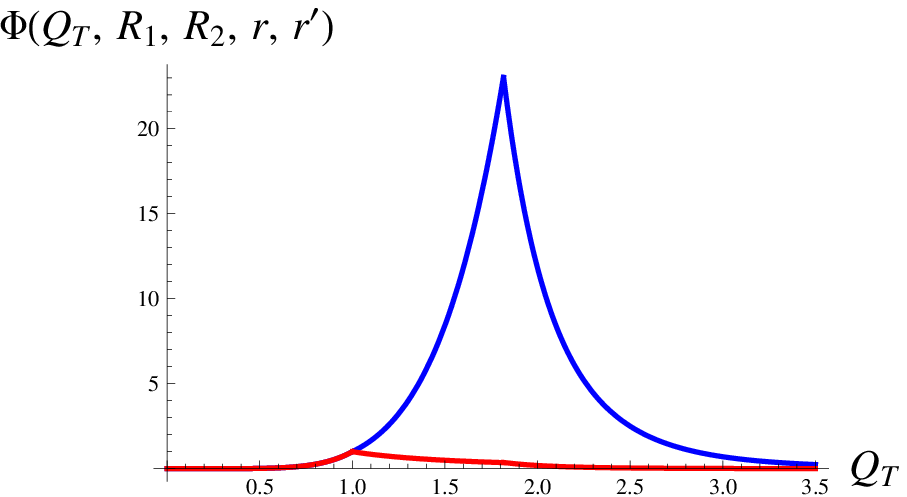} 
   \end{center}
\caption{  The integrant  $\Phi\Lb Q_T; R_1, R_2, r, r'\Rb$ (see \protect\eq{INTEGRANTQ}) versus $Q_T$. The blue curve shows the contribution of the kinematical region {2.1} while the red one corresponds to the kinematic region {\bf 2.2}. In the picture the values of $R_1$ and $R_2$ are chosen:  $R^2_1 = 1 GeV^{-2}$ and $R^2_2\,=\,0.3 \,GeV^{-2}$. }
 \label{phi}
\end{figure}

We need to re-visit the integration over $Q_T$ in \eq{RC0} and re-analyzed this integration based on \eq{AC1}-\eq{AC31}.
 Let us consider the integration over $Q_T$ in three kinematic regions assuming that  $R_1\,>\,R_2$:
 \begin{enumerate}
 \item \quad $Q_T R_1 \ll 1$ and $Q_T R_2 \ll 1$. In this kinematic region all Pomerons in the loop enter at small arguments and they  do not depend on $Q_T$. However, the azimuthal angle correlations stems from \eq{AC31} leading to the additional factor which is proportional to $\Lb \vec{Q}_T \cdot \vec{r}\Rb^2\,\Lb \vec{Q}_T \cdot \vec{r}^{\,'}\Rb^2 $. Therefore, the integration in this region leads to $Q_T \to 1/R_1$. 

  \item \quad $Q_T R_1 \sim 1$ and $Q_T R_2 \ll  1$.  As has been  discussed (see \eq{RC0})
  in this region $\nabla^2_\bot\,N_\pom\Lb Y_1 - Y''; r_\bot, R_{2}; Q_T \Rb $ and  $\nabla^2_\bot\,N_\pom\Lb Y_2 -Y''; r'_\bot, R_{2}; Q_T \Rb $ do not depend on $Q_T$ while  $\nabla^2_\bot\,N_\pom\Lb Y' - Y_1; r_\bot, R_{1}; Q_T \Rb $ and\\ $\nabla^2_\bot\,N_\pom\Lb Y' - Y_2; r'_\bot, R_{2}; Q_T \Rb $  are in the region of \eq{NPQ3} and give contribution proportional to $
  Q^{-2 (1 - \bar{\gamma})}$   each.  However, for the analysis of the integration over $Q_T$ we need to consider separately two sources of the angular correlations.
 
 {\bf 2.1} The angle correlations comes from  the Pomerons with the arguments $Q_T R_1 \,\leq 1$. In this case they stem from the contributions of \eq{AC31} leading to the $Q_T$ dependence of the following  type:
 \beq \label{ACR21}
 \int d^2 Q_T \,\Lb \vec{Q}_T \cdot \vec{r}\Rb^2\,\Lb \vec{Q}_T \cdot \vec{r}^{\,'}\Rb^2 \frac{1}{\Lb Q^2_T\Rb^{2 ( 1 - \gamma_{cr})}}\,\,\to\,\,\Lb Q^2_{T,\mbox{ max}} \Rb^{3 + 2 \gamma_{cr}}\,\,=\,\,\Lb \frac{1}{R^2_2}\Rb^{3 + 2 \gamma_{cr}}
 \eeq
 
  {\bf 2.2} The angle correlations comes from  the Pomerons with the arguments $Q_T R_2 \,>\, 1$.  The source of the angular correlations is shown in \eq{AC2}. One can see that this equation generates the contribution which is proportional to
  $\Lb Q^2_T R^2\Rb^{-2(1 - \bar{\gamma})} \Lb \vec{Q}_T \cdot \vec{r}\Rb^2\,\Lb \vec{Q}_T \cdot \vec{r}^{\,'}\Rb^2 \Big{/}\Lb  Q^2_T R^2_2 \Rb^4$ which leads to the convergent integral over $Q_T$.
    \item \quad $Q_T R_1 \gg 1$ and $Q_T R_2 \gg  1$. The integrant has the $Q_T$ dependance which is  $\Lb Q_T^2
\Rb^{-4 ( 1 - \bar{\gamma})}$ which is multiplied by the factor: $\Lb \vec{Q}_T \cdot \vec{r}\Rb^2\,\Lb \vec{Q}_T \cdot \vec{r}^{\,'}\Rb^2 \Big{/}\Lb  Q^2_T R^2_i \Rb^4$. Therefore, the integral converges.
  \end{enumerate} 
Concluding this discussion we see that the most contribution in the integral over $Q_T$ comes from the region {\bf 2.1} and 
the typical $Q_T \approx 1/R^2_2$ assuming that $R_2 < R_1$. In other words, the typical values of the impact parameters in the Pomeron loops turns out to be  about $ | \vec{b}_1 \,-\,\vec{b}_2| \, \sim \,R_2$ as in the case of rapidity correlations (see \eq{RC0} and \eq{INTQ}). One can see that $R_1 \to R_2$ from \eq{INTQ}. In \fig{phi} we plot the integrant $\Phi\Lb Q_T\Rb$ defined as
\beq \label{INTEGRANTQ}
\frac{d \sigma}{d Y_1 d Y_2 d^2 p_{\bot,1} d^2p_{\bot,2}}\,\,\propto\,\,\int d^2 Q_T\, \Phi\Lb Q_T; R_1,R_2,r,r'\Rb
\eeq
One can see that this function has a  sharp maximum at $Q_T\approx 1/R^2$.

 %%%%%%%%%%%%%%%%%%%%%%%%%%%%%%%%%%%%%%%%%%%%%%%%%%%%

%%%%%%%%%%%%%%%%%%%%%%%%%%%%%%%%%%%%%%%%%%%%%%%%%%%%

Our integral over $Q_T$  in the diagram of \fig{1disat} differs from the same integral in the case of rapidity correlations and it takes the following form for $Q_T R_2 \gg 1$
\bea \label{AC4}
\hspace{-0.5cm}&&I_{i,j,k,l}\Lb r,r',R_1,R_2\Rb\,\,=\\
\hspace{-0.5cm}&&\int d^2 Q_T\Big\{ Q_{T,i} Q_{T,j} Q_{T,k} Q_{T,l} \Big\}
 \Lb Q^2_T\Rb^{4 (1 - \bar{\gamma})} \exp\Big( i \vec{Q}_T \cdot \{ \vec{r}_{01} + \vec{r}_{02} +  \vec{r}^{\,'}_{01} + \vec{r}^{\,'}_{02}\}\Big)\,\Phi\Lb r,r';R_1,R_2,Y,Y',Y";Y_1,Y_2\Rb\nn
 \eea
where $\vec{r}_{01}= \h( \vec{r} + \vec{R}_1)$ and $\vec{r}_{02}= \h( \vec{r} + \vec{R}_2)$ while $\vec{r}^{\,'}_{01}= \h( \vec{r}^{\,'} + \vec{R}_1)$ and $ \vec{r}^{\,'}_{02}= \h( \vec{r}^{\,'} + \vec{R}_2)$ and $i,j,k,l$ are equal to 1 and 2.
$\Phi\Lb r,r';R_1,R_2,Y,Y',Y";Y_1,Y_2\Rb$ is the   function given by \eq{RCN6}.

This integral can be re-written in the form ( for brevity, we will use notation  $\Phi\Lb r,r';R_1,R_2\Rb$ instead of $ \Phi\Lb r,r';R_1,R_2,Y,Y',Y";Y_1,Y_2\Rb$).
\vspace{-0.5cm}
\bea \label{QINT}
&&I_{i,j,k,l}\Lb r,r',R_1,R_2\Rb=\\
&&= \int d^2 Q_T \Lb Q^2_T\Rb^{4 (1 - \bar{\gamma})}\Bigg(\Big\{ \nabla_{r_{01},i} \nabla_{r_{02},j} \nabla_{r'_{01},k} \nabla_{r'_{02},l} \Big\}
 \exp\Big( i \vec{Q}_T \cdot \{ \vec{r}_{01} + \vec{r}_{02} +  \vec{r}^{\,'}_{01} + \vec{r}^{\,'}_{02}\}\Big)\Bigg)\,\Phi\Lb r,r',R_1,R_2\Rb =\nn\\
 &&=\,\Lb - 1 \Rb^4\, \int d^2 Q_T \Lb Q^2_T\Rb^{4 (1 - \bar{\gamma})} \exp\Big( i \vec{Q}_T \cdot \{ \vec{r}_{01} + \vec{r}_{02} +  \vec{r}^{\,'}_{01} + \vec{r}^{\,'}_{02}\}\Big) \Bigg(\Big\{ \nabla_{r_{01},i} \nabla_{r_{02},j} \nabla_{r'_{01},k} \nabla_{r'_{02},l} \Big\}\,\Phi\Lb r,r',R_1,R_2\Rb\Bigg)\nn
 \eea 
 where function $ \Phi$ is the same as in \eq{RCN6}.  It should be noted that  the second equation in \eq{QINT} is derived using the integration by parts and taking into account that $\Phi\Lb r,r',R_1,R_2\Rb\Big{|}^{r(r') \to + \infty}_{ r(r') \to - \infty} $= 0.
 
 Considering $r$ and $r'$ being small ($\ll R_1,R_2$) we restrict ourselves by the contribution given by \eq{AC3}. \eq{QINT} can be re-written taking gradients in the form
 
 \beq \label{AC5}
 I_{i,j,k,l}\Lb r,r',R_1,R_2\Rb\,=\, \int d^2 Q_T \Lb\frac{1}{ Q^2_T}\Rb^{4 (1 - \bar{\gamma})} \exp\Big( i \vec{Q}_T \cdot \{ \vec{r}_{01} + \vec{r}_{02} +  \vec{r}^{\,'}_{01} + \vec{r}^{\,'}_{02}\}\Big)\, \frac{R_{1,i} R_{1,k}}{R^4_1}\,  \frac{R_{2,j} R_{2,l}}{R^4_2}\,\Phi\Lb r,r',R_1,R_2\Rb
\eeq

 Finally, the contribution to the double inclusive production takes the form 
 \bea \label{AC6}
&&\frac{d \sigma}{d Y_1 d Y_2 d^2 p_{\bot,1} d^2p_{\bot,2}}\,\,\propto\,\,\frac{4 \pi^2 \bas^4}{ N^2_c}\frac{8\,C_F}{\alpha_s (2\pi)^4}\,\frac{1}{p^2_{\bot,1}}\frac{8\,C_F}{\alpha_s (2\pi)^4}\,\frac{1}{p^2_{\bot,2}}\,\frac{\Lb \vec{p}_{\bot,1} \cdot \vec{p}_{\bot,2}\Rb^2}{p^2_{\bot,1}\,p^2_{\bot,2}}\nn\\ 
&&  \times \,\,
  \int\frac{r d r}{r^2} \Big(J_0\Lb p_{\bot,1} r\Rb - J_2\Lb p_{\bot,1} r\Rb \Big)\,
   \int\frac{r' d r'}{r'^2} \Big( J_0\Lb p_{\bot,2} r'\Rb - J_2\Lb p_{\bot,2} r'\Rb \Big)  \\
&&R^2_p \,R^2_p\,\bar{{\cal U}}\Lb\bar{ \gamma} \Rb\int_{r^2}^{R^2_p}  d R^2_1 \,  \delta\Lb \frac{R_1}{R_2 }- 1\Rb\,\frac{d R^2_2}{R^2_2} \int ^Y_{Y_1} d Y' \int^{Y_2}_{Y_0} d Y''\frac{1}{4} \frac{1}{R^2_1}  \frac{1}{R^2_2}\,\Phi\Lb r,r';R_1,R_2,Y,Y',Y";Y_1,Y_2\Rb\nn
 \eea

The  difference between \eq{AC6} and the calculations of rapidity correlation that has been done in the previous subsection  is in the different factor $\bar{{\cal U}}\Lb\bar{ \gamma}\Rb$ and in the extra factors $\frac{1}{4} \frac{1}{R^2_1}  \frac{1}{R^2_2}$,  However these factors do not change qualitatively  the character of integration over $R_1$ and $R_2$: the integration over $Q_T$ leads to $R_1 = R_2 = R $ and to $Q_T \approx 1/R$. Therefore, we obtain that
\beq \label{AC7}
  \frac{1}{4} \frac{1}{R^2_1}  \frac{1}{R^2_2}\,\,\longrightarrow\,\, 0.53\,\frac{1}{4} \, \frac{1}{R^4}
  \eeq
  where  the numerical factor 0.53 reflect the difference in the averaging given by \eq{RCN31}.

  \eq{AC7}  can be  translated into the following 
expression for $\frac{d \sigma}{d Y_1 d Y_2 d^2 p_{\bot,1} d^2p_{\bot,2}}$:
\bea \label{AC15}
&& \frac{d \sigma}{d Y_1 d Y_2 d^2 p_{\bot,1} d^2p_{\bot,2}}\,\,= \,\,\, I\Lb \bar{\gamma}\Rb\,N^2_0\,\bar{\phi^4_0}\frac{8\,C_F}{\alpha_s (2\pi)^4}\,\frac{8\,C_F}{\alpha_s (2\pi)^4}\frac{\Lb \vec{p}_{\bot,1} \cdot \vec{p}_{\bot,2}\Rb^2}{p^2_{\bot,1}\,p^2_{\bot,2}}\,R^2_p  \,\frac{1}{p^2_{\bot,1}\,p^2_{\bot,2}}\nn\\
&\times& \Lb \frac{Q_s^2\Lb Y - Y_1\Rb}{p^2_{\bot,1}}\Rb^{\bar{\ga}}\Lb \frac{Q_s^2\Lb Y_2 - Y_0\Rb}{p^2_{\bot,2}}\Rb^{\bar{\ga}}\Lb \frac{Q_s^2\Lb Y_1 - Y_0\Rb}{p^2_{\bot,1}}\Rb^{\bar{\ga}}\Lb \frac{Q_s^2\Lb Y - Y_2\Rb}{p^2_{\bot,2}}\Rb^{\bar{\ga}}   \nn
     \eea

In \eq{AC16} factor $ I\Lb \bar{\gamma}\Rb$ includes all numerical factors that depend on $\ga_{cr}$ .

\eq{AC15} is written in the kinematic region where all  factors $Q^2_s/p^2_{\bot,i} \,\leq\,1$.
Integrating over $p_{\bot,1}$ and $ p_{\bot,2}$ in this kinematic region we obtain for $ Y_1 < \h( Y + Y_0)$ and $Y_2 <  \h ( Y + Y_0)$ that \eq{AC15} takes the form
\bea \label{AC16}
&&\int \frac{d \sigma}{d Y_1 d Y_2 d^2 p_{\bot,1} d^2 p_{\bot,2}}  d p^2_{\bot,1}   d p^2_{\bot,1}\,\, = \\
&&\,\,\cos^2\varphi \,\,I\Lb \bar{\gamma}\Rb\,N^2_0\,
\,\phi^4_0\frac{8\,C_F}{\alpha_s (2\pi)^4}\,\frac{8\,C_F}{\alpha_s (2\pi)^4}\,R^2_p   \Lb \frac{Q_s^2\Lb Y_1 - Y_0\Rb}{Q^2_s\Lb Y - Y_1\Rb}\Rb^{\bar{\ga}}\Lb \frac{Q_s^2\Lb Y_2 - Y_0\Rb}{Q^2_s\Lb Y - Y_2\Rb}\Rb^{\bar{\ga}} \nn
     \eea 
where $\varphi$ is the azimuthal angle between vectors $\vec{p}_{\bot.1}$  and $\vec{p}_{\bot,2}$. For  the angular correlation function
in proton-proton scattering we obtain
\beq \label{AC17}
R\Lb \cos\phi\Rb =  \frac{\frac{1}{\sigma_{in}}\,\int \frac{d \sigma}{d Y_1 d Y_2 d^2 p_{\bot,1} d^2p_{\bot,2}}  d p^2_{\bot,1}   d p^2_{\bot,1}}{\frac{1}{\sigma_{in}}\,\int \frac{d \sigma}{d Y_1  d^2 p_{\bot,1}}  d p^2_{\bot,1} \,  \frac{1}{\sigma_{in}}\,\int \frac{d \sigma}{d Y_2  d^2 p_{\bot,1}}  d p^2_{\bot,2}}\,\, - \,\,1\,\,=\,\,\\cos^2\phi\,\,I\Lb \bar{\gamma}\Rb\widetilde{N}^2_0\,\frac{\sigma_{in}\Lb Y \Rb}{R^2_p}\frac{1}{R^2_p Q^2_s\Lb Y_1\Rb\,R^2_p Q^2_s\Lb Y_2\Rb}\eeq
where $\widetilde{N}_0 \,\,=\,\,N^2_0/n^2_d$ and
\beq \label{II}
I\Lb \bar{\gamma}\Rb\,\,=\,\,I_Y\Lb \bar{\gamma}\Rb\, \frac{\bar{\ga}}{2\,(3 - 2 \bar{\ga})}\,\,=\,\,0.53\,\frac{1}{2 \bar{\gamma} - 1}\,\,\frac{1}{4}\,\Bigg(\frac{1}{\Lb 3 - 2\bar{\gamma}\Rb}\Bigg)^2\frac{4}{ \pi^2}\, \Lb \frac{2^{-\h+ \bar{\gamma}}}{\Gamma\Lb 3/2 - \bar{\gamma}\Rb}\Rb^{-2} \,\,\,\approx\,\,0.068
\eeq
In \eq{II} we include factors from \eq{U} and \eq{AC31}. Comparing this factor with the rapidity correlation we see that the azimuthal angle correlations are suppressed by factor  $0.63/\Lb 2\,(3 - 2\,\bar{\gamma})\Rb^2 \approx 0.05$. In addition we consider $Y_1 \,> \,Y - Y_1$ and $ Y_2 \,>\,Y - Y_2$. $R_p = R$
is the size of the typical dipole inside of the hadron. One can see that the coefficient in front of $\cos^2\varphi$ is rather small in the angular correlation function mostly due to the large large multiplicity of the inclusively produced gluons which is proportional to $Q_s$. It worthwhile mentioning that such suppression we did not see in the rapidity correlation function (see \eq{RAPCOR}). This suppression can be easily understood. Indeed, the angular correlations appear due to factor $\Lb \vec{r} \cdot \vec{Q}_T\Rb^2\,\Lb \vec{r}^{\,'} \cdot \vec{Q}_T\Rb^2$
which has the value of the order $\Lb  \langle Q^2_T\rangle/p^2_{1,\bot}\Rb\,  \Lb\langle Q^2_T\rangle/p^2_{2,\bot}\Rb$. As has been shown the average $Q_T$ turns out to be of the order of $Q_T \sim 1/R$ where $R$ is the size of the dipole in a hadron, while $ p_\bot \sim Q_s$. Therefore, we expect that the angle correlations are suppressed as $1/(R^2Q^2_s)^2$, as we saw in \eq{AC17}.

The coefficient in front of $\cos^2 \varphi$ is not suppressed  showing the possibility to find another definition of the correlation function with  enhanced contribution of the angular correlations.

\begin{figure}[h]
\begin{center}
 \includegraphics[width=0.5\textwidth]{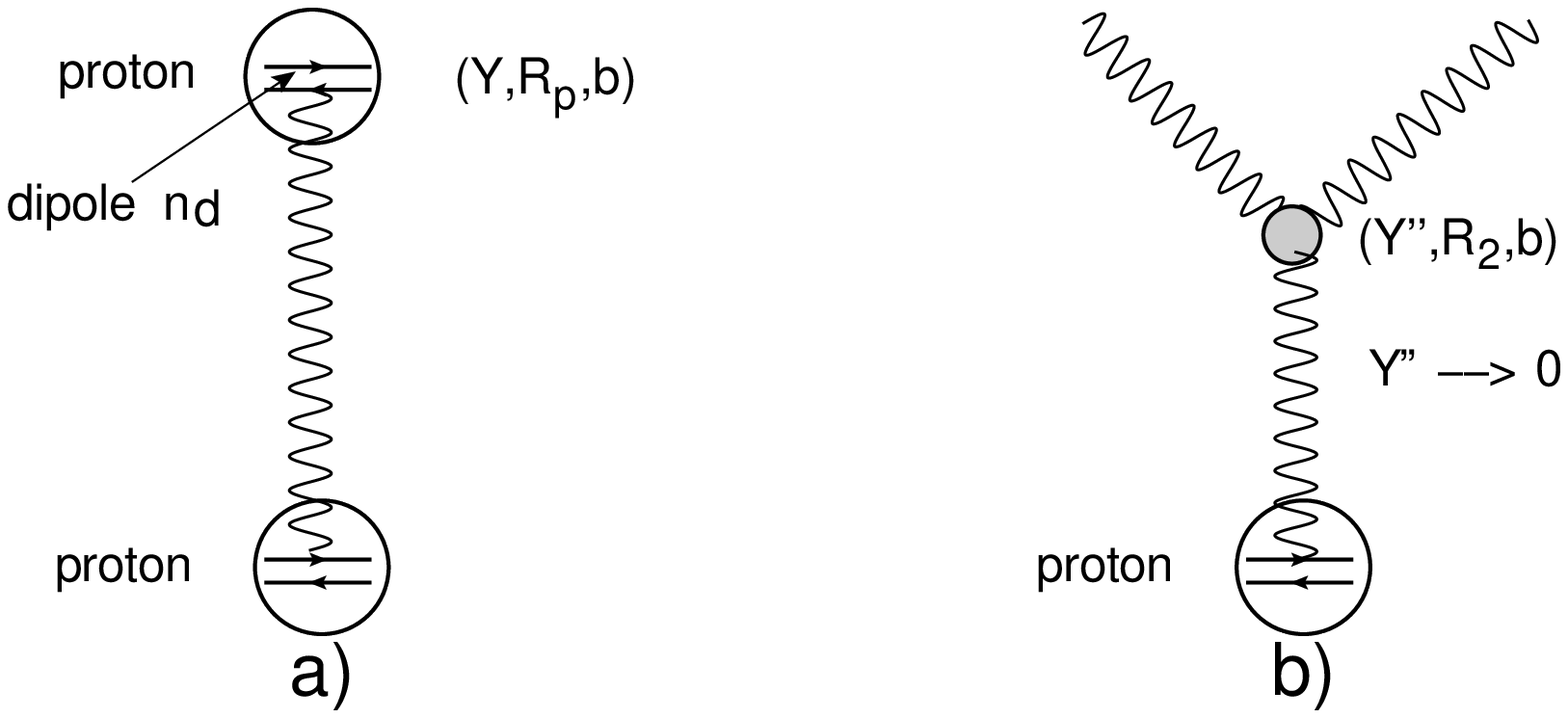}
\end{center}
 \caption{$\sigma_{in}$ (see \protect\fig{n0p}-a) and $\widetilde{N}_0$ (see \protect\fig{n0p}-b) in \protect\eq{AC17}.}
 \label{n0p}
\end{figure}

  It is interesting to note that the correlations at fixed $p_{\bot,1 } \sim Q_s$ and  $p_{\bot,2} \sim Q_s$   take  the following form
  
  \beq \label{ACPT} 
  R\Lb \cos\varphi, p_{\bot,1},p_{\bot,2}\Rb\,\,=\,\,\sigma_{in}\frac{\frac{d \sigma}{d Y_1 d Y_2 d^2 p_{\bot,1} d^2p_{\bot,2}} } {\frac{d \sigma}{d Y_1 d^2 p_{\bot,1} }\,\, \frac{d \sigma}{ d Y_2 d^2p_{\bot,2}} }\,\,=\,\,\cos^2\varphi\,  I\Lb \bar{\gamma}\Rb\widetilde{N}^2_0\,\frac{\sigma_{in}\Lb Y\Rb}{\pi R^2} \,\frac{1}{R^2_p \,p^2_{1,\bot}\,R^2_p\,p^2_{2,\bot}}
  \eeq
$\sigma_{in}$ is the inelastic cross section for dipole-dipole scattering.  From \fig{n0p} we see that $n_d$'s ( the vertices of proton-BFKL Pomeron) cancel in the ratio of \eq{AC17}.  We need the phenomenological approach for the soft high energy scattering based on CGC/saturation approach ( see Ref.\cite{LEDD} for first try to  estimate $\sigma_{in}$ in dipole-dipole scattering using a minimal phenomenological input).  In numerical evaluations  we use , in spirit of our approach to $N_0$ ( see \eq{N0}),  $\sigma_{in} = \sigma_{in}\Lb \mbox{proton-proton}\Rb/9$ assuming that we have 3 dipoles in a proton. For estimates, we consider $\sigma_{in} \,=\,\sigma_{tot} - \sigma_{el} - \sigma_{diff}$. The size of the dipole inside of the proton we chose to be equal  $ R^2  \approx 1\,GeV^{-2}$( see Ref.\cite{LEDD}).

 Substituting these values we obtain $R\Lb \cos\phi\Rb\,\,\approx\,\,\Lb 0.48/\Lb R^2_p Q^2_s\Lb Y_1\Rb\,R^2_p Q^2_s\Lb Y_2\Rb\Rb\Rb \, \cos^2 \varphi$ at the LHC energy $W= 7 GeV$ in the central region of rapidity 
 $Y_1=Y_2=Y/2$. The large coefficient in front of $\cos^2 \varphi$ indicates the enhanced diagrams can give a large contribution to the angular correlations. However, it should be noted that the accuracy of our estimates are rather low due to uncertainty in  both the value of $R$ and the number of dipoles in the proton.
  
   %%%%%%%%%%%%%%%%%%%%%%%%%%%%%%%%%%%%%%%%%%%%%%%%%%%

%%%%%%%%%%%%%%%%%%%%%%%%%%%%%%%%%%%%%%%%%%%%%%%%%%%%
We integrate \eq{INC9}over $p_\bot$ for calculating the single inclusive cross sections in the dominator of \eq{AC17}.

%%%%%%%%%%%%%%%%%%%%%%%%%%%%%%%%%%%%%%%%%%%%%%%%%%%%
\section{The first enhanced diagram in the saturation environment}
%%%%%%%%%%%%%%%%%%%%%%%%%%%%%%%%%%%%%%%%%%%%%%%%%%%%
In proton-nucleus scattering the simplest diagram of \fig{1di} cannot be considered as a good approximation for $p_{\bot} \to Q_s$ since the Pomerons interact with the dense target (with nucleus). In \fig{1dienv} we demonstrate several examples of such interactions.  Such interaction leads to the saturation of the parton (gluon) density\cite{GLR,MUQI,MV} and in this section we wish to discuss the BFKL Pomeron Green's function in the kinematic region when the Pomeron interaction with nucleus become essential.

      \subsection{Equation for  Pomeron Green's function}

\begin{figure}[h]
\begin{minipage}{11cm}
\begin{center}
 \includegraphics[width=0.50\textwidth]{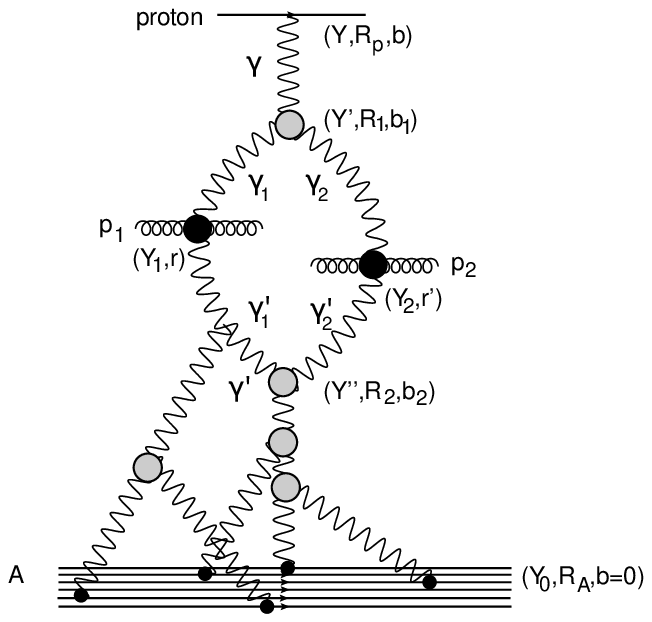}
\end{center}
\end{minipage}
 \begin{minipage}{6.5cm}
\caption{The interactions with the nucleons  of a nucleus in the first enhanced diagram.
  Wavy  lines denote the BFKL
 Pomerons.  Helix lines show the gluons. Black blob stands for the Mueller vertex for inclusive production of gluon jet with the transverse momentum $p_{\perp,1}$ ($p_{\perp, 2}$), respectively. }
 \label{1dienv}
\end{minipage}
\end{figure}

\begin{figure}[h]
\begin{center}
 \includegraphics[width=0.7\textwidth]{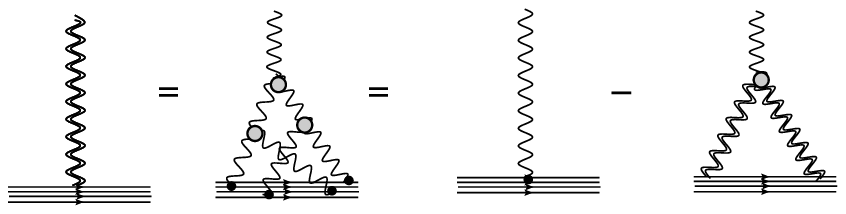}
\end{center}
\caption{ The graphic form of Balitsky-Kovchegov equation: fan diagrams. The double wavy lines denote the resulting Green's function of the BFKL  Pomeron.
  Wavy  lines denote the BFKL Pomeron. The gray circles  show the triple Pomeron vertices.}
 \label{fandi}
\end{figure}

   For dilute-dense parton systems scattering  the main contribution stem from `fan' diagrams\cite{GLR,BRN} (see \fig{1dienv}).
   In the case of the lower Pomeron in the first enhanced diagram (see \fig{1di} and \fig{1dienv}) the sum of `fan' diagrams leads to Balitsky-Kovchegov (BK) equation \cite{BK} (see \fig{fandi}), which takes the form
\bea \label{BK}
&&\frac{\partial N_\pom \Lb Y'' - Y_0; R_2,R_A,\vec{b}_2\Rb}{\partial Y''}\,\,=\,\,\bas \int d^2 R'_2 K\Lb R_2 | R'_2\Rb\Bigg\{    
N_\pom \Lb Y'' - Y_0; R'_2,R_A,\vec{b}_2 - \h ( \vec{R}_2 - \vec{R}'_2) \Rb\,\nn\\
&&+\,N_\pom \Lb Y'' - Y_0; | \vec{R}_2 - \vec{R}'_2|,R_A,\vec{b}_2 - \h\vec{R}'_2 \Rb
\,-\,N_\pom \Lb Y'' - Y_0; R_2,R_A,\vec{b}_2  \Rb\nn\\
&&\,-\,\,N_\pom \Lb Y'' - Y_0; R'_2,R_A,\vec{b}_2 - \h ( \vec{R}_2 - \vec{R}'_2) \Rb\,N_\pom\Lb Y'' - Y_0; | \vec{R}_2 - \vec{R}'_2|,R_A,\vec{b}_2 - \h\vec{R}'_2 \Rb\Bigg\}
\eea
where the vertex $ K\Lb R_2 | R'_2\Rb$ describes the decay of the dipole with the size $R_2$ to two dipoles with sizes: $R'_2$ and $| \vec{R}_2 - \vec{R}'_2|$ and it is given by \eq{K}.

%%%%%%%%%%%%%%%%%%%%%%%%%%%%%%%%%%%%%%%%%%%%%%%%%%%%

%%%%%%%%%%%%%%%%%%%%%%%%%%%%%%%%%%%%%%%%%%%%%%%%%%%%
However BK equation is not enough since we need to find Green's function of the BFKL Pomeron that  propagates from coordinate $( Y'', R_2)$ to coordinate $(Y_1, r)$ and/or $(Y_2,r')$ and interacts with the nucleus. This interactions are carried by the `fan' diagrams as one can see in \fig{eqgf}. The equation for Green's function takes the form (see the graphic representation  in \fig{eqgf}):
\bea \label{EQGF}
 G\Lb Y, r, b; Y', r'; b'\Rb\,&=&\, G^{\mbox{\tiny BFKL}}\Lb Y  - Y', r;  r'; \vec{b}  - \vec{b}^{\,'}\Rb\,-\,\int^Y_{Y'} \!\!\!d Y''\!\!\! \int\!\!\! \!\!\int\! d^2 r'' d^2 b''\,G^{\mbox{\tiny BFKL}}\Lb Y - Y'', r; r''; \vec{b}- \vec{b}^{\,''}\Rb \, \nn\\
 &\times&K\Lb r'',\hat{r}\Rb G\Lb Y'', \hat{r},\vec{b}^{\,''}-\h(\vec{r}^{\,''} - \hat{\vec{r}}),Y',r',b'\Rb N_\pom\Lb Y'', \vec{r}^{\,''} - \hat{\vec{r}}; R_A; \vec{b}^{\,''}-\h \hat{\vec{r}}\Rb
 \eea
 
 From \fig{eqgf} one can see that we can write  the second equation for Green's function summing the diagrams as an evolution in $Y'$. The equation is 
 \bea \label{EQGF1}
 G\Lb Y, r, b; Y', r'; b'\Rb\,&=&\, G^{\mbox{\tiny BFKL}}\Lb Y  -Y', r;, r'; \vec{b}-\vec{b}^{\,'}\Rb\,-\,\int^Y_{Y'} \!\!\!d Y''\!\!\! \int\!\!\! \!\!\int\! d^2 r'' d^2 b''\,G\Lb Y, r, b; Y'', r'', b''\Rb \, \\
 &\times&K\Lb r'',\hat{r}\Rb G^{\mbox{\tiny BFKL}}\Lb Y'' - Y',  \hat{r}, r';\vec{b}^{\,''} - \vec{b}^{\,'}-\h(\vec{r}^{\,''} - \hat{\vec{r}})\Rb N_\pom\Lb Y'', \vec{r}^{\,''} - \hat{\vec{r}}; R_A; \vec{b}^{\,''}-\h \hat{\vec{r}}\Rb\nn
 \eea 
 %%%%%%%%%%%%%%%%%%%%%%%%%%%%%%%%%%%%%%%%%%%%%%%%%%%%

%%%%%%%%%%%%%%%%%%%%%%%%%%%%%%%%%%%%%%%%%%%%%%%%%%%%

\eq{EQGF} can be re-written in the differential form 
 \bea \label{EQGFD} 
\frac{\partial G\Lb Y, r, b; Y', r'; b'\Rb}{\partial Y}\,\,&=&\,\,\bas \int d^2 r'' K\Lb r | r''\Rb\Bigg\{ G\Lb Y, r'',\vec{b}-\h(\vec{r} - \vec{r}^{\,''}); Y', r', b'\Rb\nn\\
&+&\, G\Lb Y, \vec{r} - \vec{r}^{\,''},\vec{b}-\h \vec{r}^{\,'};Y', r', b'\Rb\,-\,G\Lb Y, r, b; Y', r', b'\Rb\nn\\
&-&\,\,  G\Lb Y, r'',\vec{b}-\h(\vec{r} - \vec{r}^{\,''}),  Y', r', b'\Rb N_\pom\Lb Y, \vec{r} - \vec{r}^{\,''}; R_A; \vec{b}-\h \vec{r}^{\,'}\Rb
\Bigg\}
\eea
while \eq{EQGF1} takes the form
 \bea \label{EQGFD1} 
\frac{\partial G\Lb Y, r, b; Y', r', b'\Rb}{\partial Y'}&=&\,\,-\bas \int d^2 r'' K\Lb r''| r'\Rb\Bigg\{ G\Lb Y, r, b; Y', r'';\vec{b}^{\,'}-\h(\vec{r}^{\,'} - \vec{r}^{\,''})\Rb\nn\\
&+&\, G\Lb Y, r, b ; Y', \vec{r}^{\,'} - \vec{r}^{\,''}, \vec{b}^{\,'}-\h \vec{r}^{\,'}\Rb\,-\,G\Lb Y, r, b; Y', r', b'\Rb\nn\\
&-&\, G\Lb Y, r, b ; Y', r'';\vec{b}-\h(\vec{r}^{\,'} - \vec{r}^{\,''}\Rb N_\pom\Lb Y',  \vec{r}^{\,'} - \vec{r}^{\,''}; R_A; \vec{b}-\h \vec{r}^{\,'}\Rb\Bigg\}
\eea
Introducing $\bar{G}\Lb Y, r, b; Y', r', b'\Rb\,\,=\,\,r'^4\bar{G}\Lb Y, r, b; Y', r', b'\Rb$ we can re-write \eq{EQGFD1} as follows\footnote{We will use below for function $\bar{G}$ notation $G$ and , hope, it will not cause any misunderstanding.}
\bea \label{EQGFD2}
 \frac{\partial \bar{G}\Lb Y, r, b; Y', r', b'\Rb}{\partial Y'} &=&\,\,-\bas \int d^2 r'' K\Lb r' | r''\Rb\Bigg\{ \bar{G}\Lb Y, r, b; Y', r'';\vec{b}^{\,'}-\h(\vec{r}^{\,'} - \vec{r}^{\,''})\Rb\nn\\
&+&\, \bar{G}\Lb Y, r, b ; Y', \vec{r}^{\,'} - \vec{r}^{\,''}, \vec{b}^{\,'}-\h \vec{r}^{\,'}\Rb\,-\,\bar{G}\Lb Y, r, b; Y', r', b'\Rb\nn\\
&-&\, \bar{G}\Lb Y, r, b ; Y', r'';\vec{b}-\h(\vec{r}^{\,'} - \vec{r}^{\,''}\Rb N_\pom\Lb Y',  \vec{r}^{\,'} - \vec{r}^{\,''}; R_A; \vec{b}-\h \vec{r}^{\,'}\Rb\Bigg\}
\eea

One can see from \eq{EQGFD} and \eq{EQGFD2}  that $G\Lb r,Y, b; r',Y',b' \Rb$ can be factorized as
\beq \label{FACTOR}
G\Lb r,Y, b; r',Y',b' \Rb\,\,=\,\,G\Lb r,Y, b\Rb\,G\Lb r',Y',b' \Rb
\eeq

%%%%%%%%%%%%%%%%%%%%%%%%%%%%%%%%%%%%%%%%%%%%%%%%%%%%
\subsection{Solutions}
%%%%%%%%%%%%%%%%%%%%%%%%%%%%%%%%%%%%%%%%%%%%%%%%%%%%   
\subsubsection{The toy model}
%%%%%%%%%%%%%%%%%%%%%%%%%%%%%%%%%%%%%%%%%%%%%%%%%%%%
It is instructive to start the search for the solution to \eq{EQGFD} in the simple toy-model \cite{MUCD,LELU} in which all dipoles are assumed to have the same size. In this model the BFKL equation looks as follows
\beq \label{TM1}
\frac{ d N^{\mbox{\tiny BFKL}}_\pom( Y,Y')}{ d Y}\,\,=\,\,\Delta_\pom N^{\mbox{\tiny BFKL}}_\pom( Y , Y')\,\,\,\mbox{with solution}~~~~ N^{\mbox{\tiny BFKL}}_\pom\Lb Y - Y'\Rb\,= N_0\,e^{\Delta_\pom \Lb Y - Y'\Rb}
\eeq

where $N_0$ is the value of the amplitude at $Y = Y'$. Note, that Green's function of the BFKL Pomeron in this model is equal to $ G^{\mbox{\tiny BFKL}}_\pom( Y,Y') \,=\,\exp\Lb \Delta_\pom \Lb Y - Y'\Rb\Rb$.

The BK equation takes the form
\beq \label{TM2}
\frac{ d N^{\mbox{\tiny BK}}_\pom( Y,Y_0)}{ d Y}\,\,=\,\,\Delta_\pom\Big\{  N^{\mbox{\tiny BK}}_\pom( Y , Y_0)\,\,-\,\,\,\Lb N^{\mbox{\tiny BK}}_\pom( Y , Y_0)\Rb^2\Big\}
\eeq

   %%%%%%%%%%%%%%%%%%%%%%%%%%%%%%%%%%%%%%%%%%%%%%%%%%%
\begin{figure}[h]
\begin{center}
 \includegraphics[width=0.8\textwidth]{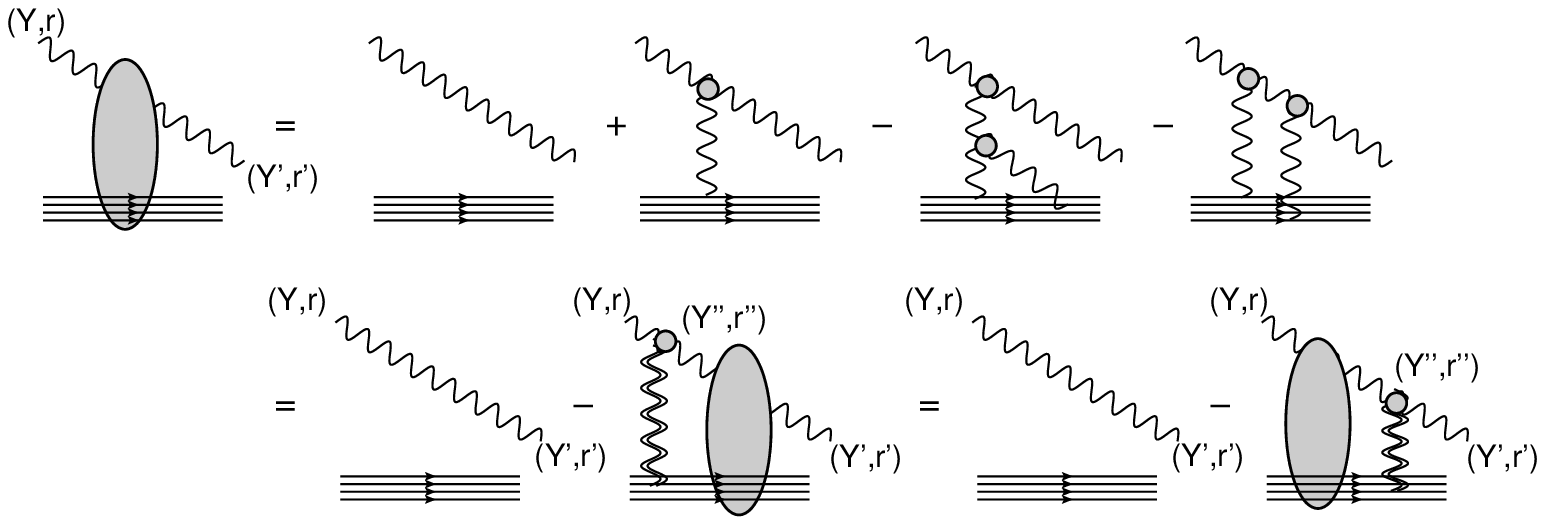}
\end{center}
\caption{ Equation for Green's function of the resulting BFKL Pomeron. The double wavy lines denote Green's function from the BK  equation ( see \protect \fig{fandi}).
  Wavy  lines denote the BFKL Pomeron. The gray circles show the triple Pomeron vertices.  The gray blobs describe the resulting Green's function.}
 \label{eqgf}
\end{figure}

%%%%%%%%%%%%%%%%%%%%%%%%%%%%%%%%%%%%%%%%%%%%%%%%%%%%
Solution to this equation can be found and it looks as follows
\beq \label{TM3}
N^{\mbox{\tiny BK}}_\pom( Y,Y_0)\,\,=\,\,\frac{N_0}{N_0   + \Lb 1 -  N_0\Rb \exp\Lb - \Delta_\pom \Lb Y - Y_0\Rb\Rb}
\eeq
where $N_0 $ is the value of the amplitude at $Y = Y_0$. One can see that $N_\pom \to 1$ at $Y \gg 1$ reproducing the saturation in this model.

\eq{EQGFD} can be re-written in the following form
 \beq \label{TM4} 
\frac{d G\Lb Y,  Y'\Rb}{d Y}\,\,=\,\,\Delta_\pom\Bigg\{ G\Lb Y,  Y'\Rb
\,\,-\,\,  G\Lb Y, Y' \Rb N^{\mbox{\tiny BK}}_\pom\Lb Y, Y_0\Rb\Bigg\}
\eeq

It has solution in the form
\beq \label{TM5}
G\Lb Y,  Y'\Rb\,\,=\,\,e^{\Delta_\pom \Lb Y - Y'\Rb}\,\frac{ N_0 e^{\Delta Y'}  + 1 - N_0}{N_0 e^{\Delta Y}  + 1 - N_0}\,\,=\,\,\frac{N^{\mbox{\tiny BK}}_\pom( Y,Y_0)}{N^{\mbox{\tiny BK}}_\pom( Y',Y_0)}
\eeq

First, at small $Y$ and $Y'$ when both $N_0  e^{\Delta Y'} \ll 1$ and $N_0  e^{\Delta Y} \ll 1$ \eq{TM5} reduces to the Green function of the BFKL Pomeron in this model $G^{\mbox{\tiny BFKL}}\Lb Y, Y'\Rb \,\,=\,\,\exp\Lb \Delta_\pom ( Y - Y')\Rb$.
Second, at $Y \gg 1$ and $Y' \gg 1$ $G\Lb Y,Y'\Rb$ is saturated reaching unity. Third, at $Y'=0$ $N_0 G(Y, 0 ) \,= \,N^{\mbox{\tiny BK}}_\pom( Y,0)$ as it should be from the diagrams of \fig{eqgf}.

%%%%%%%%%%%%%%%%%%%%%%%%%%%%%%%%%%%%%%%%%%%%%%%%%%%%
\subsubsection{Equations in the momentum representation}
%%%%%%%%%%%%%%%%%%%%%%%%%%%%%%%%%%%%%%%%%%%%%%%%%%%%%%%%%%%
 \eq{EQGFD} and \eq{EQGFD1}  look simpler in the momentum representation  defined as
 \beq \label{EQM1}
  G\Lb Y, r, b; Y', r', b'\Rb\,\,\,=\,\,r^2 \,r'^2\,\int \frac{ d^2 k}{(2 \pi)^2} \frac{ d^2 k'}{(2 \pi)^2}  G\Lb Y, k, b; Y', k', b'\Rb 
 \eeq
Considering $b\, \gg \,r $ and $b ' \, \gg \,r'$ one can see that  \eq{EQGFD} and \eq{EQGFD1}  take the forms
\bea 
&&\frac{\partial  G\Lb Y, k, b; Y', k', b'\Rb}{\partial Y}\,\,= \label{EQM21}\\
&&~~~~~~\,\,\bas \Big\{ \int d^2 k'' K^{\tiny{BFKL}}\Lb k, k''\Rb\,G\Lb Y, k'', b; Y', k', b'\Rb 
\,-\,G\Lb Y, k, b; Y', k', b'\Rb \,N\Lb Y, k, b\Rb\Big\}\,;\nn\\
&&\frac{\partial G\Lb Y, k, b; Y', k', b'\Rb}{\partial Y'}\,\,=\label{EQM22}\\
&&~~~~~~ \,\,-\bas \Big\{ \int d^2 k'' K^{\tiny{BFKL}}\Lb k'', k'\Rb\,G\Lb Y, k, b; Y', k'', b'\Rb 
\,-\,G\Lb Y, k, b; Y', k', b'\Rb \,N\Lb Y', k', b'\Rb\Big\}\,;\nn
\eea 
where the BFKL kernel   $  K^{\tiny{BFKL}}\Lb k, k''\Rb$ takes the form\cite{BFKL,REV}
\beq \label{KERBFKLM}
 K^{\tiny{BFKL}}\Lb k, k''\Rb\,\,=\,\,\frac{1}{\Lb \vec{k}\,-\,\vec{k}^{\,'}\Rb^2}\,\,-\,\,\h\int \frac{k^2\,d^2 k''}{k''^2\,\Lb \vec{k}\,-\,\vec{k}^{\,''}\Rb^2}
 \,\delta^{(2)}\Lb \vec{k}\,-\,\vec{k}^{\,'}\Rb
 \eeq

The  graphical form of these equations is the same as for  \eq{EQGFD} and \eq{EQGFD1} in \fig{eqgf} where we need to replace $r  \to k$ and $r' \to k'$. It should be stressed that for large $b$ and $b'$   $b = b'$ since the BFKL kernel does not change the impact parameters of the dipoles.

\begin{figure}[h]
\begin{center}
 \includegraphics[width=0.8\textwidth]{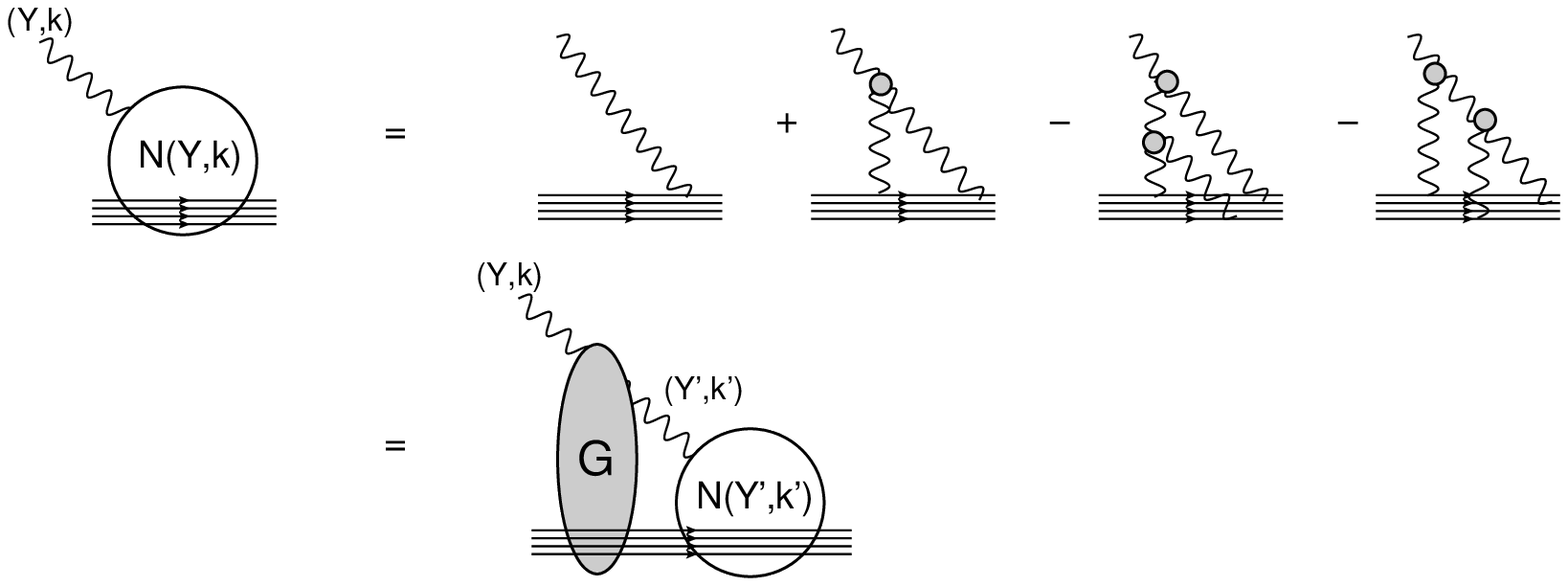}
\end{center}
\caption{ Equation for Green's function of the resulting BFKL Pomeron.  Wavy  lines denote the BFKL Pomerons. The gray circles show the triple Pomeron vertices.  The gray blobs describe the resulting Green's function:  $G\Lb Y, k, b; Y', k', b'\Rb$.
 The white blobs denote denote Green's function from the BK  equation ( see \protect \fig{fandi}).
}
 \label{eqad}
\end{figure}

In \fig{eqad} we show the graphical form of the equation for Green's function $G\Lb Y, k, b; Y', k', b'\Rb$ in a different form:
\beq \label{ADEQ}
\frac{\partial  N\Lb Y, k, b\Rb}{\partial Y}\,\,= \,\,
 \int d^2 k' G\Lb Y, k, b; Y', k', b\Rb N\Lb Y', k', b\Rb
 \eeq
In derivation opf \eq{ADEQ} we use the following property of Green's function of the BFKL Pomeron  which follows directly from t-channel unitarity\cite{GLR,MUSH}:
\beq \label{EQM3}
 G^{\tiny{BFKL}}\Lb Y, k, b; Y', k', b\Rb\,\,=\,\,\int d^2 k'' \, G^{\tiny{BFKL}}\Lb Y, k, b; Y'', k'', b\Rb\,G^{\tiny{BFKL}}\Lb Y'', k'', b; Y', k', b\Rb\,  
 \eeq
 
  %%%%%%%%%%%%%%%%%%%%%%%%%%%%%%%%%%%%%%%%%%%%%%%%%%%
  %%%%%%%%%%%%%%%%%%%%%%%%%%%%%%%%%%%%%%%%%%%%%%%%%%%
  
 All these equations should be solved with the following initial and boundary conditions which follow directly from
 \fig{eqgf} and \fig{eqad}, i.e.
 \bea \label{SOLIC}
 && G\Lb Y, k, b; Y', k', b\Rb \,\,\,\xrightarrow{Y' \to Y}\, \,\,G^{\tiny{BFKL}}\Lb Y, k, b; Y' \to Y , k', b\Rb \,\,=\,\,\delta^{(2)}\Lb \vec{k} \,-\,\vec{k}^{\,'}\Rb\,;\nn\\
&&\int d^2 k'\,  G\Lb Y, k, b; Y' = 0, k', b\Rb\,N\Lb Y' = 0, k',b\Rb\,\, \,=\,\,\,\,N\Lb Y, k, b\Rb\,\eea
 where the second equation follows from \eq{ADEQ}.
  
 %%%%%%%%%%%%%%%%%%%%%%%%%%%%%%%%%%%%%%%%%%%%%%%%%%%%
\subsubsection{Solution to the equations}

%%%%%%%%%%%%%%%%%%%%%%%%%%%%%%%%%%%%%%%%%%%%%%%%%%%%%%%%%%%
We suggest  the following ansatz for the solution
\beq \label{ANSTZ}
  G\Lb Y, k, b; Y', k', b\Rb\,\,=\,\,\frac{N\Lb Y,k, b \Rb}{N\Lb Y', k', b \Rb}\,\delta^{(2)}\Lb \vec{k} \,-\,\vec{k}^{\,'}\Rb\,;
\eeq
This ansatz is motivated by the solution fot the toy model (see \eq{TM5}) and satisfies the boundary and initial conditions of \eq{SOLIC}. Now let us checked that \eq{EQM21} and \eq{EQM22} are satisfied with this ansatz.

Substituting \eq{ANSTZ} in \eq{EQM21} we see that this equation reduces to 
\bea \label{SOLEQ1}
&&\frac{1}{N\Lb Y',k', b \Rb} \frac{\partial N\Lb Y, k,b  \Rb}{\partial Y}\delta^{(2)}\Lb \vec{k} \,-\,\vec{k}^{\,'}\Rb\,\,=\\
\,\,
&&=\bas \frac{1}{N\Lb Y',k', b \Rb}\Big\{ K^{\tiny{BFKL}}\Lb k, k'\Rb\,N\Lb Y,k'\Rb\,\,-\,\,N^2\Lb Y, k, b \Rb \delta^{(2)}\Lb \vec{k} \,-\,\vec{k}^{\,'}\Rb\Big\}\nn
\eea
Multiplying both part of the equation by $N\Lb Y',k', b \Rb$ and integrating over $k'$ one can see that \eq{SOLEQ1} takes the form
\beq \label{SOLEQ2}
\frac{\partial N\Lb Y, k,b  \Rb}{\partial Y}\,\,=\,\,
\bas \Big\{ \int d^2 k' \,K^{\tiny{BFKL}}\Lb k, k'\Rb\,N\Lb Y,k'\Rb\,\,-\,\,N^2\Lb Y, k, b \Rb\Big\}
\eeq
which is the Balitsky-Kovchegov equation  for $N$.

Plugging \eq{ANSTZ} in \eq{EQM22} leads to the following expression
\bea 
&&\frac{N\Lb Y,k,b\Rb}{N^2\Lb Y',k', b \Rb} \frac{\partial N\Lb Y', k',b  \Rb}{\partial Y'}\delta^{(2)}\Lb \vec{k} \,-\,\vec{k}^{\,'}\Rb
\,\,=\\
&&=\,\,\bas \frac{N\Lb Y,k,b\Rb}{N^2\Lb Y',k', b \Rb} \Big\{ K^{\tiny{BFKL}}\Lb k, k'\Rb\,N\Lb Y,k'\Rb\,\,-\,\,N^2\Lb Y',k'\Rb\delta^{(2)}\Lb \vec{k} \,-\,\vec{k}^{\,'}\Rb\Big\}\nn
\eea
Multiplying by $N^2\Lb Y',k', b \Rb\Big{/}N\Lb Y,k,b\Rb$ we obtain the Balitsky-Kovchegov equation for $N\Lb Y',k', b \Rb$.

It is easy to check that the ansatz of \eq{ANSTZ} satisfies \eq{ADEQ}. Therefore, we can state that \eq{ANSTZ} is the solution to \eq{EQM21} and \eq{EQM22}.

 %%%%%%%%%%%%%%%%%%%%%%%%%%%%%%%%%%%%%%%%%%%%%%%%%%%%
\subsubsection{Solution to the equation  in the vicinity of the saturation scale}
%%%%%%%%%%%%%%%%%%%%%%%%%%%%%%%%%%%%%%%%%%%%%%%%%%%% 

\eq{ANSTZ} gives the general solution but,
 as we have discussed, for the evaluation of the correlation function Green's function of the BFKL Pomeron is actually needed only in the vicinity of the saturation scale (see \fig{1di} and \fig{1dienv}).  In the vicinity of the saturation scale we can neglect the non-linear terms in \eq{EQGFD} and \eq{EQGFD1} and the solution to the linear equation takes the form of \eq{RCN9}. In other words,
 \bea\label{SOL8}
 \hspace{-0.7cm}G\Lb z, z'\Rb\,\,&=&\,\,\phi_0\Lb \vec{b} - \vec{b}^{\,'}\Rb\int^{\epsilon + i \infty}_{\epsilon - i \infty}\frac{d \ga}{2 \pi i} e^{\Lb z - z'\Rb \bar{\gamma}}\,\exp\Big( \Lb \gamma - 
\bar{\gamma}\Rb \Lb  z\,-\,z'\Rb\,\,+\,\,\h\omega''_{\ga \ga}\Lb \ga=\ga_{cr},0\Rb \,Y\Lb \ga - \bar{\ga}\Rb^2\Big)\nn\\
 &=&\,\,\,\phi_0\Lb \vec{b} - \vec{b}^{\,'} \Rb e^{\Lb z - z'\Rb\bar{\gamma}}\frac{1}{\sqrt{ 2 \pi \omega''_{\ga \ga}\Lb \ga=\ga_{cr},0\Rb \,(Y - Y')}}
\exp\Big( - \frac{\Lb z - z'\Rb^2}{2\omega''_{\ga \ga}\Lb \ga=\ga_{cr},0\Rb \,(Y - Y')}\Big)\nn
\\
&&\xrightarrow{z - z' \ll\sqrt{ 2 \pi \omega''_{\ga \ga}\Lb \ga=\ga_{cr},0\Rb \,(Y - Y')}}
\,\,\frac{\phi_0\Lb \vec{b} - \vec{b}^{\,'}\Rb}{\sqrt{ 2 \pi \omega''_{\ga \ga}\Lb \ga=\ga_{cr},0\Rb \,(Y - Y')}}\Bigg(\frac{r^2_1 \,Q^2_s\Lb Y\Rb}{r^2_2 Q^2_s\Lb Y'\Rb}\Bigg)^{\bar{\ga}} \eea 
 where $Q^2_s\Lb Y\Rb \,\,=\,\,\Lb 1/R^2_p\Rb \exp\Lb \bas \frac{\chi\Lb \ga_{cr}\Rb}{\bar{\ga}} Y\Rb$ with $R_p $ is the soft scale ( the size of the typical dipole in the proton).
 
 One can see that \eq{SOL8} is very close to the solution in the toy model (see \eq{TM5}).
 The main difference in the function $\phi_0\Lb \vec{b} - \vec{b}^{\,'}\Rb$ since the factor
 $1/\sqrt{ 2 \pi \omega''_{\ga \ga}\Lb \ga=\ga_{cr},0\Rb \,(Y - Y')}$ could be absorbed into the redefinition of the variable $z$\cite{MUTR,MUPE}. The origin of this factor is in the boundary condition: $G\Lb z, z' \to 0\Rb \to N_\pom\Lb z\Rb$.  In the vicinity of the saturation scale 
  for the scattering with the nucleus $ N_\pom\Lb z\Rb$ takes the following form
  \beq \label{POMA}
    N_\pom\Lb z\Rb\,\,=\,\,\phi_0 R^2_p\,S_A\Lb b \Rb \Lb r^2 Q^2_p\Lb Y\Rb \Rb^{\bar \ga}
    \eeq
    with $S_A\Lb b\Rb$ defined in \eq{SA}.

     Therefore, $\phi_0\Lb \vec{b} - \vec{b}^{\,'}\Rb\,=\,
     \phi_0 \,R^2_p \,S_A\Lb \vec{b} - \vec{b}^{\,'}\Rb$ where $\phi_0 $ is a constant.     
     
      %%%%%%%%%%%%%%%%%%%%%%%%%%%%%%%%%%%%%%%%%%%%%%%%%%%%
 \section{ Enhanced diagrams: correlations in hadron-nucleus scattering}
 
 %%%%%%%%%%%%%%%%%%%%%%%%%%%%%%%%%%%%%%%%%%%%%%%%%%%%
 In this section we summarize the experience in the calculation of the first enhanced diagram and will discuss the A dependence of the correlation function for hadron-nucleus scattering.
 
 First, we list the main features of the enhanced diagrams which we found instructive and  which make our explicit calculation more transparent.
 \begin{enumerate}
 \item \quad Each enhanced diagram has a Pomeron loop that starts at rapidity $Y'$ with the Pomeron splitting and ends with the Pomeron merging at rapidity $Y''$ (see \fig{1di}). The first enhanced diagram has only one Pomeron loop . The calculation of the Pomeron loop in which Pomerons produce particles  (see \fig{1di}) turns out to be quite different from the calculation of the Pomeron loop in the total cross sections. The largest contributions of the Pomerons which propagate from rapidity $Y'$ to rapidity $Y''$
 stems from the kinematic region in the vicinity of the saturation scale. This feature stems from the expression for single inclusive cross section that has been discussed in section 2.2 and is a manifestation that the typical transverse momenta of the gluons in the wave function of the fast hadron is about of the saturation scale $Q_s$.
  \item \quad Integration over the transverse momentum $Q_T$ in the loop leads to the configuration with the same sizes of the dipoles at  rapidity $Y'$ and  at rapidity $ Y''$ ( in the  triple Pomeron vertices).
    \item \quad.
     In the integration over $Q_T$    the value of the typical $Q_T$ is the maximum of the saturation momenta for four Pomeron with the rapidities: $( Y, Y')$ and $(Y'', 0)$ (see \fig{1di}).
    \beq \label{TYPQT}
    Q_T \,\,\,\approx\,\,\, \mbox{max} \Big\{ Q_s\Lb Y- Y'\Rb,  Q_s\Lb Y''\Rb ,
      \Big\}
     \eeq
    
      \item \quad    As we have discussed four Pomerons in the loop being in the vicinity of the saturation scale  lead to the following rapidity dependence 
      \bea \label{RAPDEP}
&&\frac{d \sigma}{d Y_1 d Y_2 d^2 p_{\bot,1} d^2 p_{\bot,2}}\,\, \propto \\
&&~~~~~~~~~ \,\,\exp\Big(\chi\Lb \bar \ga\Rb \Lb (Y' - Y_1) + (Y_1 - Y'') + (Y  -Y_2 )+ (Y_2 - Y'')\Rb\Big)\,=\,\exp\Big(2\chi\Lb \bar \ga\Rb \Lb Y' \,-\, Y''\Rb\Big) \nn  
\eea  
   The maximal growth of the BFKL Pomerons with rapidities $Y - Y'$ and $Y''$ in the kinematic region far away of the saturation domain is given by
   \beq \label{RAPDEP1}
   \exp\Big(\chi\Lb \ga = 1/2\Rb \Lb (Y - Y') + Y"\Rb\Big) 
   \eeq
   Collecting \eq{RAPDEP} and \eq{RAPDEP1} we obtain
   \bea \label{RAPDEP2}
   &&\frac{d \sigma}{d Y_1 d Y_2 d^2 p_{\bot,1} d^2 p_{\bot,2}}\,\, \propto \\
&&~~~~~~~~~ \,\,\,\exp\Big(2\chi\Lb \bar \ga\Rb \Lb Y' \,-\, Y''\Rb  + \chi\Lb \ga=1/2\Rb\Lb (Y - Y') + Y''\Rb\Big) \nn 
\eea    
 Since $  2\chi\Lb \bar \ga\Rb -  \chi\Lb \ga=1/2\Rb\ \,>\,0$ the integrals over $Y'$ and $Y''$ are convergent leading to $Y' \to Y$ and $Y'' \to 0$. Therefore, the integration over $Y'$ and $Y''$ have the same structure as in calculation of the contribution of the enhanced diagram in the total cross section (see Ref.\cite{LMP}).
       \item \quad   For $Y' = Y$ and $Y''=0$ \eq{TYPQT} takes the form: 
       \beq \label{QTA}
       Q_T \,\,=\,\,Q_s\Lb A, Y=0\Rb.
       \eeq
        Actually we take into account  \eq{SOL8} and \eq{POMA}  of the previous section to justify this equation.
      \end{enumerate}    
       \eq{QTA}  distinguishes the hadron-nucleus scattering from the hadron-hadron interaction and we need to re-consider the integration over $R_1$ and $R_2$ in \eq{AC6} to take into  account this equation.    Re-visiting this integration we see that \eq{RCN12} takes the form
\bea \label{PAC1}
&&\int d^2 p_{\bot,1} d^2p_{\bot,2} \frac{d \sigma}{d Y_1 d Y_2 d^2 p_{\bot,1} d^2p_{\bot,2}}\,\,=\\
 &&~~~~~~~~~~~~~~~\,\,N^2_0\,\phi^4_0\frac{8\,C_F}{\alpha_s (2\pi)^4}\,\frac{8\,C_F}{\alpha_s (2\pi)^4}\,\sigma_p  \sigma_A \frac{1}{ Q_s\Lb A, Y=0\Rb}\, Q^2_s\Lb A;  Y_1\Rb 
 Q^2_s\Lb A;  Y_2\Rb\,\Lb  \frac{Q_s^2\Lb p; Y- Y_1\Rb}{Q^2_s\Lb A; Y_1\Rb}\Rb^{2\bar{\ga}}\nn  
  \eea 
where $\sigma_p$ and $\sigma_A$ are the cross sections of the interaction of the dipole with the typical size in a hadron with the hadron and nucleus, respectively. Plugging  this equation in \eq{RAPCOR}
we obtain that
\beq \label{RAPCORA}
R\Lb A; Y_1,Y_2; Y\Rb\,\,=\,\,\tilde{N}^2_0\,I_Y\Lb \bar{\gamma}\Rb\,\frac{\sigma_{in}\Lb Y\Rb}{\sigma_A}\,\frac{1}{\sigma_p\,  Q_s\Lb A, Y=0\Rb}    
\eeq
   The factor $\sigma_{in}\Lb Y\Rb/\sigma_A$ we will discuss below, but the ratio $1/\Lb\sigma_p\,  Q_s\Lb A, Y=0\Rb\Rb$ was expected as the natural estimates for the correlations (see Ref.\cite{COREST,KOLUREV} ).
   However, in the angular correlations the extra factor $Q^2_T$ appears in the integration over $ Q_T$    which leads to 
\beq \label{AC17M}
R\Lb \cos \varphi\Rb = \frac{\frac{1}{\sigma_{in}}\,\int \frac{d \sigma}{d Y_1 d Y_2 d^2 p_{\bot,1} d^2p_{\bot,2}}  d p^2_{\bot,1}   d p^2_{\bot,1}}{\frac{1}{\sigma_{in}}\,\int \frac{d \sigma}{d Y_1  d^2 p_{\bot,1}}  d p^2_{\bot,1} \, \frac{1}{\sigma_{in}}\,\int \frac{d \sigma}{d Y_2  d^2 p_{\bot,1}}  d p^2_{\bot,2}} = \cos^2\varphi\,I\Lb \bar{\gamma}\Rb\,\,\widetilde{N}^2_0\frac{\sigma_{in}\Lb Y\Rb}{\sigma_A}\frac{Q_s\Lb A, Y=0\Rb}{Q_s\Lb A, Y_1\Rb}\frac{1}{\sigma_p \,Q_s\Lb A, Y_2\Rb} \eeq
where $I\Lb \bar{\gamma}\Rb$ is given by \eq{II}.
   
   In the region of $p_{1,\bot} \sim Q_s\Lb A, Y_1\Rb$ and $p_{2,\bot} \sim Q_s\Lb A, Y_2\Rb$    the correlation function takes the form:
   \beq \label{AC17P}
  R\Lb \cos\varphi, p_{\bot,1},p_{\bot,2}\Rb\,\,=\,\,\sigma_{in}\frac{\frac{d \sigma}{d Y_1 d Y_2 d^2 p_{\bot,1} d^2p_{\bot,2}} } {\frac{d \sigma}{d Y_1 d^2 p_{\bot,1} }\,\, \frac{d \sigma}{ d Y_2 d^2p_{\bot,2}} }\,\,=\,\,
   \cos^2\varphi\,I\Lb \bar{\gamma}\Rb\,\,\widetilde{N}^2_0\frac{\sigma_{in}\Lb Y\Rb}{\sigma_A}\frac{Q_s\Lb A, Y=0\Rb}{p_{1,\bot}^2}\frac{1}{\sigma_p p^2_{2,\bot}}  
   \eeq
  
One can see that the contribution of the enhanced diagrams do not depend on  $\nabla^2 S_A\Lb b \Rb$ as in \eq{COR} and, therefore, the density variation mechanism, suggested in Ref.\cite{LERECOR}, does not lead to any suppression for the case of hadron-nucleus scattering.

  %%%%%%%%%%%%%%%%%%%%%%%%%%%%%%%%%%%%%%%%%%%%%%%%%%%
\begin{figure}[h]
\begin{center}
 \includegraphics[width=0.5\textwidth]{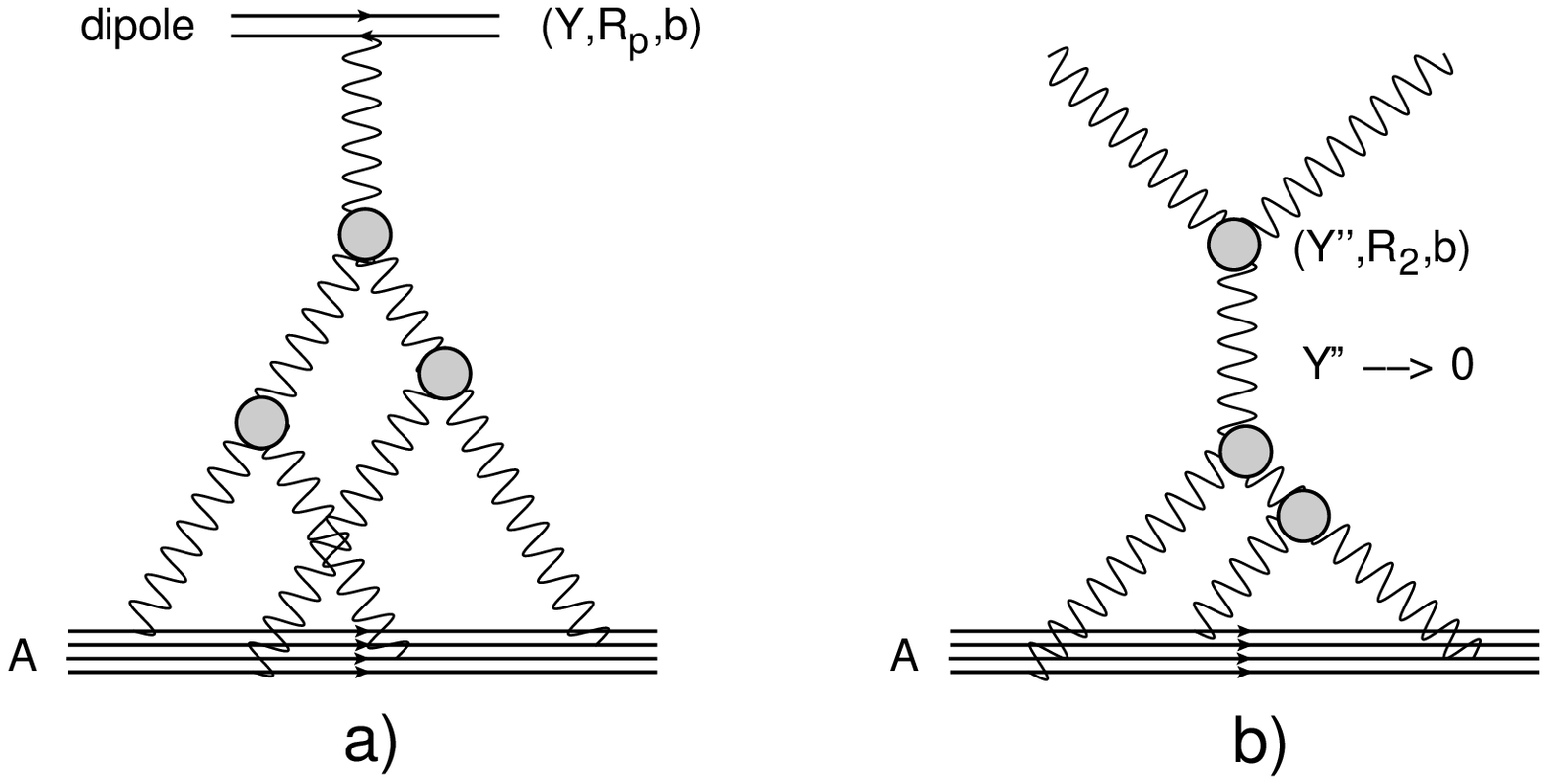}
\end{center}
\caption{ $\sigma_{in}$ (see \protect\fig{si}-a) and $\widetilde{N}_0$ (see \protect\fig{si}-b) in \protect\eq{AC17M}.}
 \label{si}
\end{figure}
  %%%%%%%%%%%%%%%%%%%%%%%%%%%%%%%%%%%%%%%%%%%%%%%%%%%
In the definition of the correlation function in \eq{AC17M} $\sigma_{in}\Lb Y - Y_0\Rb$ is the dipole-nucleus cross section shown in \fig{si}-a.  From this figure one can see that $\sigma_{in} \,=\,\int d^2 b \Lb 2 N_A\Lb r, Y  - Y_0  ; b\Rb -  N_A^2\Lb r, Y  - Y_0 ; b\Rb\Rb\,\,\propto\,\,A^{2/3}$ where $N\Lb r, Y ; b\Rb$ is the solution of the Balitsky-Kovchegov equation ( see \eq{BK} and Ref.\cite{BK}) with the initial condition given by the McLerran-Venugopalan formula \cite{MV}:
\beq \label{MVF}
N_A\Lb r, Y= Y_0; b\Rb\,\,=\,\,1\,\,-\,\,\exp \Big( - r^2 Q^2_A\Lb Y =  Y_0,b \Rb/2\Big)
~~~~~\mbox{where}~~~~ Q^2_A\Lb Y =  Y_0,b \Rb\,\,\,=\,\,\,Q^2_p\Lb Y = Y_ 0\Rb S_A\Lb b \Rb
\eeq
with $Q_p\Lb Y= Y_0\Rb$ is the saturation momentum of proton at low energy and $S_A\Lb b \Rb$ is given by \eq{SA}. It should be mentioned that we write the  McLerran-Venugopalan formula for the inelastic cross section.

On the other hand $N_0$ shown in \fig{si}-b is proportional to $\int d^2 b N_A\Lb r, Y=0; b\Rb\,\,\propto\,\,A^{2/3}$ given by \eq{MVF}.  Each factor in \eq{AC17M}
is proportional to the ratio
\beq \label{RA}
\frac{\sigma_A}{\sigma_{in}}\,=\,\frac{\int d^2 b\, N_A\Lb r, Y= Y_0; b  | \mbox{McLerran-Venugopalan formula}\Rb}{\int d^2 b  \Lb 2 N^{\mbox{\tiny BK}}_A\Lb r, Y  - Y_0  ; b\Rb -  \Lb N^{\mbox{\tiny BK}}_A\Lb r, Y  - Y_0 ; b\Rb\Rb^2\Rb}
\eeq
This ration does not depend on $A$ and it's value depends  on the the value of $Q^2_p\Lb Y = Y_ 0\Rb$  since the dominator in \eq{RA} is equal to $\pi R^2_A$ where $R_A $ is the nucleus radius.  Only at ultra high energies the dominator of \eq{RA} starts depend on  $Y$ demonstrating the Froissart - type behaviour of the interaction radius. For $Q^2_p\Lb Y = Y_ 0\Rb\,\,=\,\,0.2\,GeV^2$ we obtain $\sigma_{in}\Lb Y - Y_0\Rb/\sigma_A\,\,\approx\,\,14.5$ at $W = 5.5 TeV$ for gold.
$R^2$ in \eq{AC17M} is the size of the dipole in the proton, which is found to be about $1/GeV^2$ in  Ref.\cite{LEDD}.  It  gives  $R\Lb \cos \varphi\Rb\,\,=\,\,1.45 \cos^2 \varphi \Lb Q_s\Lb A, Y=0\Rb/ Q_s\Lb A, Y_1\Rb\Rb\,\Lb 1/\Lb\sigma_p \,Q_s\Lb A, Y_2\Rb\Rb\Rb$. In other words 
\beq \label{RAPT}
\frac{  R_A\Lb \cos\varphi, p_{\bot,1},p_{\bot,2}\Rb}{  R_p\Lb \cos\varphi, p_{\bot,1},p_{\bot,2}\Rb} \,\approx \,\,2 \frac{\sigma_p}{\sigma_{in}\Lb Y\Rb}\,R^2_p Q^2_s\Lb A,Y=0\Rb\,\,\approx \,\,\Lb 0.8 \div 1\Rb A^{1/3}~~\mbox{$\leftarrow$ at LHC energy}
\eeq
In \eq{RAPT} we denote by $R_A$ the correlation function of \eq{AC17P} and by $R_p$ the correlation function of \eq{ACPT}.

\section{Conclusions}
As it has been mentioned the main goal of this paper is to developed a more quantitative approach for the density variation mechanism suggested in Ref.\cite{LERECOR}.  Calculating the first BFKL Pomeron loop diagrams in the dense parton environment  in which the density variations have been expected to be large,  we learned the following lessons. 

First, the calculation of the enhanced diagram for the double inclusive productions shows  the same pattern as for the total cross section: $Y' \to Y$ and $Y'' \to 0$ where $Y'$ and $Y''$ are rapidities of the upper and low vertices, respectively.

Second, the correlation function generated by this diagram turns out to be large , viz,
\bea \label{CON1}
R\Lb Y_1,Y_2; Y\Rb\,\,&=&\,\,\frac{\frac{1}{\sigma_{in}(Y)}\int d^2 p_{1 ,T}\, d^2 p_{2 ,T} \frac{d^2 \sigma}{d Y_1 d^2 p_{1 ,T}\,d Y_2 d^2 p_{2 ,T} }}{
\frac{1}{\sigma_{in}(Y)}\int d^2 p_{1 ,T} \frac{d^2 \sigma}{d Y_1 d^2 p_{1 ,T}}\,\frac{1}{\sigma_{in}(Y)}\int d^2 p_{1 ,T} \frac{d^2 \sigma}{d Y_1 d^2 p_{1 ,T}}\,}\,\,-\,\,1\,\,\propto\,\,\frac{\sigma_{in}\Lb Y\Rb}{\pi R^2}    \eea
  where $R$ is the size of the typical dipole inside the proton.    
  
     In \eq{CON1} $\sigma_{in} = \sigma_{tot} - \sigma_{el} - \sigma_{sd} - \sigma_{dd}$ where $\sigma_{el},\sigma_{sd} $ and $\sigma_{dd}$ are cross section for elastic scattering and single and double diffractive production. All rapidities are denoted in \fig{1di}. This estimate looks natural since  $1/R^2$ is the typical momentum in the Pomeron loop in \fig{1di} and $\sigma_{in}$ stems from the definition of $R\Lb Y_1, Y_2, Y\Rb$. The production of the gluon jet by the Pomerons in the loop cancel with the dominators in \eq{CON1}.
     
  Third, the azimuthal correlations induced by the enhanced  Pomeron diagram, are suppressed  in comparison with the rapidity correlations   by factor $0.53/\Lb 2 (1 + 2 \gamma_{cr})\Rb^2  \approx 0.045/\Lb R^2 \,Q_s\Lb Y_1\Rb  R^2 \,Q_s\Lb Y_1\Rb\Rb$.
   
Fourth, it turns out the the typical sizes of dipoles in two triple Pomeron vertices in the Pomeron loop are the same and they are both of the order of the size of the typical dipole inside the proton $R $.
   
   The main observation of the papers that the correlations generated by the enhanced diagrams depend on the processes, in which they are measured. In particular, the rapidity correlations for the hadron-nucleus collisions take the following form:
    \beq \label{RAPCORA1}
R\Lb A; Y_1,Y_2; Y\Rb\,\propto\,\,\,\frac{1}{\sigma_p\,  Q_s\Lb A, Y=0\Rb}   \,\,\propto 1/A^{1/3} 
\eeq  
 The suppression of the angular correlations turns out to be more pronounced, viz.
 \beq \label{RATCOR}
   \frac{ R\Lb \cos\varphi; pA\Rb}{ R\Lb \cos\varphi; pp\Rb}\,\,\propto\,\,A^{- 1/3}
   \eeq
  
  In general the enhanced diagrams generate rather small angular correlation functions but the coefficient in front of $\cos^2 \varphi$ turns out to be  rather large. The smallness of the correlation function stems from the large values of the single  inclusive cross section which is proportional to the multiplicity of produced gluons. In particular, for hadron-nucleus collisions
  
  \beq \label{DICON}
  \int d^2 p_{\bot,1} d^2p_{\bot,2} \frac{d \sigma\Lb p A\Rb}{d Y_1 d Y_2 d^2 p_{\bot,1} d^2p_{\bot,2}}\,\propto\,\cos^2 \varphi\,A^{1/3}
  \eeq
  
  \eq{DICON}  is most pronounced in the ratio
  
  \beq \label{RCON}
\frac{  R_A\Lb \cos\varphi, p_{\bot,1},p_{\bot,2}\Rb}{  R_p\Lb \cos\varphi, p_{\bot,1},p_{\bot,2}\Rb} \,\,\propto\,\,A^{1/3}\eeq
         
     In general we demonstrate that the density variation mechanism suggested in Ref.\cite{LERECOR} generates sizable contributions to the coefficient in front of $\cos^2\varphi$ in the double inclusive cross sections. The value of this coefficient is proportional $A^{1/3}$ , does not include $\nabla^2 S_A(b)$ as in \eq{COR} and, therefore, density variation mechanism leads to substantial contribution for hadron-nucleus collision.

            %%%%%%%%%%%%%%%%%%%%%%%%%%%%%%%%%%%%%%%%%%%%%%%%%%%%
      \section{Acknowledgements}
  
    %%%%%%%%%%%%%%%%%%%%%%%%%%%%%%%%%%%%%%%%%%%%%%%%%%%%%%%%%
 We thank our    colleagues at UTFSM and Tel Aviv university for encouraging discussions.   This research was supported by the BSF grant 2012124  and by the  Fondecyt (Chile) grants  1140842  and 1120920.


\begin{thebibliography}{99}
\bibitem{CMSPP}
  V.~Khachatryan {\it et al.}  [CMS Collaboration],
  %``Observation of Long-Range Near-Side Angular Correlations in Proton-Proton Collisions at the LHC,''
  JHEP {\bf 1009} (2010) 091
  [arXiv:1009.4122 [hep-ex]].
  %%CITATION = ARXIV:1009.4122;%%

\bibitem{STARAA}
  J.~Adams {\it et al.}  [STAR Collaboration],
  %``Distributions of charged hadrons associated with high transverse momentum particles in pp and Au + Au collisions at s(NN)**(1/2) = 200-GeV,''
  Phys.\ Rev.\ Lett.\  {\bf 95} (2005) 152301
  [nucl-ex/0501016].
  %%CITATION = NUCL-EX/0501016;%%
\bibitem{PHOBOSAA}
  B.~Alver {\it et al.}  [PHOBOS Collaboration],
  %``High transverse momentum triggered correlations over a large pseudorapidity acceptance in Au+Au collisions at s(NN)**1/2 = 200 GeV,''
  Phys.\ Rev.\ Lett.\  {\bf 104} (2010) 062301
  [arXiv:0903.2811 [nucl-ex]].
  %%CITATION = ARXIV:0903.2811;%%
\bibitem{STARAA1}
  H.~Agakishiev {\it et al.}  [STAR Collaboration],
  %``Measurements of Dihadron Correlations Relative to the Event Plane in Au+Au Collisions at $\sqrt{s_{NN}}=200$ GeV,''
  arXiv:1010.0690 [nucl-ex].
  %%CITATION = ARXIV:1010.0690;%%
\bibitem{CMSPA}
 S.~Chatrchyan {\it et al.}  [CMS Collaboration],
  %``Observation of long-range near-side angular correlations in proton-lead collisions at the LHC,''
  Phys.\ Lett.\ B {\bf 718} (2013) 795
  [arXiv:1210.5482 [nucl-ex]].
  %%CITATION = ARXIV:1210.5482;%%
\bibitem{CMSAA}
  S.~Chatrchyan {\it et al.}  [CMS Collaboration],
  %``Centrality dependence of dihadron correlations and azimuthal anisotropy harmonics in PbPb collisions at $\sqrt{s_{NN}}=2.76$ TeV,''
  Eur.\ Phys.\ J.\ C {\bf 72} (2012) 2012
  [arXiv:1201.3158 [nucl-ex]].
  %%CITATION = ARXIV:1201.3158;%%
\bibitem{ALICE}
 A.~R.~Timmins [ALICE Collaboration],
  %``Untriggered di-hadron correlations in Pb-Pb collisions at s(NN)**(1/2) = 2.76-TeV from ALICE,''
  J.\ Phys.\ G {\bf 38} (2011) 124093;\,\,\,
  %%CITATION = JPHGB,G38,124093;%%
 A.~R.~Timmins [ALICE Collaboration],
  %``Untriggered di-hadron correlations in Pb-Pb collisions at $\sqrt{s_{NN}} =$ 2.76 TeV from ALICE,''
  arXiv:1106.6057 [nucl-ex];\,\,\,
  %%CITATION = ARXIV:1106.6057;%% A.~Adare,
  %``Triggered di-hadron correlations in Pb-Pb collisions from the ALICE experiment,''
  J.\ Phys.\ G {\bf 38} (2011) 124091
  [arXiv:1107.0285 [nucl-ex]].
  %%CITATION = ARXIV:1107.0285;%%
\bibitem{CAUSALITY}
A.~Dumitru, F.~Gelis, L.~McLerran and R.~Venugopalan,
  %``Glasma flux tubes and the near side ridge phenomenon at RHIC,''
  Nucl.\ Phys.\ A {\bf 810} (2008) 91
  [arXiv:0804.3858 [hep-ph]].
  %%CITATION = ARXIV:0804.3858;%%
\bibitem{BFKL}
 E. A. Kuraev, L. N. Lipatov, and F. S. Fadin, {\it  Sov. Phys.
JETP}
                {\bf 45}, 199 (1977); \,\,\,
Ya. Ya. Balitsky and L. N. Lipatov,
               {\it   Sov. J. Nucl. Phys.}\, {\bf 28}, 822 (1978).
\bibitem{LIREV}
L. N. Lipatov,
Phys. Rep. {\bf 286} (1997) 131; 
Sov. Phys. JETP {\bf 63} (1986) 904   [Zh.\ Eksp.\ Teor.\ Fiz.\  {\bf 90}, 1536 (1986)].



\bibitem{GLR}
L. V. Gribov, E. M. Levin and M. G. Ryskin,
Phys. Rep. {\bf 100} (1983) 1.
  %%CITATION = PRPLC,100,1;%%

\bibitem{MUQI}
A. H. Mueller and J. Qiu,
Nucl. Phys. {\bf B268} (1986) 427.

%3
\bibitem{MV}
L. McLerran and R. Venugopalan,
Phys. Rev. {\bf D49} (1994) 2233, 3352; {\bf D50} (1994) 2225;
{\bf D53} (1996) 458;\\ {\bf D59} (1999) 09400.

%4
\bibitem{MUCD}
 A.~H.~Mueller,
  %``Soft Gluons In The Infinite Momentum Wave Function And The Bfkl Pomeron,''
  Nucl.\ Phys.\  B {\bf 415}, 373 (1994);
 %``Unitarity and the BFKL pomeron,''
  Nucl.\ Phys.\  B {\bf 437} (1995) 107
  [arXiv:hep-ph/9408245].
  %%CITATION = NUPHA,B415,373;%%
  %%CITATION = NUPHA,B437,107;%

\bibitem{BK}
I.~Balitsky,
[arXiv:hep-ph/9509348];\,\,
%%CITATION = HEP-PH 9509348;%%
{\it Phys.\ Rev.} {\bf D60}, 014020 (1999)
[arXiv:hep-ph/9812311]\,\,\,\,
%%CITATION = HEP-PH 9812311;%%
Y.~V.~Kovchegov,
{\it Phys.\ Rev.}  {\bf D60}, 034008  (1999),
[arXiv:hep-ph/9901281].
%%CITATION = HEP-PH 9901281;%%
\bibitem{JIMWLK}
~J.~Jalilian-Marian, A.~Kovner, A.~Leonidov and H.~Weigert,
{\it  Phys.\ Rev.}\,  {\bf D59}, 014014 (1999),
[arXiv:hep-ph/9706377];\,\,  {\it Nucl.\ Phys.}\,{\bf B504}, 415
(1997),
[arXiv:hep-ph/9701284]; \,\,\,
J.~Jalilian-Marian, A.~Kovner and H.~Weigert,
  {\it Phys.\ Rev.}  {\bf D59}, 014015 (1999),
  [arXiv:hep-ph/9709432];\,\,\,
  %%CITATION = HEP-PH 9709432;%%
 A.~Kovner, J.~G.~Milhano and H.~Weigert,
 {\it  Phys.\ Rev.}  {\bf D62}, 114005 (2000),
  [arXiv:hep-ph/0004014]\,; \,\,\,
  %%CITATION = HEP-PH 0004014;%%
E.~Iancu, A.~Leonidov and L.~D.~McLerran,
{\it  Phys.\ Lett.}\,  {\bf B510}, 133 (2001);
[arXiv:hep-ph/0102009];\,\, {\it  Nucl.\ Phys.}\,  {\bf A692}, 583
(2001),
[arXiv:hep-ph/0011241];\,\,\,
E.~Ferreiro, E.~Iancu, A.~Leonidov and L.~McLerran,
 {\it  Nucl.\ Phys.}\  {\bf A703}, 489 (2002),
  [arXiv:hep-ph/0109115];\,\,\,
  %%CITATION = HEP-PH 0109115;%%
H.~Weigert,
{\it  Nucl.\ Phys.}  {\bf A703}, 823 (2002),
[arXiv:hep-ph/0004044].
\bibitem{REV}
Yuri V Kovchegov and Eugene Levin, {\it `` Quantum Choromodynamics at High Energies"}, Cambridge Monographs on Particle Physics, Nuclear Physics and Cosmology, Cambridge University Press, 2012 and references therein.
\bibitem{AKLL}
T.~Altinoluk, A.~Kovner, E.~Levin and M.~Lublinsky,
  {\it ``Reggeon Field Theory for Large Pomeron Loops,''}
  arXiv:1401.7431 [hep-ph] and references therein.

\bibitem{FINSTATE}
E.~V.~Shuryak,
  %``On the origin of the 'Ridge' phenomenon induced by jets in heavy ion collisions,''
  Phys.\ Rev.\ C {\bf 76} (2007) 047901
  [arXiv:0706.3531 [nucl-th]];\,\,\,S.~A.~Voloshin,
  %``Transverse radial expansion in nuclear collisions and two particle correlations,''
  Phys.\ Lett.\ B {\bf 632} (2006) 490
  [nucl-th/0312065];\,\,\,S.~Gavin, L.~McLerran and G.~Moschelli,
  %``Long Range Correlations and the Soft Ridge in Relativistic Nuclear Collisions,''
  Phys.\ Rev.\ C {\bf 79} (2009) 051902
  [arXiv:0806.4718 [nucl-th]].
  %%CITATION = ARXIV:0804.3858;%%
  %%CITATION = ARXIV:0806.4718;%%
  %%CITATION = NUCL-TH/0312065;%%
  %%CITATION = ARXIV:0706.3531;%%
\bibitem{KOVT}
A. Kovner, {\it ``How to build a ridge in  pA
collisions"}, talk at Low x WS, May 30 - June 4, 2013,Rehovot-Eilat, Israel.~~$
http://www.weizmann.ac.il/MaKaC/getFile.py/access?contribId=60\&sessionId=25\&resId=0\&materialId=slides\&confId=16$

\bibitem{DDGJLR}
  A.~Dumitru, K.~Dusling, F.~Gelis, J.~Jalilian-Marian, T.~Lappi and R.~Venugopalan,
  %``The Ridge in proton-proton collisions at the LHC,''
  Phys.\ Lett.\ B {\bf 697} (2011) 21
  [arXiv:1009.5295 [hep-ph]].
  %%CITATION = ARXIV:1009.5295;%%


\bibitem{KOLUCOR}
A.~Kovner and M.~Lublinsky,
  %``Angular Correlations in Gluon Production at High Energy,''
  Phys.\ Rev.\ D {\bf 83} (2011) 034017
  [arXiv:1012.3398 [hep-ph]].
  %%CITATION = ARXIV:1012.3398;%%

\bibitem{KOLUREV}
A.~Kovner and M.~Lublinsky,
  %``Angular and long range rapidity correlations in particle production at high energy,''
  Int.\ J.\ Mod.\ Phys.\ E {\bf 22} (2013) 1330001
  [arXiv:1211.1928 [hep-ph]].
  %%CITATION = ARXIV:1211.1928;%%


\bibitem{LERECOR}
E.~Levin and A.~H.~Rezaeian,
  %``The Ridge from the BFKL evolution and beyond,''
  Phys.\ Rev.\ D {\bf 84}, 034031 (2011)
  [arXiv:1105.3275 [hep-ph]].
  %%CITATION = ARXIV:1105.3275;%%


\bibitem{RAY}
  R.~L.~Ray,
  {\it ``Azimuthal quadrupole correlation from gluon interference in 200 GeV p+p collisions,''}
  arXiv:1406.2736 [hep-ph].
  %%CITATION = ARXIV:1406.2736;%%
\bibitem{REV1}
 W.~Li,
  %``Observation of a 'Ridge' correlation structure in high multiplicity proton-proton collisions: A brief review,''
  Mod.\ Phys.\ Lett.\ A {\bf 27} (2012) 1230018
  [arXiv:1206.0148 [nucl-ex]].
  %%CITATION = ARXIV:1206.0148;%%
\bibitem{REV2}
A.~Kovner,
  %``Particle production and angular correlations at high energy,''
  Acta Phys.\ Polon.\ B {\bf 42} (2011) 2717.
  %%CITATION = APPOA,B42,2717;%%
\bibitem{REV3}
C.~J.~Horowitz,
  %``Multi-messenger observations of neutron rich matter,''
  Int.\ J.\ Mod.\ Phys.\ E {\bf 20} (2011) 1
  [arXiv:1106.1661 [astro-ph.SR]].
  %%CITATION = ARXIV:1106.1661;%%
\bibitem{REV4}
 T.~Lappi,
  %``Small x physics and RHIC data,''
  Int.\ J.\ Mod.\ Phys.\ E {\bf 20} (2011) 1
  [arXiv:1003.1852 [hep-ph]].
  %%CITATION = ARXIV:1003.1852;%%
\bibitem{REV5}
E.~Iancu,
  %``QCD in heavy ion collisions,''
  arXiv:1205.0579 [hep-ph].
  %%CITATION = ARXIV:1205.0579;%%
\bibitem{REV6}
R.~Venugopalan,
  %``The dynamics of strongly correlated gluons at high energies,''
  PoS QNP {\bf 2012} (2012) 019
  [arXiv:1208.5731 [hep-ph]].
  %%CITATION = ARXIV:1208.5731;%%
\bibitem{MUDI}
A. H. Mueller, Phys. Rev. D 2 (1970) 2963.
\bibitem{NAPE}
~H.~Navelet ~ and ~R.~B.~ Peschanski,  Nucl. Phys. {\bf B 507}, 35 (1997) [hep-ph/9703238];\,
  Phys.\ Rev.\ Lett.\  {\bf 82} (1999) 1370
  [hep-ph/9809474];\,\,Nucl.\ Phys.\ B {\bf 634} (2002) 291
  [hep-ph/0201285].
  %%CITATION = HEP-PH/9809474;%%


\bibitem{RY}
I. Gradstein and I. Ryzhik, {\it  Table of Integrals, Series, and Products},
Fifth Edition, Academic Press, London, 1994.
\bibitem{BRN}
 M.~Braun,
  Eur.\ Phys.\ J.\ C {\bf 16}, 337 (2000)
  [hep-ph/0001268];\,\,Phys.Lett. B 483 (2000) 115, [arXiv:hep-ph/0003004]; \,\,
Eur.Phys.J C { \bf 33} (2004) 113  [arXiv:hep-ph/0309293]
;\,\,
  Phys.\ Lett.\  B {\bf 632} (2006) 297,
  [Eur.\ Phys.\ J.\  C {\bf 48} (2006) 511],
  [arXiv:hep-ph/0512057].





\bibitem{MUTR}
A.~H.~Mueller and D.~N.~Triantafyllopoulos,
{\it Nucl.\ Phys.} \, {\bf B640} (2002) 331
[arXiv:hep-ph/0205167];\,\,D.~N.~Triantafyllopoulos,
{\it Nucl.\ Phys.}\,  {\bf B648} (2003) 293
[arXiv:hep-ph/0209121].
\bibitem{MUPE}
S.~Munier and R.~B.~Peschanski,
  Phys.\ Rev.\  D {\bf 70} (2004) 077503
  [arXiv:hep-ph/0401215];\,\,
Phys.\ Rev.\  D {\bf 69} (2004) 034008
  [arXiv:hep-ph/0310357];\,\,
  Phys.\ Rev.\ Lett.\  {\bf 91} (2003) 232001
  [arXiv:hep-ph/0309177].
  %%CITATION = PRLTA,91,232001;%%
\bibitem{IIM}
 E.~Iancu, K.~Itakura, L.~McLerran,
  Nucl.\ Phys.\  {\bf A708 } (2002)  327.
  [hep-ph/0203137]
\bibitem{IIMU}
E.~Iancu, K.~Itakura and S.~Munier,
  %``Saturation and BFKL dynamics in the HERA data at small x,''
  Phys.\ Lett.\ B {\bf 590}, 199 (2004)
  [hep-ph/0310338].
  %%CITATION = HEP-PH/0310338;%%
  %301 citations counted in INSPIRE as of 19 May 2014


\bibitem{KTF}
S.~Catani, M.~Ciafaloni and F.~Hautmann,
  %``High-energy factorization and small x heavy flavor production,''
  Nucl.\ Phys.\ B {\bf 366} (1991) 135;
  %%CITATION = NUPHA,B366,135;%%
 Nucl. Phys. Proc. Suppl. 29A (1992) 182;\,\,\,J.~C.~Collins and R.~K.~Ellis,
  %``Heavy quark production in very high-energy hadron collisions,''
  Nucl.\ Phys.\ B {\bf 360} (1991) 3;\,\,\,
  %%CITATION = NUPHA,B360,3;%%
E.~M.~Levin, M.~G.~Ryskin, Y.~.M.~Shabelski and A.~G.~Shuvaev,
  %``Heavy quark production in semihard nucleon interactions,''
  Sov.\ J.\ Nucl.\ Phys.\  {\bf 53} (1991) 657
   [Yad.\ Fiz.\  {\bf 53} (1991) 1059].
  %%CITATION = SJNCA,53,657;%%

  \bibitem{KTINC}
   Y.~V.~Kovchegov and K.~Tuchin,
  %``Inclusive gluon production in DIS at high parton density,''
  Phys.\ Rev.\   {\bf D65} (2002) 074026
  [arXiv:hep-ph/0111362].
  %%CITATION = PHRVA,D65,074026;%%


\bibitem{KOJA}
J.~Jalilian-Marian and Y.~V.~Kovchegov,
  %``Inclusive two-gluon and valence quark-gluon production in DIS and pA,''
  Phys.\ Rev.\ D {\bf 70} (2004) 114017
   [Erratum-ibid.\ D {\bf 71} (2005) 079901]
  [hep-ph/0405266];\,\,
%``Saturation physics and deuteron-Gold collisions at RHIC,''
  Prog.\ Part.\ Nucl.\ Phys.\  {\bf 56} (2006) 104
  [hep-ph/0505052].
  %%CITATION = HEP-PH/0505052;%%
  %%CITATION = HEP-PH/0405266;%%
\bibitem{LMP}
 E.~Levin, J.~Miller and A.~Prygarin,
  %``Summing Pomeron loops in the dipole approach,''
  Nucl.\ Phys.\ A {\bf 806}, 245 (2008)
  [arXiv:0706.2944 [hep-ph]].
  %%CITATION = ARXIV:0706.2944;%%

\bibitem{LELU}
E.~Levin and M.~Lublinsky,
  %``A Linear evolution for nonlinear dynamics and correlations in realistic nuclei,''
  Nucl.\ Phys.\ A {\bf 730}, 191 (2004)
  [hep-ph/0308279];\,\,\,Phys.\ Lett.\ B {\bf 607} (2005) 131
  [hep-ph/0411121].
  %%CITATION = HEP-PH/0411121;%%
  %%CITATION = HEP-PH/0308279;%%
  %64 citations counted in INSPIRE as of 19 Jun 2013


\bibitem{MATH}
I. N.	Sneddon, “ Elements of partial differential equations”, Mc-Graw-Hill, New York,1957.


\bibitem{GS}
J.~Bartels, E.~Levin,
  %``Solutions to the Gribov-Levin-Ryskin equation in the nonperturbative region,''
  Nucl.\ Phys.\  {\bf B387 } (1992)  617-637;\,\,
 A.~M.~Stasto, K.~J.~Golec-Biernat, J.~Kwiecinski,
  %``Geometric scaling for the total gamma* p cross-section in the low x region,''
  Phys.\ Rev.\ Lett.\  {\bf 86 } (2001)  596-599,
  [hep-ph/0007192];\,\,\,L.~McLerran, M.~Praszalowicz,
  %``Saturation and Scaling of Multiplicity, Mean p_T and p_T Distributions from 200 GeV < sqrt{s} < 7 TeV - Addendum,''
  Acta Phys.\ Polon.\  {\bf B42 } (2011)  99,
  [arXiv:1011.3403 [hep-ph]]  {\bf B41 } (2010)  1917-1926,
  [arXiv:1006.4293 [hep-ph]].\,\,\, M.~Praszalowicz, [arXiv:1104.1777 [hep-ph]],
  %``Improved Geometrical Scaling at the LHC,''
  [arXiv:1101.0585 [hep-ph]].

\bibitem{LT}
  E.~Levin and K.~Tuchin,
  %``Solution to the evolution equation for high parton density QCD,''
  Nucl.\ Phys.\  B {\bf 573} (2000) 833
  [arXiv:hep-ph/9908317].
  %%CITATION = NUPHA,B573,833;%%

\bibitem{MUSH}
 A.~H.~Mueller and A.~I.~Shoshi,
  %``Small-x physics beyond the Kovchegov equation,''
  Nucl.\ Phys.\  B {\bf 692} (2004) 175
  [arXiv:hep-ph/0402193].
  %%CITATION = NUPHA,B692,175;%%

\bibitem{LEDD}
 E.~Levin,
  %``Dipole-dipole scattering in CGC/saturation approach at high energy: summing Pomeron loops,''
  JHEP {\bf 1311}, 039 (2013)
  [arXiv:1308.5052 [hep-ph]].
  %%CITATION = ARXIV:1308.5052;%%
\bibitem{KLN}
 D. Kharzeev and M. Nardi, Phys. Lett. {\bf B 507} (2001) 121;\,\,\,
 %%CITATION = NUCL-TH 0012025;%%
D. Kharzeev and E. Levin, Phys. Lett. {\bf B 523} (2001) 79;\,\,\,
%%CITATION = NUCL-TH 0108006;%%
D.~Kharzeev, E.~Levin and M.~Nardi,
  Phys.\ Rev.\ C {\bf 71}, 054903 (2005)
  [hep-ph/0111315];  Nucl.\ Phys.\ A {\bf 747}, 609 (2005)
  [hep-ph/0408050];\,\,\,
%%CITATION = HEP-PH 0111315;%%
D.~Kharzeev, E.~Levin and L.~McLerran,
Phys.\ Lett.\ B {\bf 561} (2003) 93;\,\,\,
  A.~Dumitru, D.~E.~Kharzeev, E.~M.~Levin and Y.~Nara,
  %``Gluon Saturation in $pA$ Collisions at the LHC: KLN Model Predictions For Hadron Multiplicities,''
  Phys.\ Rev.\ C {\bf 85}, 044920 (2012)
  [arXiv:1111.3031 [hep-ph]].
  %%CITATION = ARXIV:1111.3031;%%
  %%CITATION = HEP-PH/0408050;%%
%%CITATION = HEP-PH 0210332;%%
\bibitem{COREST}
K. ~Dusling, F. ~Gelis, T. ~ Lappi and R. ~Venugopalan, Nucl. Phys. A 836, 159 (2010),
\,\,\, F.~ Gelis, T. ~Lappi and R. ~Venugopalan, Nucl. Phys. A 830, 591C (2009).

\end{thebibliography}
 \end{document}